\documentclass[3p]{elsarticle}

\pdfoutput=1 

\usepackage[utf8]{inputenc}
\usepackage[OT4]{fontenc}
\usepackage{graphicx}
\usepackage{ifpdf}
\ifpdf
    \DeclareGraphicsExtensions{.pdf, .png}
\else
    \DeclareGraphicsExtensions{.eps}
\fi

\usepackage{enumitem}
\usepackage{textcomp}
\usepackage{amsmath}
\usepackage{amssymb}
\usepackage{amsfonts}
\usepackage{amsthm}
\usepackage{stmaryrd}
\usepackage{booktabs}
\usepackage{ifthen}
\usepackage{booktabs}
\usepackage{slashbox}
\usepackage{color}
\usepackage{tabularx}
\usepackage{multirow}
\usepackage{subfig}
\usepackage{caption}
\usepackage{array}
\usepackage[percent]{overpic}
\usepackage[unicode,
					  colorlinks,
						linkcolor={red},
						citecolor={blue},
						urlcolor={blue},
            pdfauthor={Dariusz Brzezinski, Jerzy Stefanowski, Robert Susmaga, Izabela Szczęch},
            pdftitle={Visual-Based Analysis of Classification Measures with Applications to Imbalanced Data},
            bookmarks=false,
            pdfsubject={Classification Measure Visualization and Analysis},
            pdfkeywords={classification performance measures, class imbalance, barycentric system, visualization}]{hyperref}

\usepackage{background}

\newcommand\Header{%
\noindent\parbox{\textwidth}{\textbf{\footnotesize NOTICE: This is a preliminary version of an article submitted to Information Sciences}}}

\SetBgColor{black}
\SetBgScale{1}
\SetBgOpacity{1}
\SetBgAngle{0}
\SetBgContents{%
\begin{tikzpicture}[remember picture,overlay]
\node at (0,0.6\textheight) {\Header};
\end{tikzpicture}}

\makeatletter
\newcommand\EqLabel[1]{&\refstepcounter{equation}(\theequation)\ltx@label{#1}&}
\makeatother
\newcolumntype{R}{>{\raggedleft\arraybackslash}X}
\newcolumntype{C}{>{\centering\arraybackslash}X}
\newcolumntype{V}{>{\centering\arraybackslash} m{.45\linewidth} }

\journal{Information Sciences}

\theoremstyle{plain}
\newtheorem{thm}{Theorem}

\newtheorem{prop}[thm]{Proposition}

\begin{document}
\begin{frontmatter}

\title{Visual-Based Analysis of Classification Measures with Applications to Imbalanced Data}

\author{Dariusz Brzezinski\corref{cor}}
\ead{dariusz.brzezinski@cs.put.poznan.pl}
\cortext[cor]{Tel.: +48 61 665 30 57}
\author{Jerzy Stefanowski}
\ead{jerzy.stefanowski@cs.put.poznan.pl}
\author{Robert Susmaga}
\ead{robert.susmaga@cs.put.poznan.pl}
\author{Izabela Szczęch}
\ead{izabela.szczech@cs.put.poznan.pl}

\address{Institute of Computing Science, Poznan University of Technology,\\ ul. Piotrowo 2, 60--965 Poznan, Poland}

\begin{abstract}
With a plethora of available classification performance measures, choosing the right metric for the right task requires careful thought. To make this decision in an informed manner, one should study and compare general properties of candidate measures. However, analysing measures with respect to complete ranges of their domain values is a difficult and challenging task. In this study, we attempt to support such analyses with a specialized visualization technique, which operates in a barycentric coordinate system using a 3D tetrahedron. Additionally, we adapt this technique to the context of imbalanced data and put forward a set of properties which should be taken into account when selecting a classification performance measure. As a result, we compare 22 popular measures and show important differences in their behaviour. Moreover, for parametric measures such as the \textit{F$_{\beta}$} and \textit{IBA$_\alpha$}(\textit{G-mean}), we analytically derive parameter thresholds that change measure properties. Finally, we provide an online visualization tool that can aid the analysis of complete domain ranges of performance measures.
\end{abstract}


\end{frontmatter}

\section{Introduction}
\label{sec:introduction}

Classification is one of the most important machine learning tasks, commonly applied to many real-world problems. One of the crucial ingredients of this supervised learning task is the selection of a performance measure that allows the user to discern good classifiers from bad ones. An appropriate measure should support choosing the best classifier among several candidates and help tune its parameters. As a result, the selected performance measure is responsible for the optimization of the learning process~\cite{DomingosUsefulThings}.

Although researchers often consider performance measures that promote predicting correctly the highest number of instances, many applications require other ways of handling errors referring to particular subsets of examples. This is especially true for imbalanced data~\cite{HeBook,DBLP:journals/isci/LopezFGPH13}, where classifiers are biased towards the majority classes yet the under-represented minority class is usually of more value to human experts. 

Since typical performance measures, such as classification accuracy, are not appropriate for imbalanced data~\cite{Ferri,ProvostAucAccuracy}, several more relevant measures have been considered. The most popular ones include \textit{precision}, \textit{recall} (\textit{sensitivity}), \textit{spe\-ci\-fi\-ci\-ty}, and their aggregates, e.g. \textit{G-mean} or \textit{F$_1$-score}. These and other measures for imbalanced data are typically defined on the basis of confusion matrices summarizing the predictions of a binary classifier. Looking into the related studies, one can notice that the number of such measures is relatively high and each represents different aspects of classification performance, often leading to quite different interpretations~\cite{Japkowiczbook}. This shows that there is no one measure that would be the best choice in all situations. However, which measure is used in a given problem seems to be, to a large extent, dictated simply by the measure's popularity rather than a thorough discussion of its properties.

Although there are a few systematic studies on different properties of classifier performance measures~\cite{HuD14,Gu,Flach,Sokolova-2009}, we still postulate the need for thorough analysis of the measures' behaviour. In particular, methods for: interpreting and comparing measures with respect to whole domain ranges, analysing their nature for different class and prediction distributions, and detecting the presence of unusual values are much needed. Theoretical investigations of these aspects are often very laborious and time consuming, especially when multi-dimensional aspects, provided by the confusion matrices, need to be taken into account. Due to these difficulties, such an analysis could be alternatively carried out with visual techniques to aid the understanding and interpretability of various measure properties.

In this paper, we put forward a new visualization technique for analysing entire domains of classification performance measures, which depicts all possible configurations of predictions in a confusion matrix, regardless of the used classifier. For this purpose, we adapt an approach originally created for rule interestingness measures to the context of classification~\cite{AMCS2015}. Contrary to existing performance measure visualizations, such as ROC space~\cite{Flach}, the proposed approach presents measures in a space which is defined directly on elements of the confusion matrix, is easily interpretable in 3D, and remains defined for all elements of the domain. Moreover, based on the devised visualization, we propose ten properties which should be taken into account while selecting evaluation measures, particularly for class imbalanced data. Consequently, we compare 22 popular classifier performance measures (both non-parametric and parametric) and highlight important differences in their behaviour. Finally, we demonstrate that the proposed approach can lead to concrete results by deriving property thresholds for the parametrized \textit{F$_\beta$} and \textit{IBA$_\alpha$}(\textit{G-mean}) measures.

The main contributions of our paper are as follows:
\begin{itemize}
	\item In Section~\ref{sec:visualization-technique}, we adapt a technique for visualizing classification performance measures using the barycentric coordinate system and discuss its characteristics. Additionally, we present an online tool that implements the proposed technique and allows for the analysis of several predefined and custom user-defined 4D measures.
	\item In Section~\ref{sec:desirable-properties}, we put forward ten properties, providing knowledge on the behaviour of the classifier performance measures for class biased problems. The introduced properties involve analysing maxima, minima, elements of symmetry, monotonicity, and undefined values.
	\item In Section~\ref{sec:non-parametric}, using the proposed visualization technique we analyse and compare 22 classification measures with respect to the proposed properties.
	\item In Section~\ref{sec:parametric}, we analyse how the proposed properties change for parametric measures. More precisely, we study the effect of internal parametrization on the \textit{F$_\beta$} measure and external parametrization for \textit{IBA$_\alpha$}(\textit{G-mean}). Apart from visual inspection, we analytically derive threshold parameter values for the selected measures.
	\item In Section~\ref{sec:conclusions}, we discuss the most important issues in analysing classification performance measures and draw lines of further investigations.
\end{itemize}

\section{Related Works}
\label{sec:preliminaries-related-works}

\subsection{Classifier performance measures}

Classifiers can be assessed in many aspects, such as their predictive ability, training time, memory usage, model complexity, interpretability, or other criteria~\cite{Japkowiczbook}. In this paper, we consider predictive performance only and focus on measures that evaluate crisp binary classifier predictions; measures specific to only rankers or probabilistic classifiers are out of the scope of this study. Furthermore, we concentrate mainly on measures which take into account the binary class imbalance problem.

As discussed in \cite{HeBook}, when dealing with imbalanced data  measures should focus on the more interesting minority class. Such measures are defined as functions of the confusion matrix for two-class problems, with the minority class typically referred to as \textit{positive} ($P$), while the remaining majority class as
\textit{negative} ($N$) \cite{Japkowiczbook,He_2009} (multiple non-positive classes, if present, are usually aggregated into one).

\def\arraystretch{1.4}
\begin{table}[!ht]
\caption{Confusion matrix for two-class classification}
\label{tab:cmatrix}
\centering
\begin{tabular}{|c|@{\quad}c@{\quad}c@{\quad}|@{\quad}c@{\quad}|}
\hline
\backslashbox{Actual}{Predicted}  & Positive & Negative & total\\
\hline
Positive & $\mathit{TP}$ & $\mathit{FN}$ & $P$ \\
Negative & $\mathit{FP}$ & $\mathit{TN}$ & $N$\\
\hline
  total  & $\widehat{P}$ & $\widehat{N}$ & $n$\\
\hline
\end{tabular}
\end{table}
\def\arraystretch{1}

Table \ref{tab:cmatrix} illustrates a two-class confusion matrix, which may be regarded as a special case of a contingency table that can be multi-dimensional in general. The $\mathit{TP}$ (\textit{True Positive}) and $\mathit{TN}$ (\textit{True Negative}) entries denote the number of examples classified correctly by the classifier as positive and negative, while the $\mathit{FN}$ (\textit{False Negative}) and $\mathit{FP}$ (\textit{False Positive}) indicate the number of misclassified positive and negative examples, respectively. Based
on these values, the most common performance measures are defined as:
\begin{small}
\begin{align*}
  \text{\textit{accuracy}} &= \frac{\mathit{TP}+\mathit{TN}}{\mathit{TP}+\mathit{TN}+\mathit{FP}+\mathit{FN}}\EqLabel{eq:accuracy}&
	\text{\textit{precision}} &= \frac{\mathit{TP}}{\mathit{TP}+\mathit{FP}} \EqLabel{eq:precision} \\
	\text{\textit{specificity}} &= \frac{\mathit{TN}}{\mathit{FP}+\mathit{TN}} \EqLabel{eq:specificitiy} &
  \text{\textit{sensitivity} (\textit{recall})} &=  \frac{\mathit{TP}}{\mathit{TP}+\mathit{FN}} \EqLabel{eq:recall}
\end{align*}
\end{small}

Many other classification performance measures were proposed based on values from the confusion matrix; for their reviews see~\cite{HeBook,Japkowiczbook,HuD14,Gu,Baldi2000}. In this study, we analyse the properties of 22 measures, listed and defined in the supplementary material.\footnote{\url{https://dabrze.shinyapps.io/Tetrahedron/}} Below, we highlight four measures,
 chosen for diversity of their characteristics, 
 which we will analyse and compare in more detail:
\begin{small}
\begin{align}
   \text{\textit{F$_\beta$}} &= \frac{(1+ \beta) \cdot \text{\textit{precision}} \cdot \text{\textit{recall}}}{\beta \cdot \text{\textit{precision}} + \text{\textit{recall}}}\text{, where $\beta \geq 0$,}\\
   \text{\textit{G-mean}} &=  \sqrt{\text{\textit{sensitivity}} \cdot \text{\textit{specificity}}}\text{,}  \\
	 \textit{MCC} &= \frac{\mathit{TP} \cdot \mathit{TN} - \mathit{FP} \cdot \mathit{FN}}{\sqrt{\widehat{P} \cdot P \cdot N \cdot \widehat{N}}}\text{,} \\
   \textit{OP} &= \text{\textit{accuracy}} - \frac{|\text{\textit{specificity}} - \text{\textit{sensitivity}}|}{\text{\textit{specificity}} + \text{\textit{sensitivity}}}\text{.}
\end{align}
\end{small}%

The \textit{F$_\beta$} combines \textit{precision} and \textit{recall} as a weighted harmonic mean, with the $\beta$ parameter as their relative weight. Commonly $\beta = 1$ and then the measure is referred to as \textit{F$_1$-score}. \textit{G-mean}~\cite{Kubat} is the geometric mean of \textit{sensitivity} and \textit{specificity}, which takes into account the relative balance of recognition of both positive and negative classes. The \textit{Matthews Correlation Coefficient} (\textit{MCC}) expresses a correlation between the actual and predicted classification and returns a value between $-1$ (total disagreement) and $+1$ (perfect classification). 
We highlight \textit{MCC} in our study as it was considered by some authors as one of the recommended measures for imbalanced data~\cite{Baldi2000,Bekkar}. \textit{Optimized precision} (\textit{OP}) combines \textit{sensitivity} and \textit{specificity} in a more complex way, also producing values in the $[-1,+1]$ range~\cite{RP_06}. 

Apart from these ``closed-formula'' measures, we shall also analyze in more detail a representative of what may be thought of as ``open-formula'' measures, in this case $\textit{IBA$_\alpha$}(M)$. This particular measure-wrapper is aimed at applying more weight to minority class predictions in a given measure $M$, according to a user-defined parameter $\alpha$~\cite{Garcia09,Garcia10}.

These and other measures were compared in such surveys as e.g. \cite{HeBook,Gu,He_2009,Baldi2000}, however usually with respect
to discussing the main differences in their definitions. Additionally, the \textit{F$_1$-score} was thoroughly analysed by Powers~\cite{Powers}
who claimed that some of its properties, such as focusing only on the minority class and assuming that actual and predicted distributions are identical, may be critical flaws. Another theoretical study showed that aggregating \textit{sensitivity} and \textit{specificity} 
presented more suitable behaviour than measures aggregating \textit{precision} and \textit{recall}~\cite{HuD14}. Nevertheless, theoretical analyses of measures with respect to complete ranges of domain values are very laborious and have been done only for a few classifier performance measures.

\subsection{Visualization of measures}

In this paper, we focus on visualizing measures defined on a binary confusion matrix. We note that this should not be confused with visualizations of classifier performance, e.g. using ROC graphs~\cite{DBLP:journals/prl/Fawcett06}, precision-recall curves~\cite{DBLP:conf/icml/DavisG06}, lift charts~\cite{DBLP:conf/kdd/Piatetsky-ShapiroM99}, or other attempts to graphically present experimental comparisons of classifiers~\cite{DBLP:journals/ida/VanderlooySSH09,DBLP:conf/pkdd/Alaiz-RodriguezJT08,Caruana:2004}. Our intention is to study general properties of measures rather than visualize the predictive performance of a classifier on a given dataset.

The 3D visualizations of $2 \times 2$ sum-constrained matrices, applicable in particular to confusion matrices, have already been considered in different papers.
Below, we recapitulate shortly three approaches, which bear some relation to the (regular) tetrahedron visualization used throughout this paper~\cite{LeBras2012,Celotto17,Flach}.


Le Bras et al.~\cite{LeBras2012} introduce a system of 3D spaces (referred to as the \textit{Formal Framework}), in which the contents of sum-constrained $2 \times 2$ matrices can be represented. Because of the three actual degrees of freedom of a sum-constrained $2 \times 2$ matrix, domains consisting of three variables are required and sufficient to express the matrix entries. However, the choice of a particular domain, with three particular variables, may vary depending on the application at hand. 

While the representations with three variables might be used to produce 3D visualizations of measures, the paper of Le Bras et al.~\cite{LeBras2012} does not exploit this fact in too much a detail, as its focus lies elsewhere. The authors introduce three very particular, application-driven, 3D domains referred to as: \textit{confidence}, \textit{examples} and \textit{counterexamples}. 
In its central part, the paper recalls 38 measures related to association rules and defines them consistently in terms of the matrix entries, as well as in terms of the three proposed domains. This allows for conducting dedicated analyses of the measures (e.g. expressing the Piatetsky-Shapiro recommendations \cite{Piatetsky91} in the \textit{examples} domain), with the main objective of identifying measures most relevant to association rule pruning. The introduced and in detail scrutinized properties include: all-monotonicity, generalized universal existential upward closure, and opti-monotonicity~\cite{LeBras2012}.

As far as the tetrahedron-based visualization is concerned, the \textit{examples} and \textit{counterexamples} 3D spaces introduced in \cite{LeBras2012} assume the shapes of tetrahedra. However, contrary to the approach presented in our paper, the domains are designed for analysing rule interestingness measures. Moreover, the tetrahedra of the \textit{examples} and \textit{counterexamples} domains are irregular, since these domains are assumed to have two orthogonal variables each, implying shapes with two orthogonal edges incident with one vertex, a feature unattainable in the regular tetrahedron.


Celotto~\cite{Celotto17} has introduced 2D visualization spaces that are very natural to the considered measures, i.e., Bayesian confirmation measures. The primary space, suitably referred to as the \textit{confirmation space}, consists of: $P(H|E)$ ($x$-axis) and $P(H)$ ($y$-axis).
As noted within the paper, the 2D representation of $2 \times 2$ sum-constrained matrices is incomplete, and thus aptly called a \textit{fingerprint} of the measure. The incompleteness results from the fact that the fingerprint changes as some third parameter which defines the third dimension, in this case chosen to be $P(E)$, is varied.

The confirmation space is initially set side by side with its analogue, denoted as \textit{dual confirmation space}, and another 2D space, i.e. the ROC space, which consists of false positive rate $\mathit{fpr} = \mathit{FP}/N$ ($x$-axis) and true positive rate $\mathit{tpr} = \mathit{TP}/P$ ($y$-axis). However, because confirmation measures remain the main focus of the study of Celotto~\cite{Celotto17}, presented analyses are basically confined to the confirmation space and its dual, which are used to analyse 19 measures. The measure analyses, principally concerned with identifying measures most relevant to classification rule pruning, include visualizations of some ordinal equivalence aspects and a multitude of symmetry aspects. The latter also include visually-assisted design and synthesis of measures possessing desired symmetries.

The 2D confirmation spaces introduced by Celotto correspond to rectangular cross-sections of the 3D tetrahedron presented in this paper. However, contrary to the presented approach, in~\cite{Celotto17} these originally non-independent variables are presented as orthogonal and of unified ranges, which thus requires some amount of orthonormalization.


Flach~\cite{Flach} mentions several possible definitions of variables suitable for 3D visualizations of $2 \times 2$ confusion matrices, but focuses primarily on \textit{3D ROC space}, a generalization of traditional 2D ROC space~\cite{Japkowiczbook}. The 3D ROC space consists of the false positive rate $\mathit{fpr} = \mathit{FP}/N$ ($x$-axis) and true positive rate $\mathit{tpr} = \mathit{TP}/P$ ($y$-axis), which basically constitute traditional ROC space, together with the frequency of positives $\mathit{pos} = P/n$ ($z$-axis). 
This choice had been dictated by the general topic of the paper, which was the analysis of classifier performance measures and their behaviour in ROC spaces. Notice that the three variables are selected so that the resulting $XY$-plane hosts the ROC space,
while the third co-ordinate varies with the actual class distribution. In result, the 3D ROC space is thus a collection of stacked-up ROC spaces, with the $z$-coordinate 
 corresponding to the proportion of the positive class. 
Owing to the variable mutual orthogonality and similar ranges ($[0,1]$ for $x$ and $y$ and $(0,1)$ for $z$) the total domain shape is thus a $[0,1] \times [0,1] \times (0,1)$ pseudo-cube, i.e. a cube with both the lowermost layer, corresponding to $FN+TP = 0$, and the uppermost layer, corresponding to $FP+TN = 0$, removed.

In the cited study~\cite{Flach}, Flach combines the proportion of classes with misclassification costs, generally referred to as skew, and focuses on analysing 8 selected measures in terms of sensitivity to skew. The considered key notions involve: skew-equivalence and weak/strong skew-insensitivity of the measures. We also note that a similar techniques have been used to analyse rule quality measures. The most well known are coverage spaces, introduced by Fürnkrantz and Flach \cite{DBLP:conf/icml/FurnkranzF03}, which plot the number of positive training examples and negative ones covered by the rule in the given data. Coverage spaces can be considered similar to ROC spaces in analysing isometrics of evaluation measures.

The stacked 2D spaces considered by Flach~\cite{Flach} basically correspond to the rectangle-shaped cross-sections of the tetrahedron presented in this paper. However, ROC spaces are presented in square form, which requires some amount of orthogonal rescaling compared to the approach presented in this paper. Furthermore, contrary to the visualization technique introduced in this paper, 3D ROC space remains undefined for confusion matrices with $FN+TP = 0$ or $FP+TN = 0$.

\section{The barycentric visualization technique}
\label{sec:visualization-technique}

As presented in Table~\ref{tab:cmatrix}, a confusion matrix for binary classification consists of four entries: $\mathit{TP}$, $\mathit{FP}$, $\mathit{FN}$, $\mathit{TN}$. However, for a dataset of $n$ examples these four entries are constrained, as $n = \mathit{TP} + \mathit{FP} + \mathit{FN} + \mathit{TN}$. Therefore, for a given constant $n$, any three values in the confusion matrix uniquely define the fourth value. This property allows to visualize any classification performance measure based on the two-class confusion matrix using a 4D barycentric coordinate system \cite{W_03}.

\begin{figure}[!ht]
 \centering
  \includegraphics[width=0.38\textwidth]{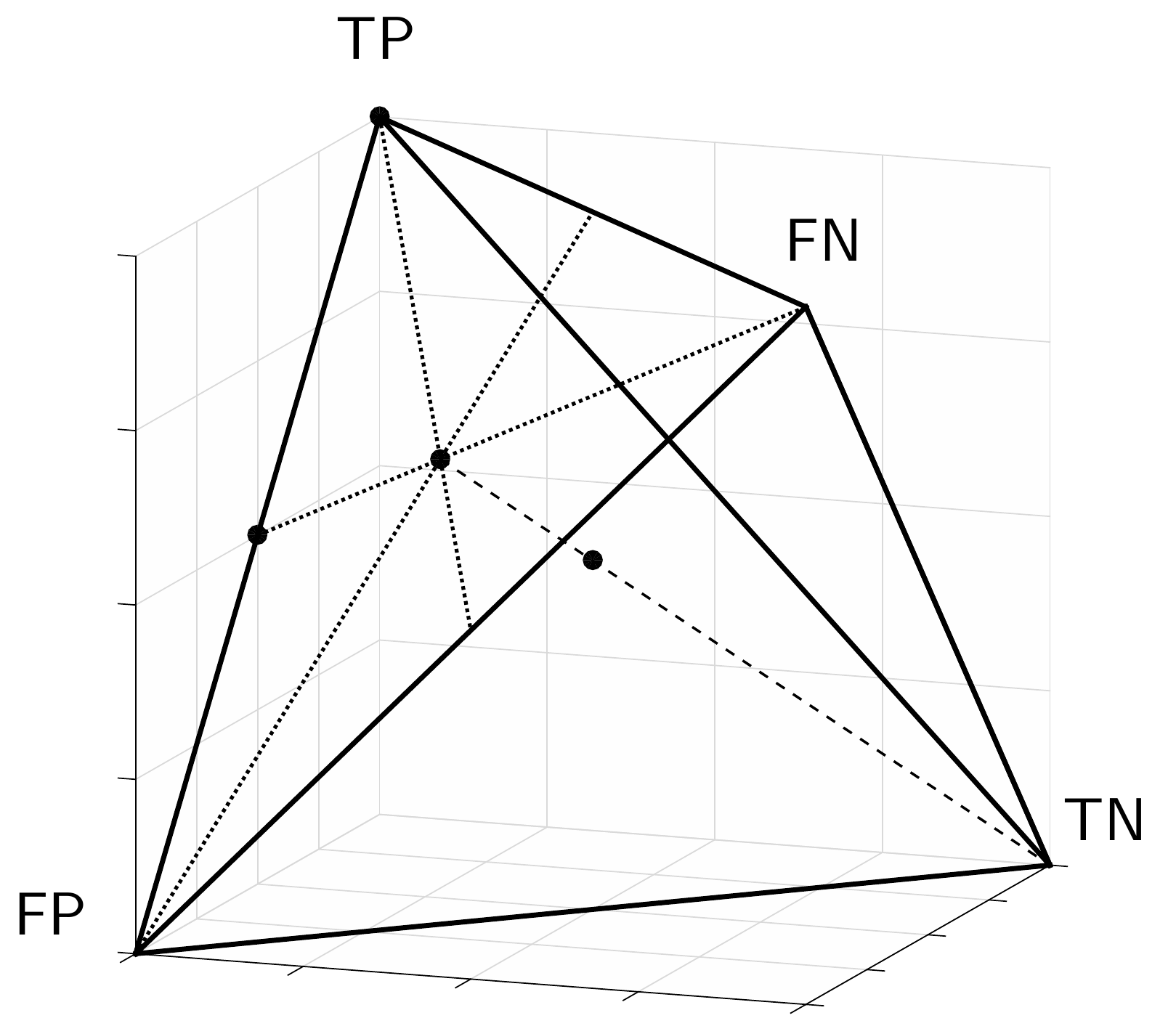}
  \caption{A skeleton visualization of the tetrahedron with four exemplary points}
  \label{fig:visualization-simplex-skeleton-points}
\end{figure}

In the \textit{barycentric coordinate system} point locations are specified relatively to vertices of a simplex (a triangle, tetrahedron, etc.). A 4D barycentric coordinate system is a tetrahedron, where each dimension is represented as one of the four vertices. Choosing vectors that represent $\mathit{TP}$, $\mathit{FP}$, $\mathit{FN}$, $\mathit{TN}$ as vertices of a regular tetrahedron in a 3D space, one arrives at a barycentric coordinate system as in Fig.~\ref{fig:visualization-simplex-skeleton-points}.

In this system, every confusion matrix $\left[ \begin{smallmatrix} \mathit{TP} & \mathit{FN} \\ \mathit{FP} & \mathit{TN} \end{smallmatrix} \right]$ is represented as a point of the tetrahedron. Let us illustrate this fact with a few examples. Figure~\ref{fig:visualization-simplex-skeleton-points} shows a skeleton of a tetrahedron with 4 exemplary points:
\begin{itemize}[noitemsep]
	\item one located in vertex $\mathsf{TP}$, which represents $\left[ \begin{smallmatrix} n & 0 \\ 0 & 0 \end{smallmatrix} \right]$,
	\item one located in the middle of edge $\mathsf{TP}$--$\mathsf{FP}$, which represents $\left[ \begin{smallmatrix} n/2 & 0 \\ n/2 & 0 \end{smallmatrix} \right]$,
	\item one located in the middle of face $\triangle\mathsf{TP}$--$\mathsf{FP}$--$\mathsf{FN}$, which represents $\left[ \begin{smallmatrix} n/3 & n/3 \\ n/3 & 0 \end{smallmatrix} \right]$,
	\item one located in the middle of the tetrahedron, which represents $\left[ \begin{smallmatrix} n/4 & n/4 \\ n/4 & n/4 \end{smallmatrix} \right]$.
\end{itemize}
One way of understanding this representation is to imagine a point in the tetrahedron as the centre of mass of the examples in a confusion matrix. If all $n$ examples are true positives, then the entire mass of the predictions is at $\mathit{TP}$ and the point coincides with vertex $\mathsf{TP}$. If all examples are false negatives, the point lies on vertex $\mathsf{FN}$, etc. Generally, whenever $a > b$ ($a, b \in \{\mathit{TP},\mathit{FN},\mathit{FP},\mathit{TN}\}$) then the point is closer to the vertex corresponding to $a$ rather than $b$.

\begin{figure}[!ht]
	\centering
		\includegraphics[width=0.6\textwidth]{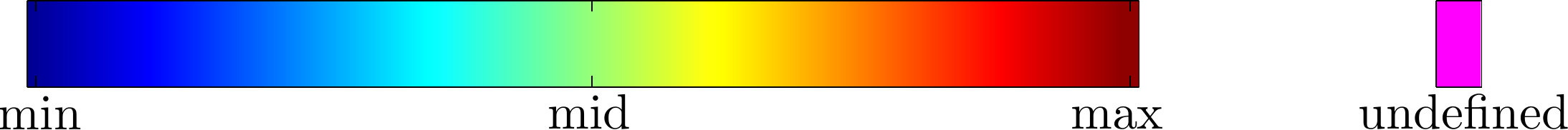}
	\caption{The color map}
	\label{fig:color-map}
\end{figure}

Using the barycentric coordinate system makes it possible to depict the originally 4D data (two-class confusion matrices) as points in 3D. Moreover, as in \cite{AMCS2015,BullPass2015}, an additional variable based on the depicted four values may be rendered as colour. 
Although any colour map can be used, in the following paragraphs we utilize 
the map shown in Fig.~\ref{fig:color-map}: dark blue --- minimum values, to dark brown --- maximum values. Areas of the same colour signify then the same values of the variable. The shape of such areas is determined by the nature of the visualized variable and usually occurs as lines in 2D (isoliness) and surfaces in 3D (isosurfaces)
Undefined values of the measures will be rendered in magenta, i.e., a colour not occurring in the map.

Here, we adapt this procedure to colour-code the values of classification performance measures, which remain the principal focus of this paper. In this respect, the presented approach is different from \cite{AMCS2015} and  \cite{BullPass2015}, in which Bayesian confirmation measures were mainly addressed. In particular, this paper introduces and discusses those aspects of the tetrahedron-based visualization that are especially useful for the analysis of classification performance measures.

The described visualization technique has been implemented as an interactive web application, available at: \url{https://dabrze.shinyapps.io/Tetrahedron/}. The application can visualize 86 predefined 4D measures, including the 22 classification performance measures described further. The user can also visualize custom measures by providing their formulae. For the remainder of the paper, the reader is encouraged to use this tool to interactively analyse the described properties of various classification measures.
 
\begin{figure}[!ht]
  \centering
	\subfloat[External view]{
		\centering
		\includegraphics[width=0.25\textwidth]{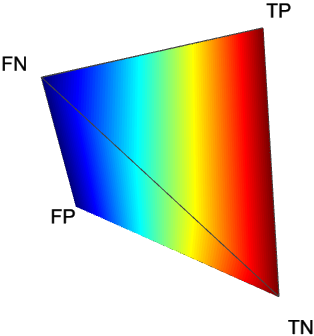}
		\label{fig:CA-external}}
	\quad\quad
	\subfloat[Point padding]{
		\centering
			\includegraphics[width=0.25\textwidth]{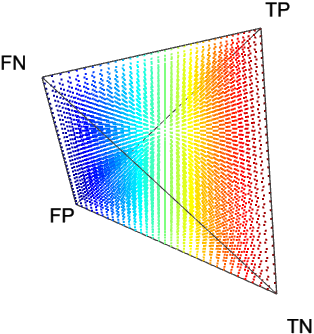}
		\label{fig:CA-padding}}
	\quad\quad
		\subfloat[Internal view]{
		\centering
			\includegraphics[width=0.25\textwidth]{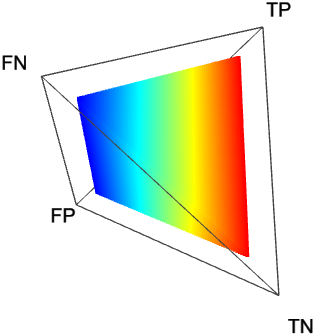}
		\label{fig:CA-internal}}
  \caption{Visualizations of classification \textit{accuracy}}
  \label{fig:CA-visualizations}
\end{figure}

\begin{figure}[!ht]
  \centering
	\subfloat[External view]{
		\centering
		\includegraphics[width=0.3\textwidth]{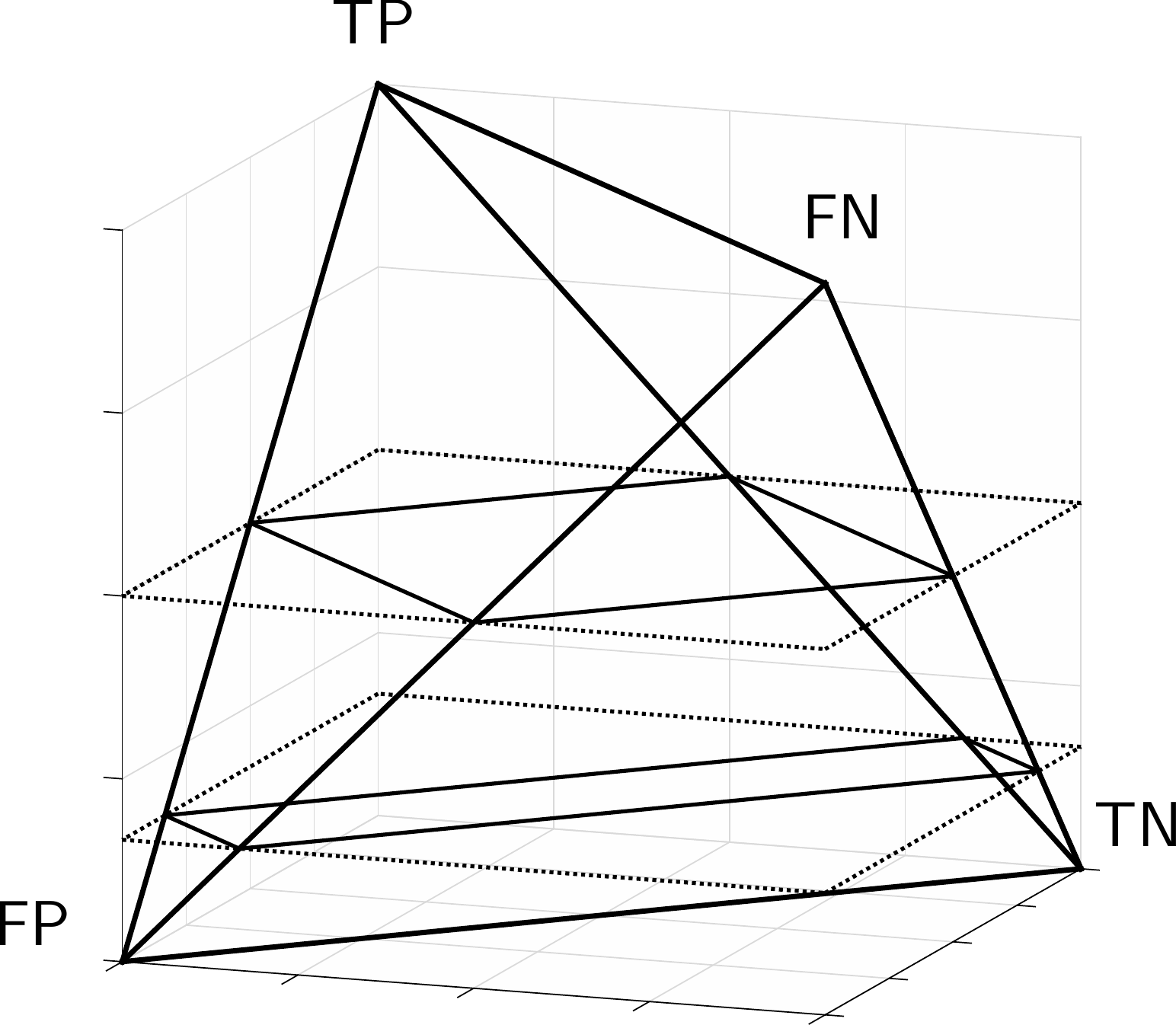}
		\label{fig:cross-sections-skeleton}}
	\subfloat[$P/n = 1/6$]{
		\centering
			\includegraphics[width=0.3\textwidth]{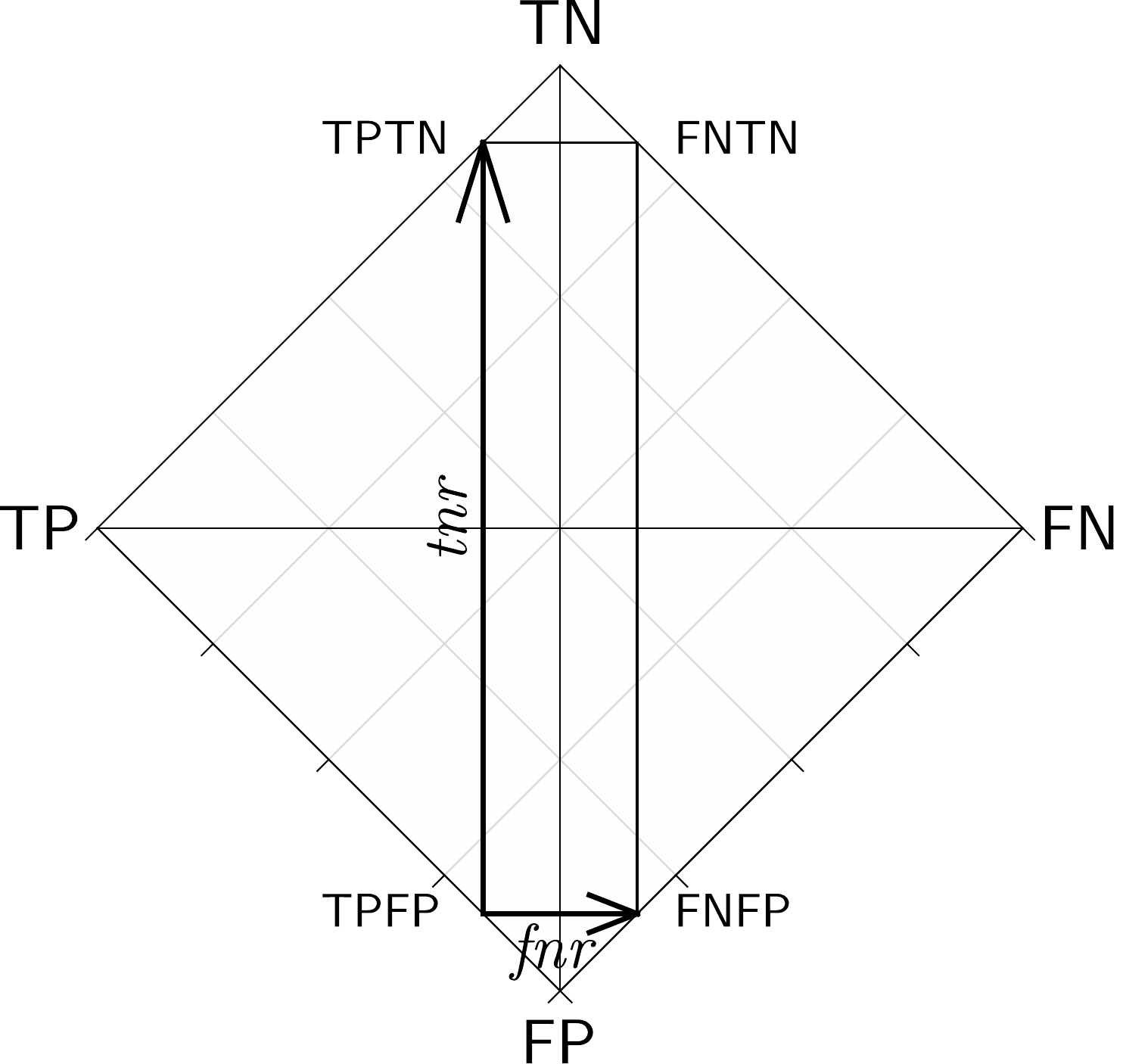}
		\label{fig:cross-sections-16}}
		\subfloat[$P/n = 1/2$]{
		\centering
			\includegraphics[width=0.3\textwidth]{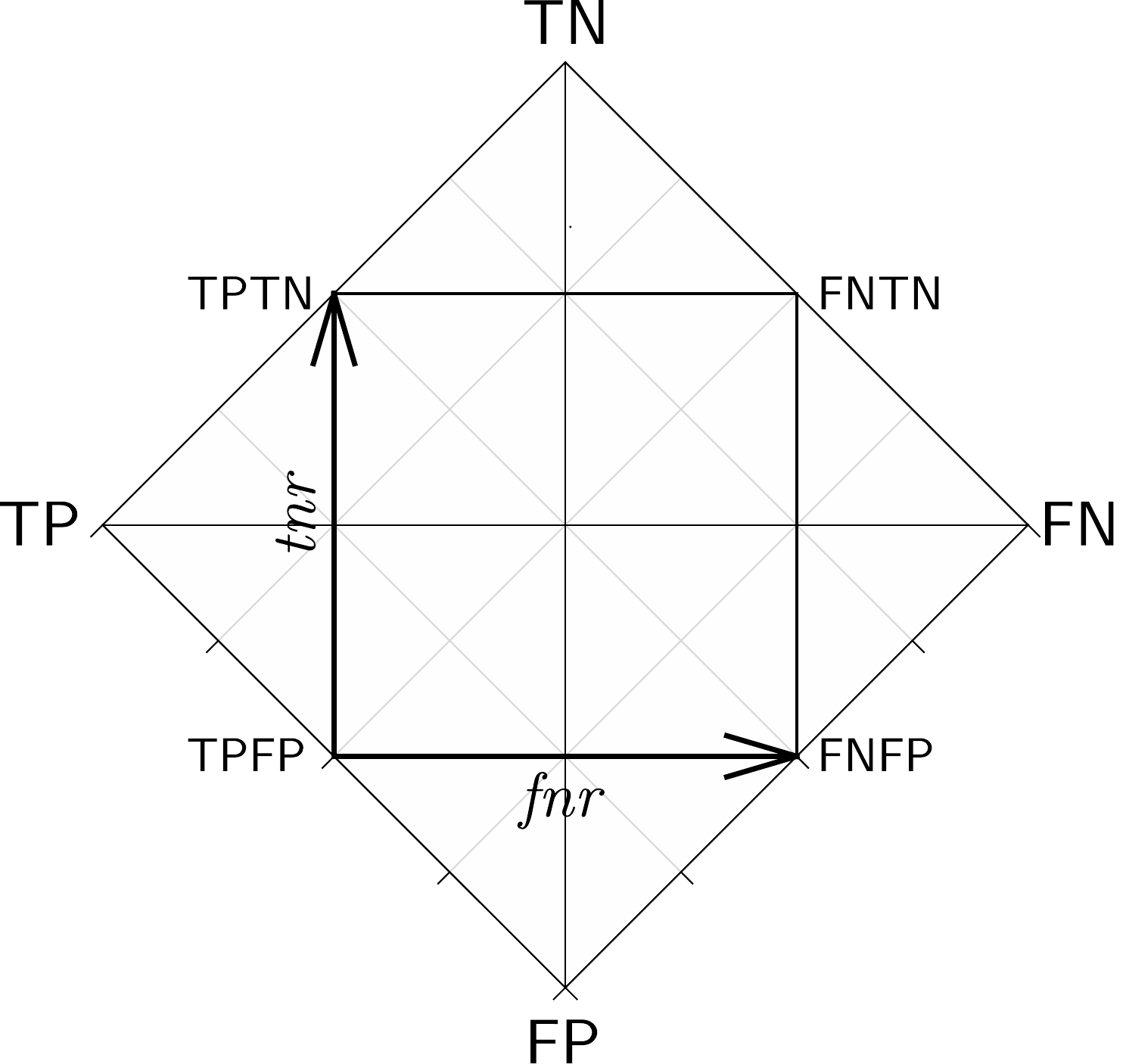}
		\label{fig:cross-sections-12}}
	\caption{Skeleton visualizations of the tetrahedron and top-down view depictions of rectangular cross-sections for two selected values of positive class rate ($P/n$)}%
	\label{fig:cross-sections}
\end{figure}

Since classification \textit{accuracy} is one of the simplest and most often used performance measures, let us use it for an exemplary visualization in Fig.~\ref{fig:CA-visualizations}. Its values range from $0$ to $1$, and there are no undefined ones. One can notice that confusion matrices with a high number of $\mathit{FP}$ and $\mathit{FN}$ result in low \textit{accuracy}, whereas high $\mathit{TP}$ and $\mathit{TN}$ yield high \textit{accuracy}. The visualization in Fig.~\ref{fig:CA-external} is only partially comprehensive, as it only shows the externals of the tetrahedron which correspond to very specific confusion matrices. However, both external as well as internal areas can be shown, e.g. by padding tetrahedron points (Fig.~\ref{fig:CA-padding}), using ``under the skin'' views (Fig.~\ref{fig:CA-internal}) or performing cross-sections (Fig.~\ref{fig:cross-sections}).

The indicated cross-sections are of particular interest in the context of analysing measures for class imbalance problems. Notice that traversing the tetrahedron alongside the vertical axis (up-down in Fig.~\ref{fig:cross-sections-skeleton}) corresponds to changing the proportions between sums $\mathit{TP}+\mathit{FN} = P$ and $\mathit{FP}+\mathit{TN} = N$, which specify the cardinalities of the actual classes. If $P = N$, then a situation of balanced classes is reproduced; otherwise the classes are imbalanced.

How a measure behaves for a particular class proportion may be visualized by producing a cross-section of the tetrahedron with a horizontal plane that cuts its vertical height. Figures~\ref{fig:cross-sections-16}~and~\ref{fig:cross-sections-12} show the two cross-sections visible in Fig.~\ref{fig:cross-sections-skeleton}), one at $P/n = 1/6$ (positive class as the minority class) and one at $P/n = 1/2$ (class balance), as seen from above the tetrahedron. Cutting the shape with a horizontal plane at $P/n = 1/6$ produces the lower, rectangular cross-section (\ref{fig:cross-sections-16}), while at $P/n = 1/2$ --- the upper, square one (\ref{fig:cross-sections-12}). In their corresponding figures, the cross-sections are oriented so that their sides incident with face $\triangle\mathsf{TP}$--$\mathsf{FP}$--$\mathsf{FN}$ of the tetrahedron are positioned at the bottom, while those incident with face $\triangle\mathsf{TP}$--$\mathsf{FN}$--$\mathsf{TN}$ --- at the top.
It is additionally worth noting that at every section the proportion of the rectangle's side lengths follows that of $P$ (the horizontal side) and $N$ (the vertical side), i.e. the class cardinalities.

Accordingly to the notation of the vertices of the tetrahedron, the sides and vertices of a cross-section rectangle are labelled as follows: 
\begin{itemize}
	\item sides: $\overline{\mathsf{TP}}$ (left), $\overline{\mathsf{TN}}$ (upper), $\overline{\mathsf{FN}}$ (right), $\overline{\mathsf{FP}}$ (lower),
	\item vertices: $\mathsf{TPTN}$ (upper-left), $\mathsf{FNTN}$  (upper-right), $\mathsf{FNFP}$ (lower-right), $\mathsf{TPFP}$ (lower-left).
\end{itemize}

The two axes, $\mathit{fnr}$ and $\mathit{tnr}$, of the 2D space in which all cross-sections are represented (including those for $P/n = 1/6$ and $P/n = 1/2$), correspond to the false negative rate, 
$\mathit{fnr} = \frac{\mathit{FN}}{\mathit{FN}+\mathit{TP}} = 1 -$ \textit{recall}, 
and the true negative rate, 
$\mathit{tnr} = \frac{\mathit{TN}}{\mathit{TN}+\mathit{FP}} =$ \textit{specificity}. 
The orientation of the axes results from the fact that traversing the rectangle left-to-right corresponds to increasing $\mathit{fnr}$ from $0$ to $1$, whereas traversing the rectangle down-up corresponds to increasing $\mathit{tnr}$ from $0$ to $1$. 
The resulting 2D space of the presented cross-section is thus an analogue of 2D ROC space, where, somewhat reversely, the false positive rate, 
$\mathit{fpr} = \frac{\mathit{FP}}{\mathit{FP}+\mathit{TN}} = 1 -$ \textit{specificity},
and the true positive rate,
$\mathit{tpr} = \frac{\mathit{TP}}{\mathit{FP}+\mathit{TP}} =$ \textit{recall},
are used as $x$ and $y$ axes, respectively. 

The presented rectangular cross-sections and 2D ROC space constitute the same, though seen from different angles, cross-sections of the tetrahedron. However, contrary to 3D ROC space~\cite{Flach}, the presented technique does not involve any non-linear transformations of the elements of the confusion matrix and remains defined for all elements of the domain. Furthermore, because the proposed barycentric coordinates directly correspond to elements of the confusion matrix, the visualization is easily interpretable also in 3D, which helps analysing the whole range of possible domain values. 

In the following sections, we demonstrate the usage of the visualization technique in some analyses of the considered classifier performance measures for imbalanced data. The technique, including the cross-sections, was particularly used to visualize several postulated properties of the measures.

\section{Properties of Measures for Imbalanced Data}
\label{sec:desirable-properties}

With a visualization technique at hand, it is much easier to define and interpret potentially desirable measure properties. In this section we put forward and discuss ten properties designed to highlight characteristic features of classifier performance measures designed for imbalanced data. The proposed properties can aid researchers in the selection of measures suitable for a given context and raise much needed discussion on the applicability of measures in certain domains.

\begin{figure}[!htb]
		\centering
	\subfloat[External view]{
		\centering
		\includegraphics[width=0.3\textwidth]{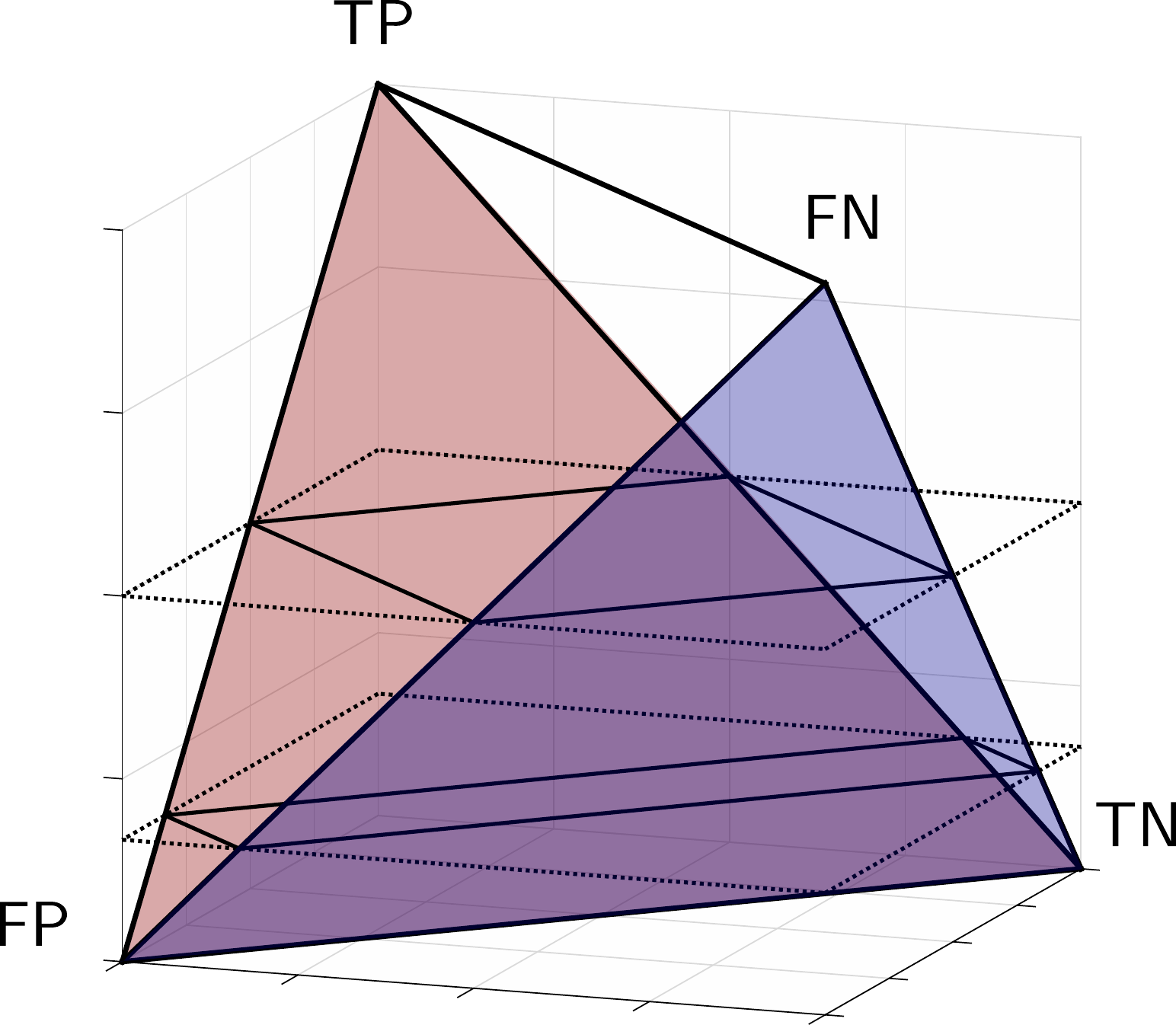}
		\label{fig:skeleton-with-planes-(P)}}
	\subfloat[$P/n = 1/6$]{
		\centering
			\includegraphics[width=0.3\textwidth]{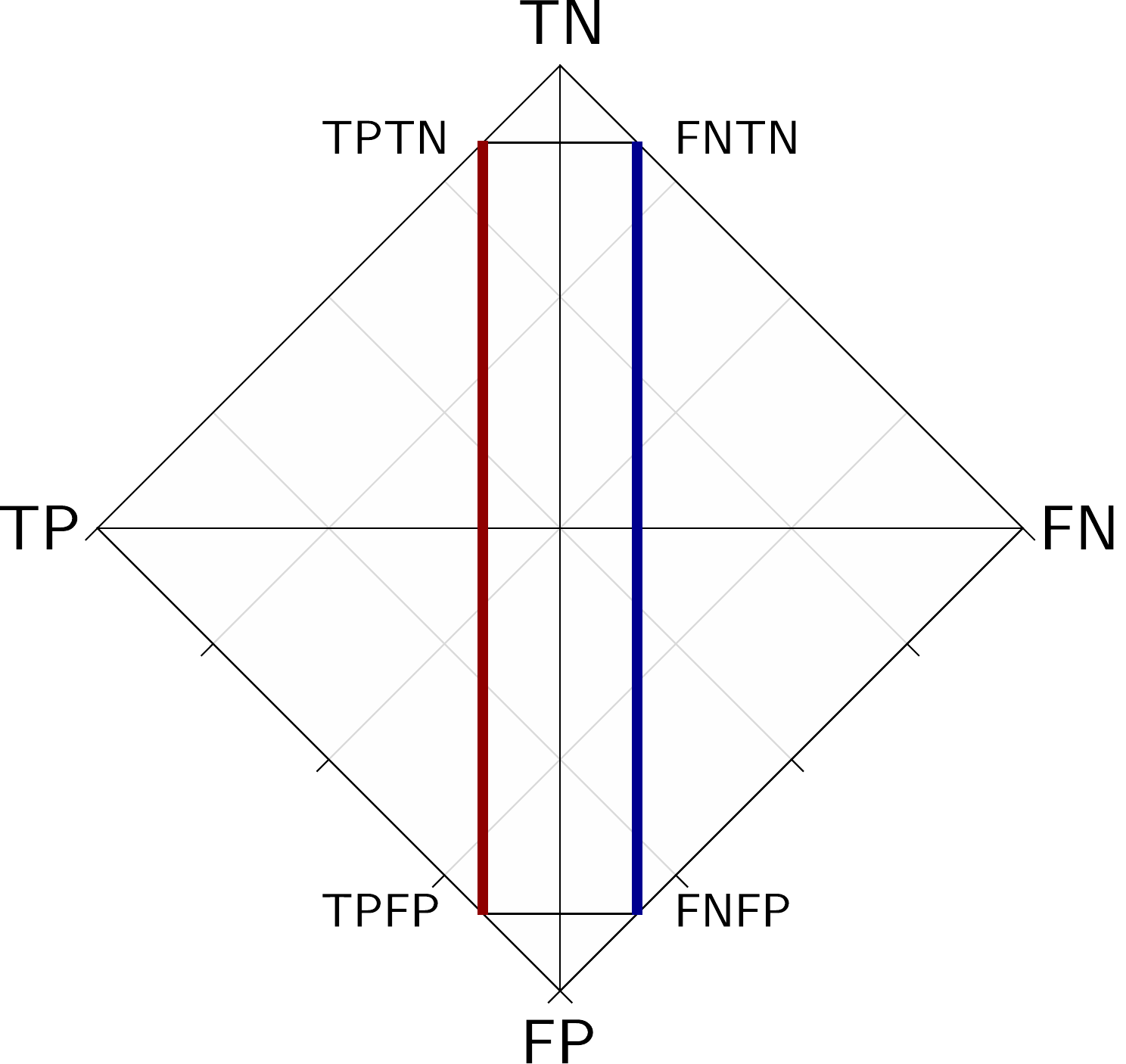}
		\label{fig:cross-sections-16-(P)}}
		\subfloat[$P/n = 1/2$]{
		\centering
			\includegraphics[width=0.3\textwidth]{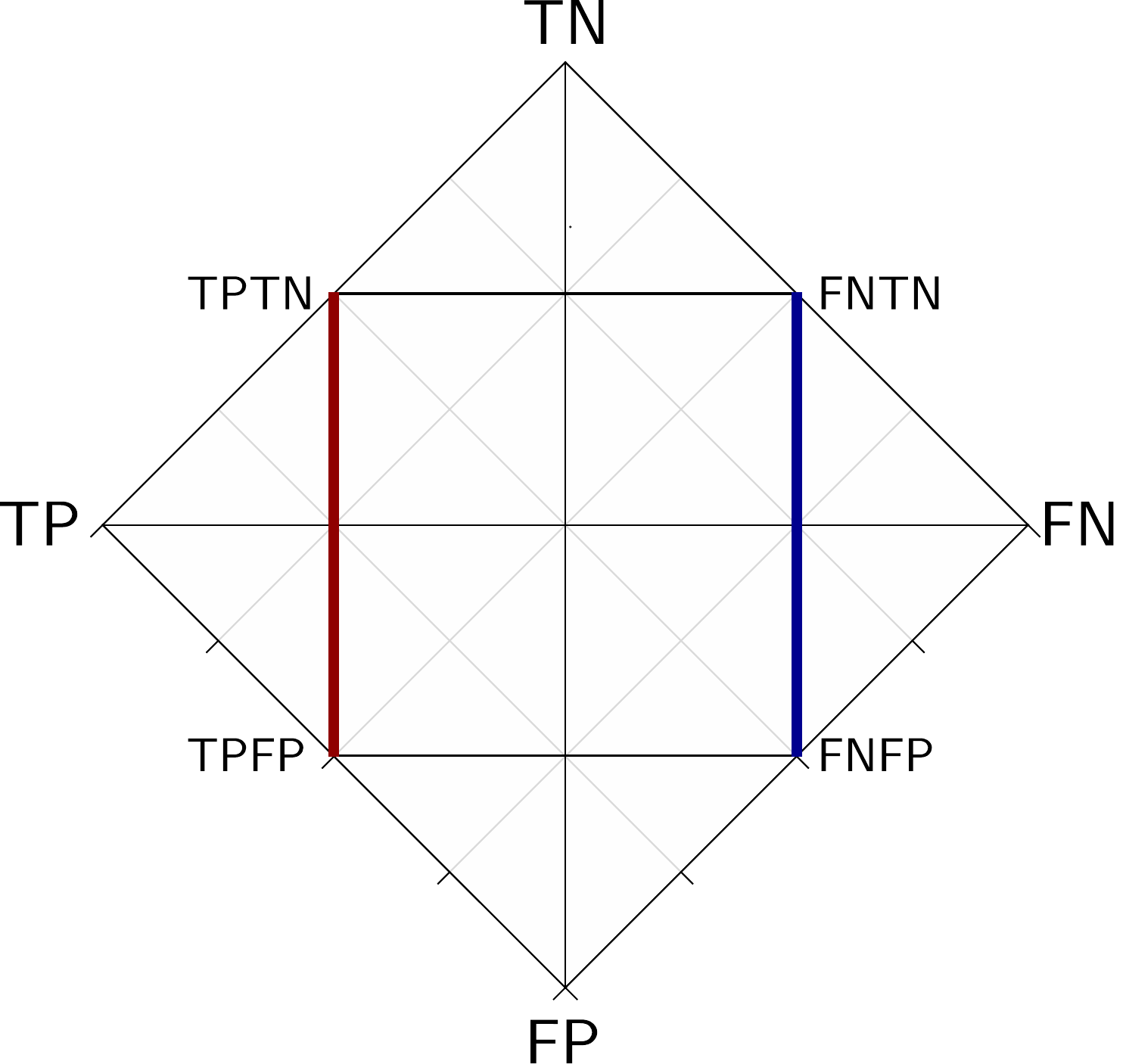}
		\label{fig:cross-sections-12-(P)}}
	\caption{Illustration of full/null recognition of the positive class}%
	\label{fig:cross-sections-(P)}
\end{figure}

\begin{figure}[!htb]
		\centering
	\subfloat[External view]{
		\centering
		\includegraphics[width=0.3\textwidth]{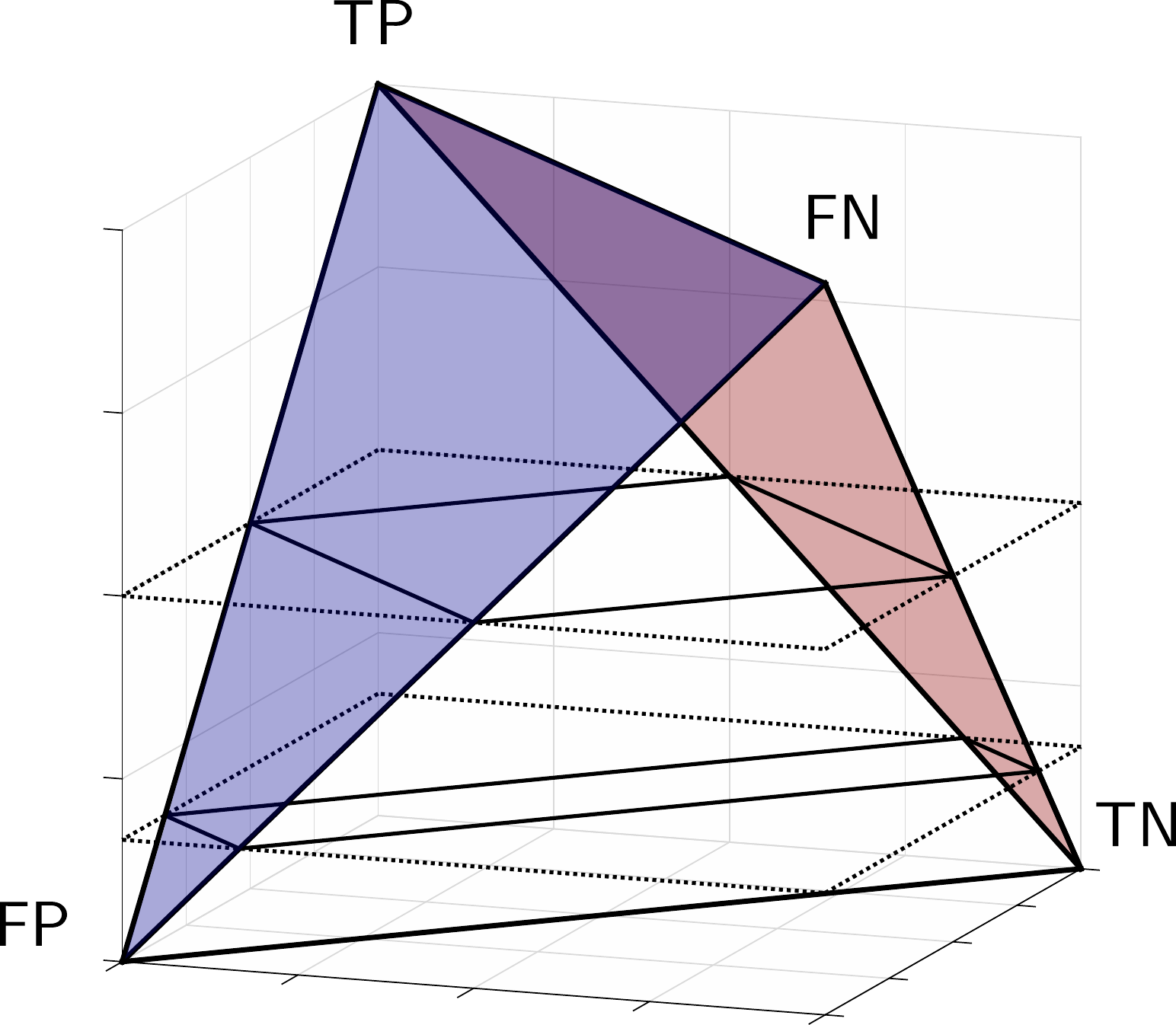}
		\label{fig:skeleton-with-planes-(N)}}
	\subfloat[$P/n = 1/6$]{
		\centering
			\includegraphics[width=0.3\textwidth]{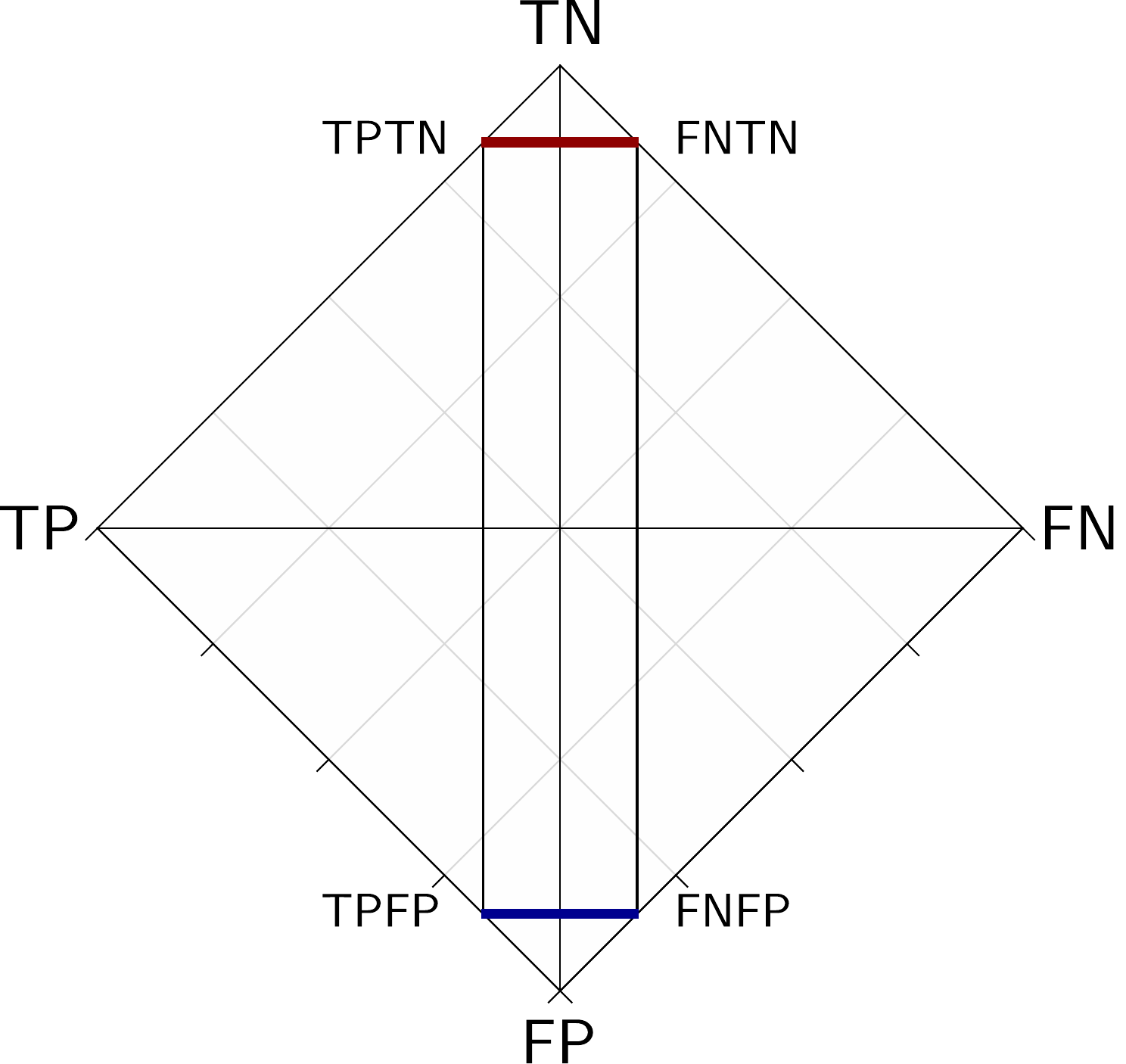}
		\label{fig:cross-sections-16-(N)}}
		\subfloat[$P/n = 1/2$]{
		\centering
			\includegraphics[width=0.3\textwidth]{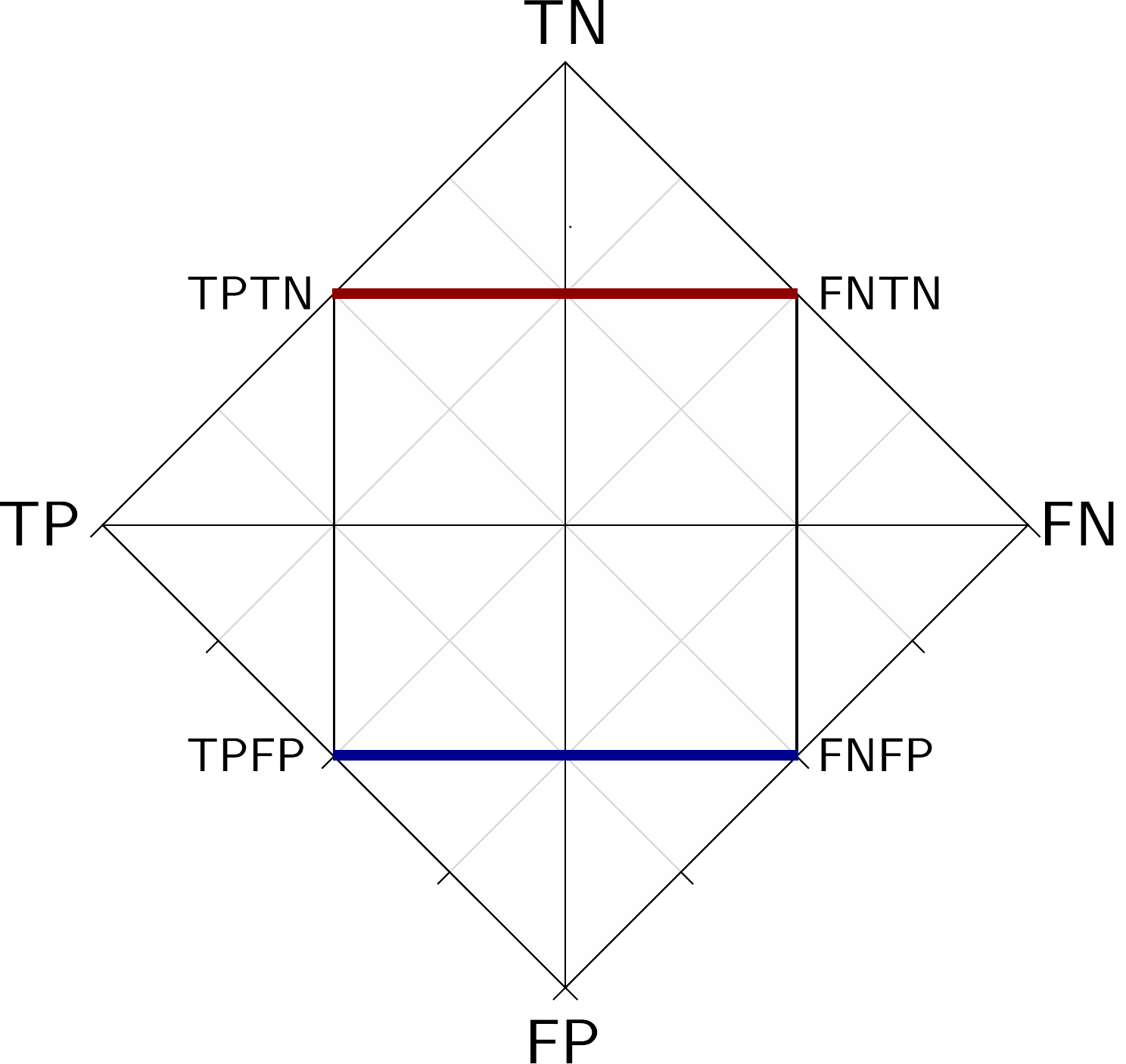}
		\label{fig:cross-sections-12-(N)}}
	\caption{Illustration of full/null recognition of the negative class}%
	\label{fig:cross-sections-(N)}
\end{figure}

Recall that the interpretation of the rectangular cross-section discussed in Section \ref{sec:visualization-technique} is as follows.
\begin{itemize}[noitemsep]
	\item side $\overline{TP}$ / $\overline{FN}$: full/null recognition of the positive class (Fig.~\ref{fig:cross-sections-(P)}),
	\item side $\overline{TN}$ / $\overline{FP}$: full/null recognition of the negative class (Fig.~\ref{fig:cross-sections-(N)}).
\end{itemize}
\vspace{-10mm}
\vspace{-8mm}

In this context, we postulate to analyse classifier performance measures with respect to ten properties:
\begin{description}[align=right,labelwidth=1.65cm,labelindent=10pt,leftmargin=2.18cm,font=\normalfont]
	\item[$\mathsf{TPTN}_{max}$:] vertex $\mathsf{TPTN}$ maximal value,
	\item[$\overline{\mathsf{FN}}_{min}$:] side $\overline{\mathsf{FN}}$ minimal value,
	\item[$\overline{\mathsf{FP}}_{min}$:] side $\overline{\mathsf{FP}}$ minimal value,
	\item[$TP_{\nearrow}$:] horizontal lines weakly monotonic value growth (from $\overline{\mathsf{FN}}$ to $\overline{\mathsf{TP}}$),
	\item[$TN_{\nearrow}$:] vertical lines weakly monotonic value growth (from $\overline{\mathsf{FP}}$ to $\overline{\mathsf{TN}}$),
	\item[$\overline{\mathsf{TN}}_{\neq max}$:] side $\overline{\mathsf{TN}}$ less than maximal value except for vertex $\mathsf{TPTN}$,
	\item[$\overline{\mathsf{TP}}_{\neq max}$:] side $\overline{\mathsf{TP}}$ less than maximal value except for vertex $\mathsf{TPTN}$,
	\item[$\mathit{ACE}$:] for any two corresponding points on sides $\overline{\mathsf{TP}}$ and $\overline{\mathsf{TN}}$ (e.g. middle points) the value on side $\overline{\mathsf{TP}}$ is greater or equal to that on $\overline{\mathsf{TN}}$,
	\item[$\mathit{ACH}$:] values invariant under exchange of $\mathit{TP}$ with $\mathit{TN}$ and $\mathit{FN}$ with $\mathit{FP}$,
	\item[$\mathit{UnDefs}$:] the existence (and the location) of undefined values.
\end{description}
If present, undefined measure values are excluded from the above considerations, except for the last property, which is directly concerned with those values.
Similarly, all but the last two properties are analysed only for `non-degenerated' rectangular cross-sections, i.e. cross-sections corresponding to $P > 0$ and $N > 0$. On the other hand, the `degenerated' cross-section, i.e. cross-sections that result in rectangles of either zero breadth or zero width, 
are taken into account only in the $\mathit{ACH}$ and $\mathit{UnDefs}$ properties. The presented properties may be regarded as a basic `check-list', providing knowledge on the behaviour of classifier performance measures for imbalanced data. 

Notice that when all feasible rectangular cross-sections of the considered type are taken into account, the properties naturally extend from 2D in the rectangles to 3D in the tetrahedron. For example, points $\mathsf{TPTN}$ of all rectangles form edge $\mathsf{TP}$--$\mathsf{TN}$ of the tetrahedron, sides $\overline{\mathsf{TP}}$ of all rectangles form face $\triangle\mathsf{FP}$--$\mathsf{TP}$--$\mathsf{TN}$ of the tetrahedron, etc. This multidimensional nature of the measures renders the analytical process of their property verification harder, emphasizing the usefulness of the introduced visual-based 3D analyses.

Recall that the analysed measures are functions of $TP \geq 0$, $FN \geq 0$, $FP \geq 0$ and $TN \geq 0$, $TP+FN+FP+TN=n$, which constitute the elements of the confusion matrix $\left[ \begin{smallmatrix} \mathit{TP} & \mathit{FN} \\ \mathit{FP} & \mathit{TN} \end{smallmatrix} \right]$ (see Table~\ref{tab:cmatrix}). 
In this context, $f(\left[ \begin{smallmatrix} \mathit{TP} & \mathit{FN} \\ \mathit{FP} & \mathit{TN} \end{smallmatrix} \right])$ denotes the value of any of the considered classification performance measures.

\begin{figure}[!htb]
		\centering
	\subfloat[External view]{
		\centering
		\includegraphics[width=0.3\textwidth]{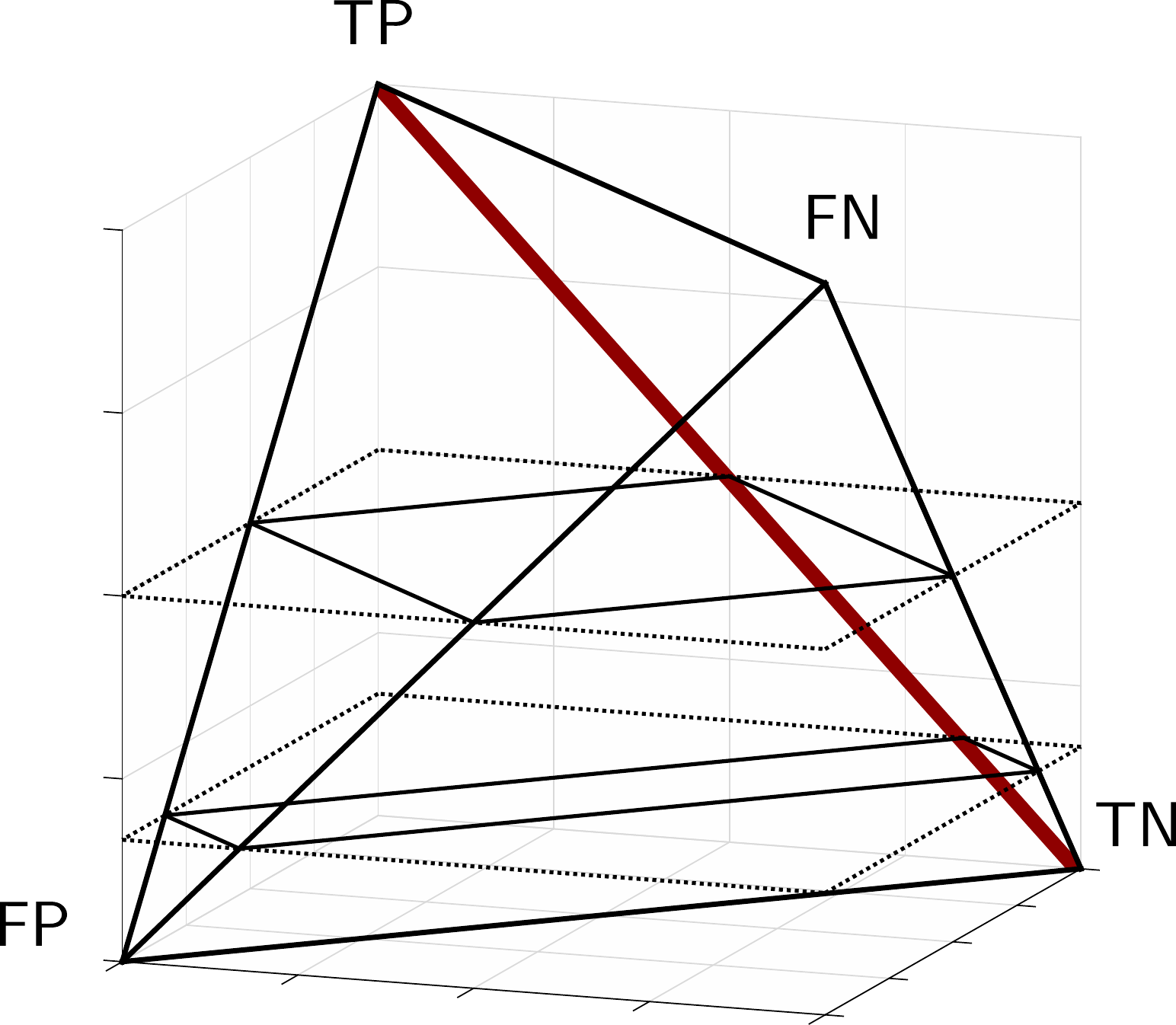}
		\label{fig:skeleton-with-planes-(1)}}
	\subfloat[$P/n = 1/6$]{
		\centering
			\includegraphics[width=0.3\textwidth]{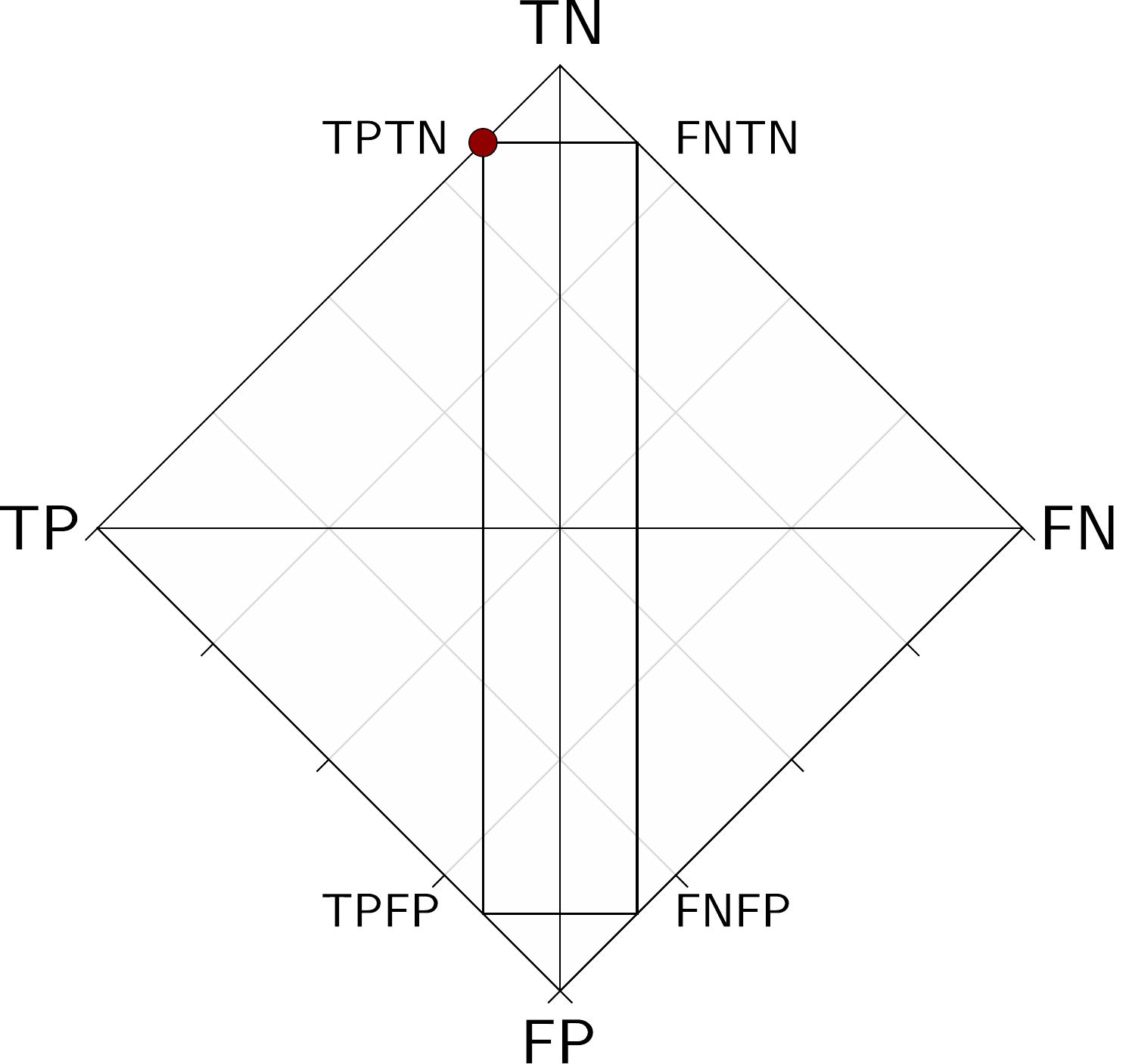}
		\label{fig:cross-sections-16-(1)}}
		\subfloat[$P/n = 1/2$]{
		\centering
			\includegraphics[width=0.3\textwidth]{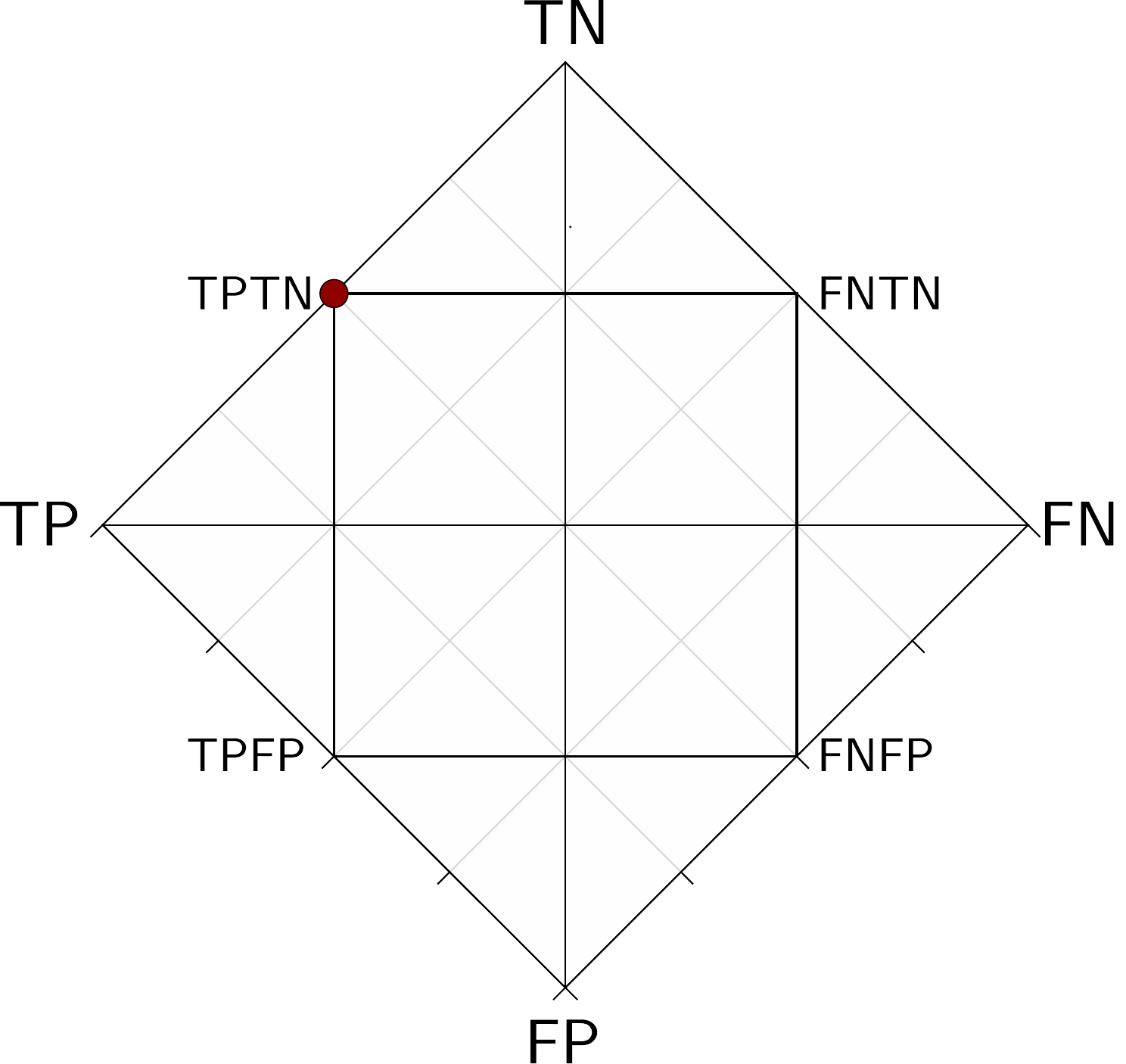}
		\label{fig:cross-sections-12-(1)}}
	\caption{Illustration of property $\mathsf{TPTN}_{max}$}%
	\label{fig:cross-sections-(1)}
\end{figure}

Property $\mathsf{TPTN}_{max}$ ensures that perfect predictions of both classes always render the best measure value (see Fig.~\ref{fig:cross-sections-(1)}). 
Notice that vertex $\mathsf{TPTN}$, being the common part of both side $\overline{\mathsf{TP}}$ and side $\overline{\mathsf{TN}}$, is actually the only point of full recognition of both the positive and the negative class.
Because  $\mathsf{TPTN}$ corresponds to 
$\left[ \begin{smallmatrix} \mathit{P} & \mathit{0} \\ \mathit{0} & \mathit{N} \end{smallmatrix} \right]$, this implies $f(\left[ \begin{smallmatrix} \mathit{P} & \mathit{0} \\ \mathit{0} & \mathit{N} \end{smallmatrix} \right]) = max$.

\begin{figure}[!ht]
		\centering
	\subfloat[External view]{
		\centering
		\includegraphics[width=0.31\textwidth]{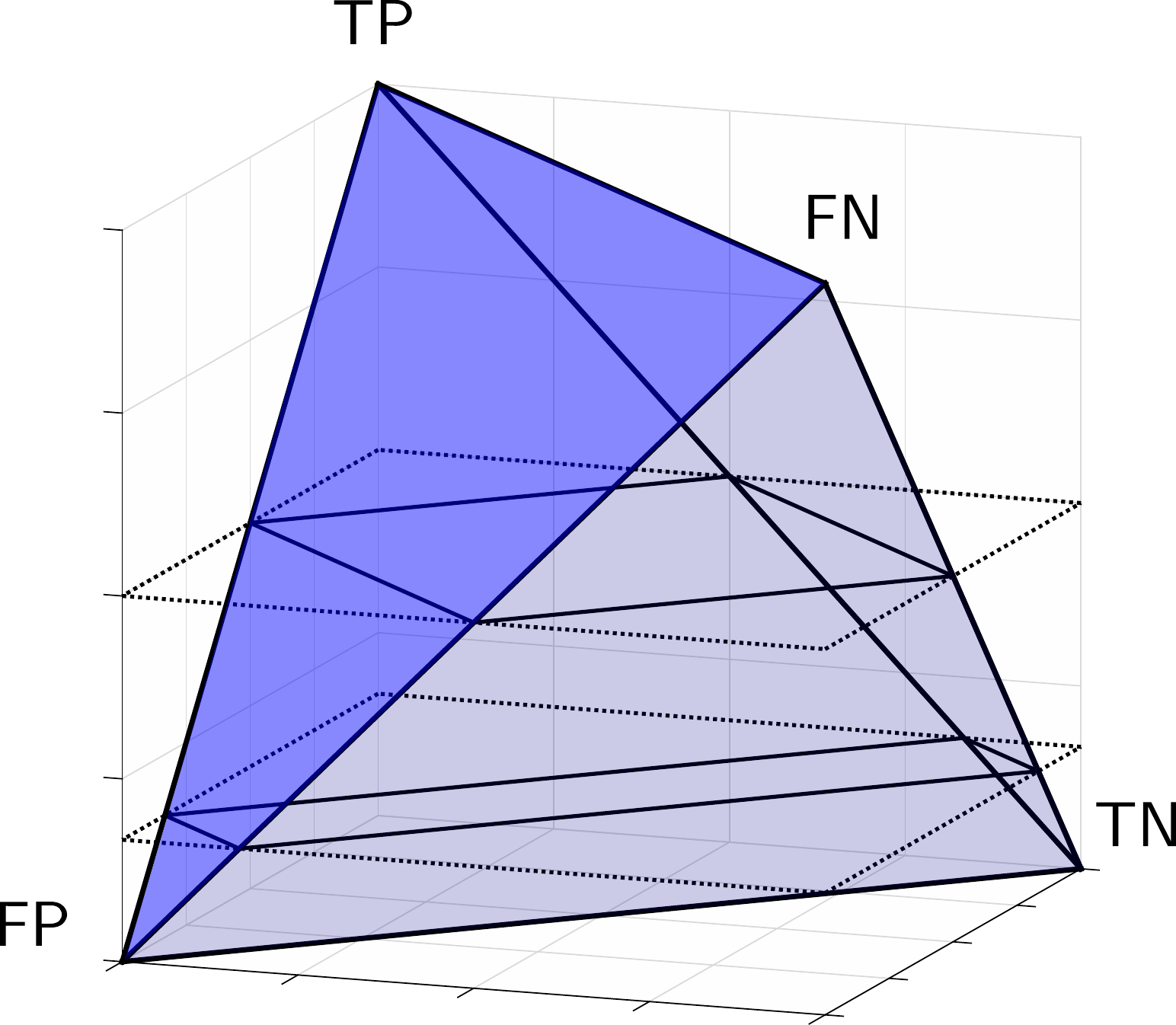}
		\label{fig:skeleton-with-planes-(23)}}
	\subfloat[$P/n = 1/6$]{
		\centering
			\includegraphics[width=0.31\textwidth]{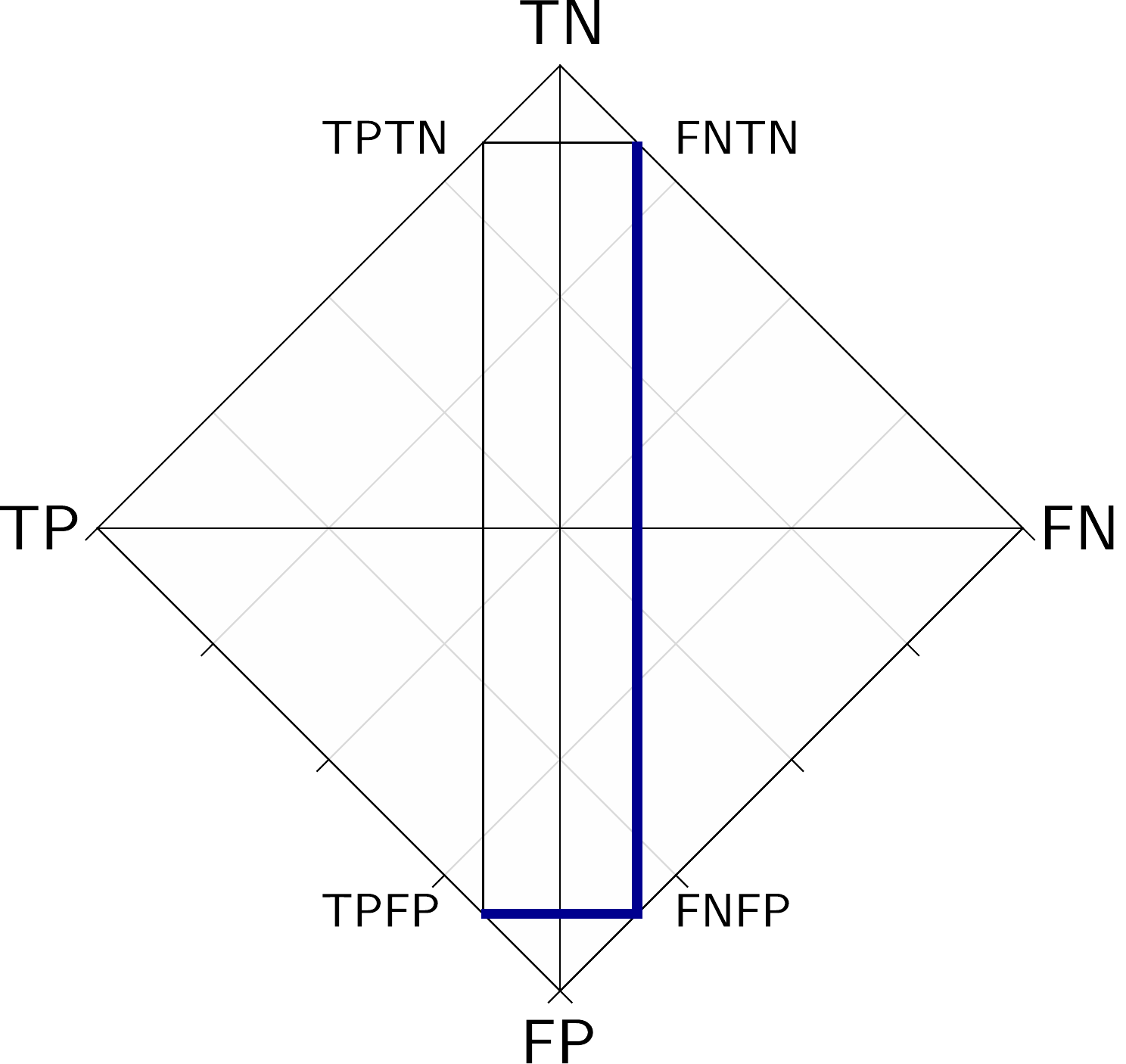}
		\label{fig:cross-sections-16-(23)}}
		\subfloat[$P/n = 1/2$]{
		\centering
			\includegraphics[width=0.31\textwidth]{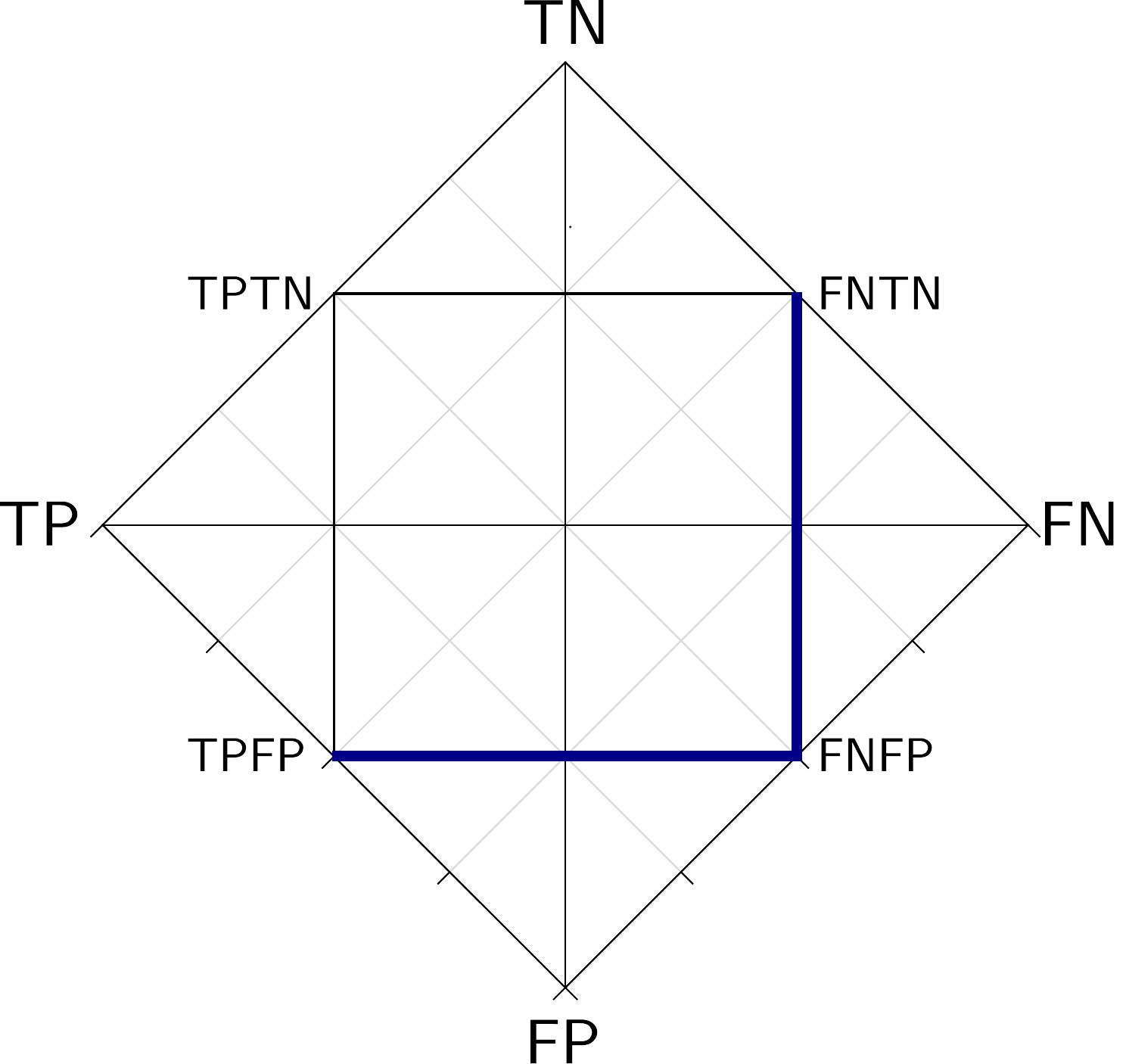}
		\label{fig:cross-sections-12-(23)}}
	\caption{Illustration of properties $\overline{\mathsf{FN}}_{min}$ and $\overline{\mathsf{FP}}_{min}$}%
	\label{fig:cross-sections-(23)}
\end{figure}

Properties $\overline{\mathsf{FN}}_{min}$ and $\overline{\mathsf{FP}}_{min}$ state that not recognizing one of the classes should correspond to the worst possible measure value (see Fig.~\ref{fig:cross-sections-(23)}). 
Recall that side $\overline{\mathsf{FN}}$ and side $\overline{\mathsf{FP}}$ correspond to null recognition of the positive and the negative class, respectively. 
In binary classification, a null recognition of any of the two classes (which concerns the minority class in most cases) is certainly insufficient. 
Thus, it is naturally required that measures should obtain minimal values on sides $\overline{\mathsf{FN}}$ and $\overline{\mathsf{FP}}$. 
This boils down to $f(\left[ \begin{smallmatrix} \mathit{0} & \mathit{P} \\ \mathit{FP} & \mathit{TN} \end{smallmatrix} \right]) = min$ and $f(\left[ \begin{smallmatrix} \mathit{TP} & \mathit{FN} \\ \mathit{N} & \mathit{0} \end{smallmatrix} \right]) = min$.

\begin{figure}[!ht]
		\centering
	\subfloat[External view]{
		\centering
		\includegraphics[width=0.31\textwidth]{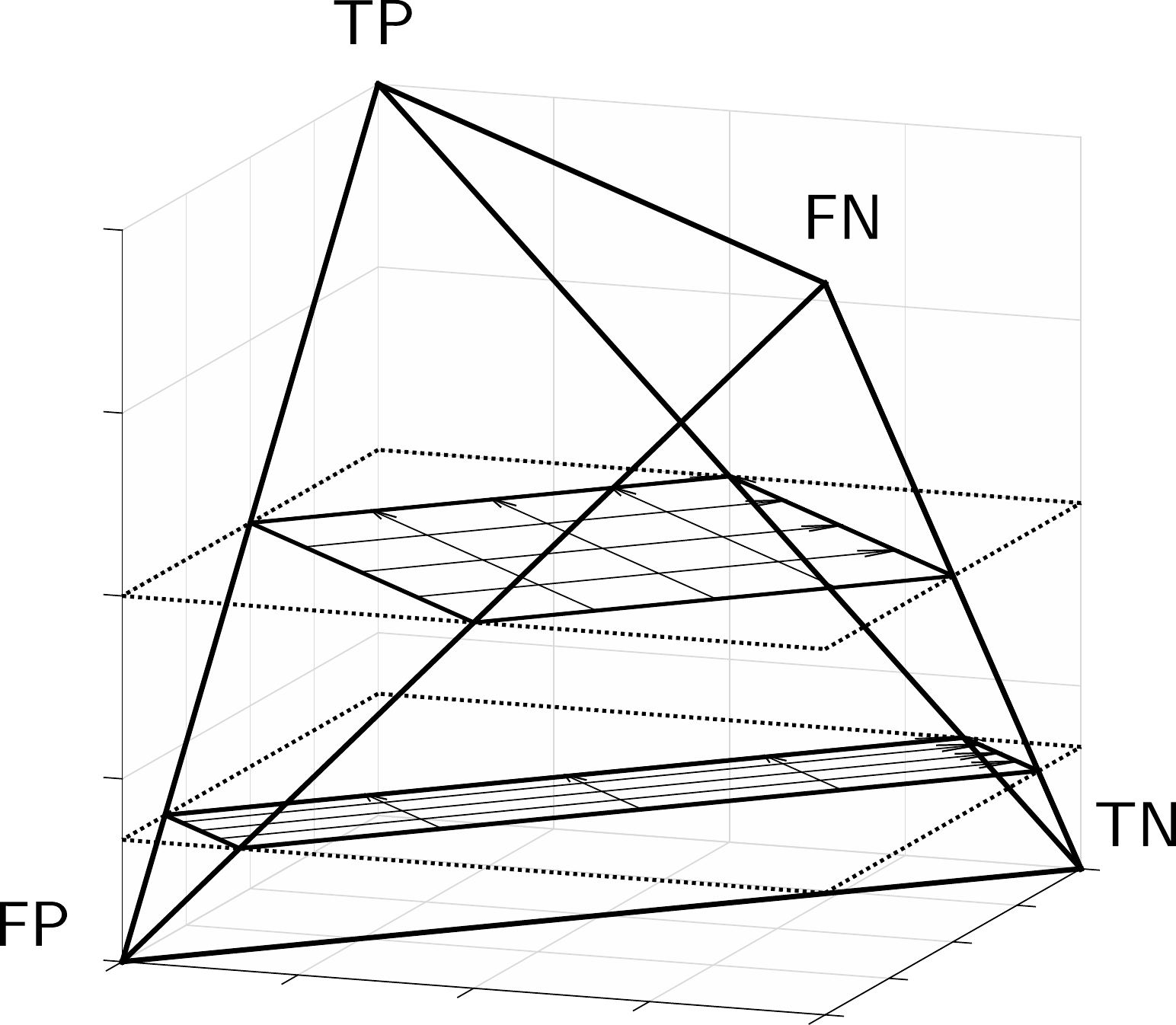}
		\label{fig:skeleton-with-planes-(45)}}
	\subfloat[$P/n = 1/6$]{
		\centering
			\includegraphics[width=0.31\textwidth]{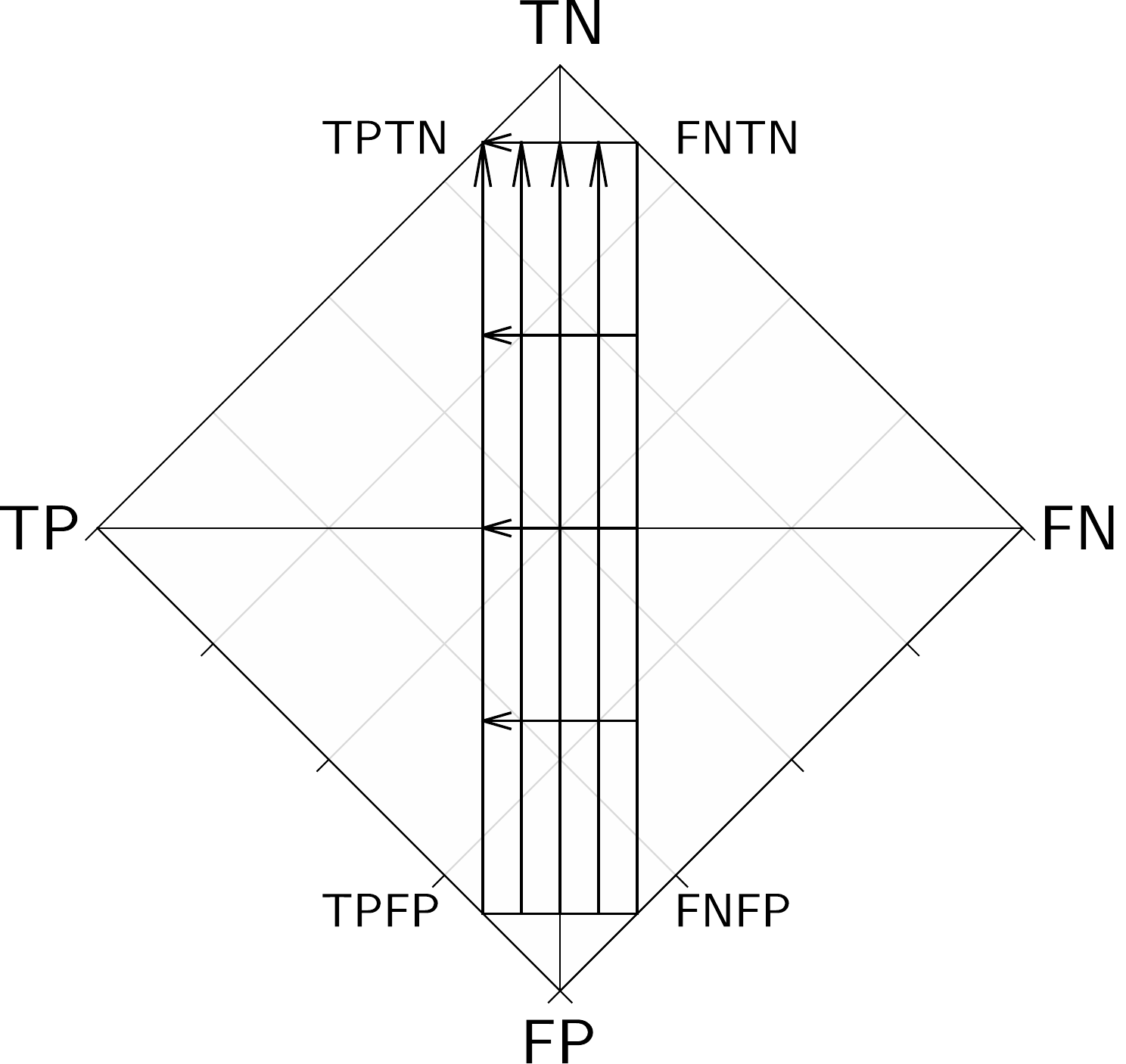}
		\label{fig:cross-sections-16-(45)}}
		\subfloat[$P/n = 1/2$]{
		\centering
			\includegraphics[width=0.31\textwidth]{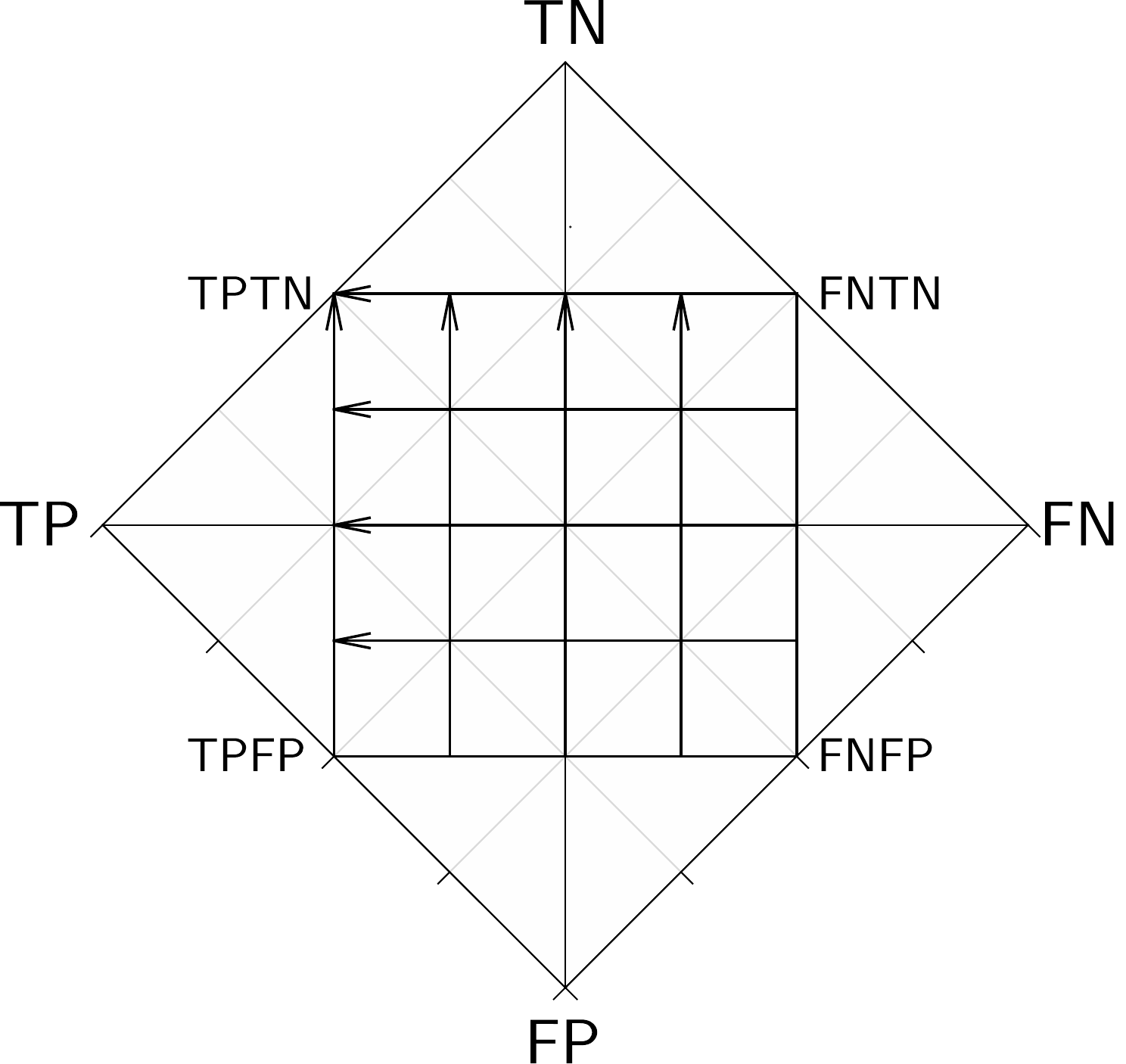}
		\label{fig:cross-sections-12-(45)}}
	\caption{Illustration of properties $TP_{\nearrow}$ and $TN_{\nearrow}$}%
	\label{fig:cross-sections-(45)}
\end{figure}

Properties $TP_{\nearrow}$ and $TN_{\nearrow}$ require that growing $\mathit{TP}$ and $\mathit{TN}$ values should coincide with a weakly monotonic growth of the measure's value (see Fig.~\ref{fig:cross-sections-(45)}). 
As far as $TP_{\nearrow}$ is concerned, observe that the greater $TP$ is in the confusion matrix, the closer we move from side $\overline{\mathsf{FN}}$ to side $\overline{\mathsf{TP}}$ in the rectangular cross-section, which translates directly to increased recognition of the positive class. 
Naturally, it would be counter-intuitive if such increased recognition resulted in decreasing values of the measure. Thus, its weakly monotonic growth is expected. As opposed to requirements $\overline{\mathsf{FN}}_{min}$ and $\overline{\mathsf{FP}}_{min}$, which concern merely the borders of the cross-section, $TP_{\nearrow}$ concerns the entirety of the cross-section. In particular, also side $\overline{\mathsf{FP}}$, where the value is required to be minimal (according to property $\overline{\mathsf{FP}}_{min}$), satisfies the weak monotonicity.
Property $TP_{\nearrow}$ boils down to the following condition: if $TP_1 \geq TP_2$, then $f(\left[ \begin{smallmatrix} \mathit{TP_1} & \mathit{FN_1} \\ \mathit{FP} & \mathit{TN} \end{smallmatrix} \right]) \geq f(\left[ \begin{smallmatrix} \mathit{TP_2} & \mathit{FN_2} \\ \mathit{FP} & \mathit{TN} \end{smallmatrix} \right])$. 
Analogously property $TN_{\nearrow}$, which boils down to the condition: if $TN_1 \geq TN_2$, then $f(\left[ \begin{smallmatrix} \mathit{TP} & \mathit{FN} \\ \mathit{FP_1} & \mathit{TN_1} \end{smallmatrix} \right]) \geq f(\left[ \begin{smallmatrix} \mathit{TP} & \mathit{FN} \\ \mathit{FP_2} & \mathit{TN_2} \end{smallmatrix} \right])$.

Properties $\overline{\mathsf{TN}}_{\neq max}$ and $\overline{\mathsf{TP}}_{\neq max}$ tackle the problem of maximal values of the measure. 
Observe that in a two-class problem, the full recognition of just one class (only positive or only negative), which can be achieved trivially, should not render the highest value of the measure. Only the full recognition of both classes should be rewarded with the maximum, as stated by $\mathsf{TPTN}_{max}$. Thus, properties 
$\overline{\mathsf{TN}}_{\neq max}$ and $\overline{\mathsf{TP}}_{\neq max}$ require that
the measure's values on sides $\overline{\mathsf{TN}}$ and $\overline{\mathsf{TP}}$ should be less than maximal, except for the very vertex $\mathsf{TPTN}$. If a classification measure fulfils this property, a simple majority or minority stub will never be mistaken with the best possible classifier. 
This boils down to: if $FN+FP>0$, then $f(\left[ \begin{smallmatrix} \mathit{TP} & \mathit{FN} \\ \mathit{FP} & \mathit{TN} \end{smallmatrix} \right]) < max$. 

\begin{figure}[!ht]
	\subfloat[External view]{
		\centering
		\includegraphics[width=0.31\textwidth]{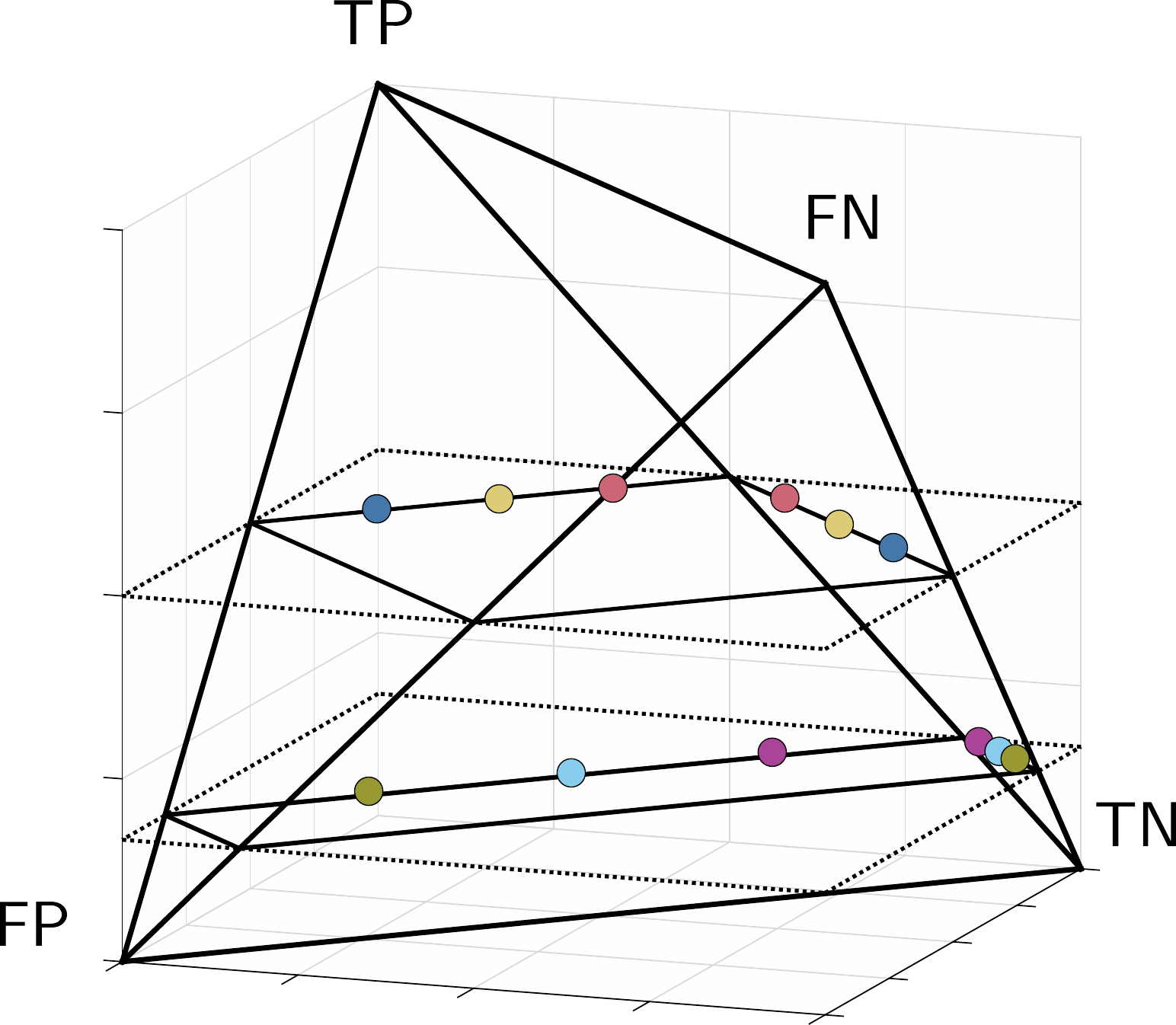}
		\label{fig:skeleton-with-planes-(67)}}
	\subfloat[$P/n = 1/6$]{
		\centering
			\includegraphics[width=0.31\textwidth]{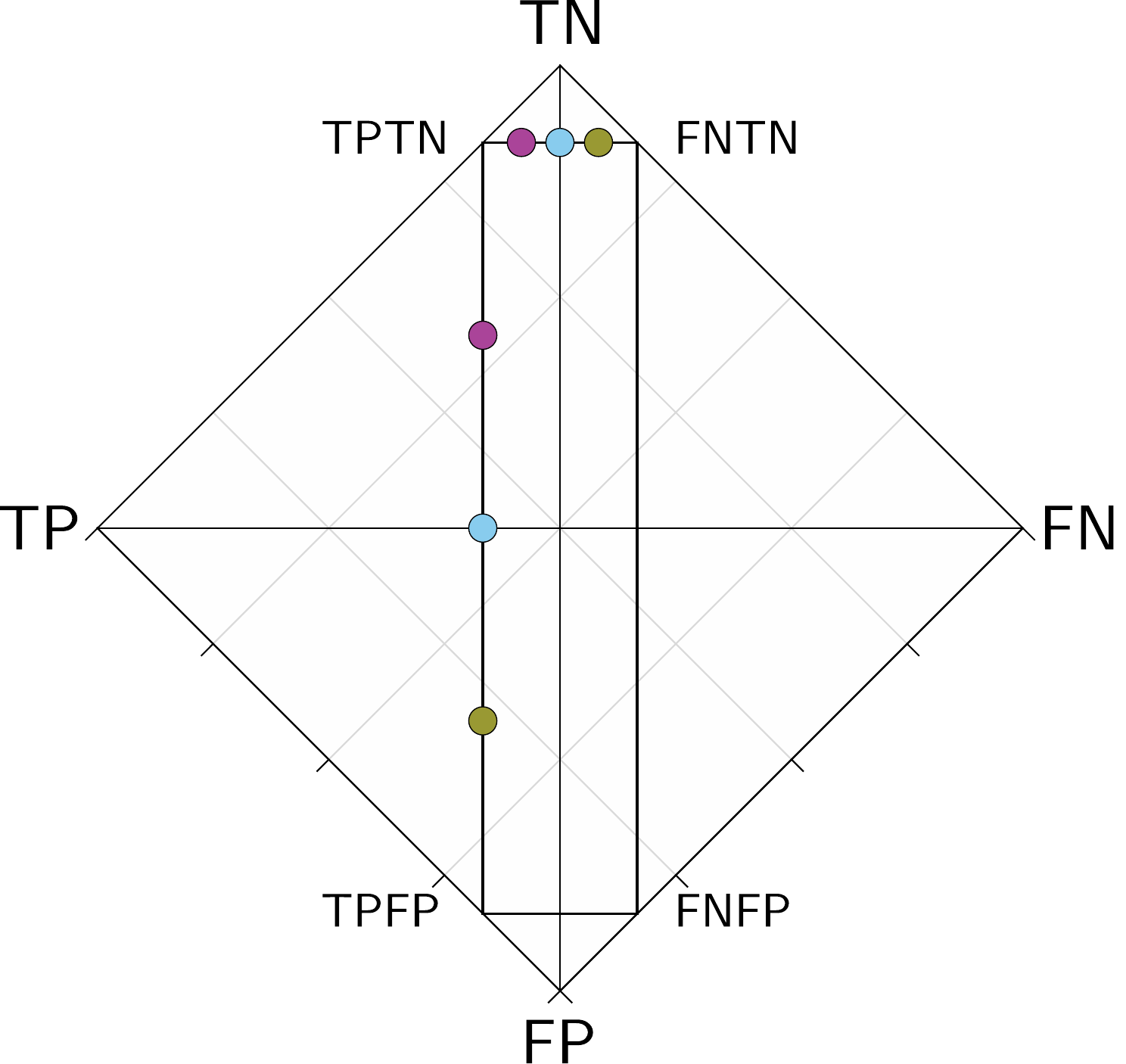}
		\label{fig:cross-sections-16-(67)}}
		\subfloat[$P/n = 1/2$]{
		\centering
			\includegraphics[width=0.31\textwidth]{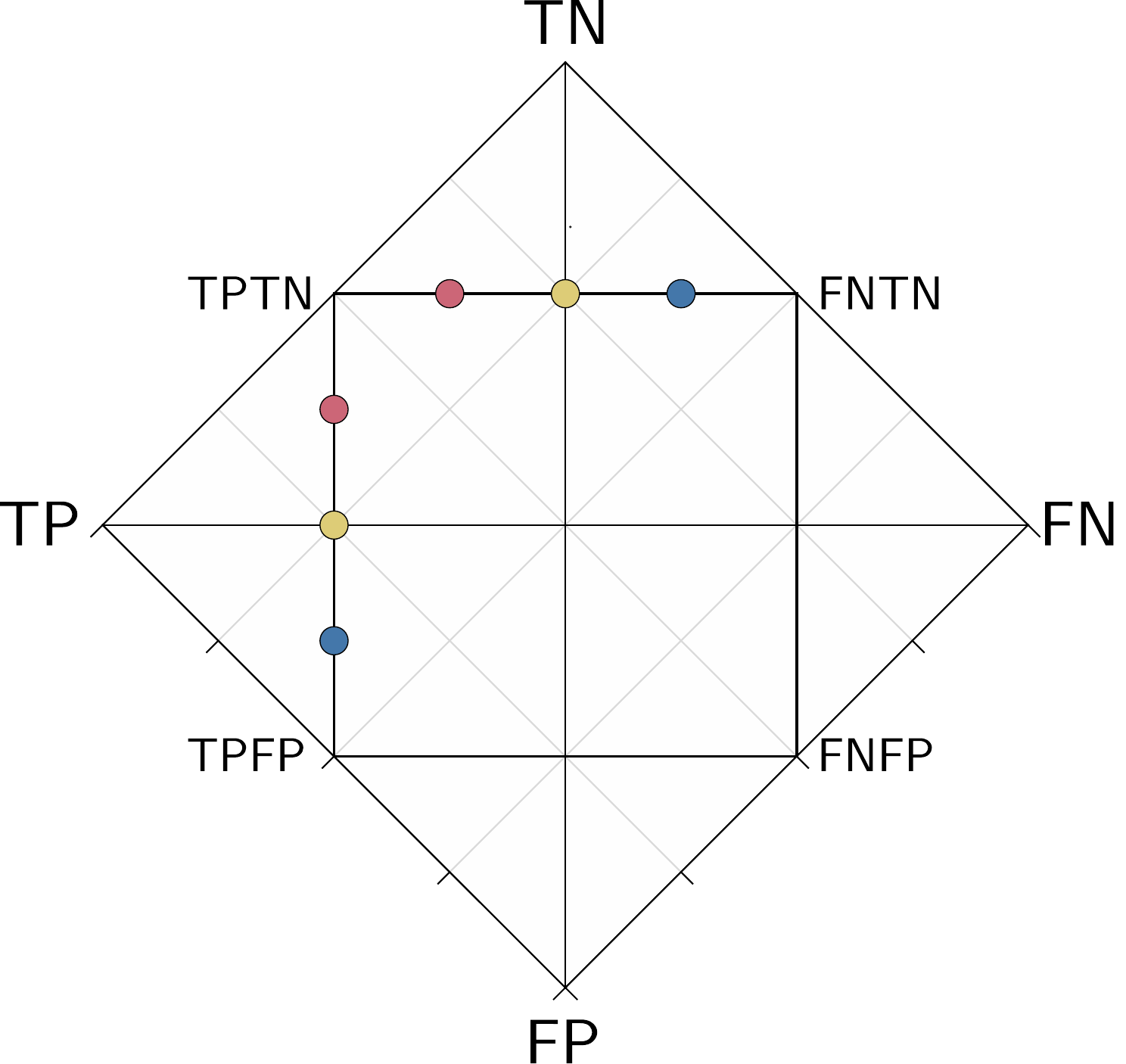}
		\label{fig:cross-sections-12-(67)}}
	\caption{Illustration of property $ACE$}%
	\label{fig:cross-sections-(67)}
\end{figure}

Property $\mathit{ACE}$ reveals the class bias resulting from \textit{asymmetric class evaluation}, typical for class imbalance problems.
It is introduced to guarantee that full recognition of only the negative class 
is never rewarded with a higher value than the full recognition of only the positive one (assuming the respective other class is recognized to the same degree). 
In particular, since the recognition of the positive class is of high importance, 
the middle point of side $\overline{\mathsf{TP}}$ (i.e. when the whole positive and half of the negative class is recognized) should not be assessed with a lower value than the middle point of side $\overline{\mathsf{TN}}$ (i.e. when the whole negative and half of the positive class is recognized). Similarly for all other pairs of corresponding points on sides $\overline{\mathsf{TP}}$ and $\overline{\mathsf{TN}}$ (three of which are depicted in Fig.~\ref{fig:cross-sections-(67)}). In terms of the entries of the confusion matrix, property $\mathit{ACE}$ boils down to: $f(\left[ \begin{smallmatrix} \mathit{P} & \enskip \mathit{0} \\ \mathit{\gamma N} & \enskip \mathit{(1-\gamma)N} \end{smallmatrix} \right]) \geq  f(\left[ \begin{smallmatrix} \mathit{(1-\gamma) P} & \enskip \mathit{\gamma P} \\ \mathit{0} & \enskip \mathit{N} \end{smallmatrix} \right])$, where $\gamma \in [0,1]$ (in Fig.~\ref{fig:cross-sections-(67)} $\gamma$ takes on values $1/4$, $2/4$ and $3/4$). Notice that the weak nature of the property is implied by the fact that it does not specify by how much the full recognition of the positive class should be favoured over the full recognition of the negative class. On the other hand the unsatisfied $\mathit{ACE}$ reveals instantly, however, that the measure favours (in the above sense) the negative over the positive.

Much the same, property $\mathit{ACH}$ deals with the issue of \textit{asymmetric class handling}.
It tests if the classes can be exchanged without influencing the measure's behaviour. This could be especially relevant in highly dynamic situations, e.g. in data streams plagued by concept drift~\cite{PreqAucKAIS}, in which the percentage of the positive class may increase to make it actually (albeit temporarily) the majority class~\cite{DBLP:journals/tkde/WangMY15}. 
Expressed with the confusion matrix, $\mathit{ACH}$ boils down to: $f(\left[ \begin{smallmatrix} \mathit{TP} & \mathit{FN} \\ \mathit{FP} & \mathit{TN} \end{smallmatrix} \right]) =  f(\left[ \begin{smallmatrix} \mathit{TN} & \mathit{FP} \\ \mathit{FN} & \mathit{TP} \end{smallmatrix} \right])$. 

Finally, property $\mathit{UnDefs}$ pinpoints the existence and the location of undefined values ($+\infty$ whenever a positive value is divided by $0$, $-\infty$ whenever a negative value is divided by $0$, and $NaN$ whenever $0$ is divided by $0$).
As such, it highlights potential numerical pitfalls that can arise when calculating the measure values. While occurring fairly seldom with real life data, such undefined values are needed to fully characterize and thoroughly compare the considered measures. 

In the following section, we use the proposed ten properties to compare various classification measures.

\section{Visual-based Analysis of Selected Measures}
\label{sec:visual-based-analysis}

Having presented the visualization technique in Section~\ref{sec:visualization-technique} and having defined the properties to be researched in Section~\ref{sec:desirable-properties}, now we use the proposed tools to analyse 22 classification measures. The selected set of measures includes the most popular ones defined using elements from a two-class confusion matrix, and comprises non-parametric as well as parametric indices. Table~\ref{tab:measure-properties} presents the analysis results for the selected measures, whereas their definitions are available in the supplementary data\footnote{\url{https://dabrze.shinyapps.io/Tetrahedron/}}.

\begin{table}[ht!]
  \centering
  \caption{Properties of selected classification measures; $^{*}$: contains NaN (undefined value); $^{\dagger}$: NaN side, $^{s}$: strong monotonicity}
	\label{tab:measure-properties}%
  \begin{tabularx}{\textwidth}{@{}l@{}C@{}C@{}C@{}C@{}C@{}C@{}C@{}C@{}C@{}p{2.5cm}@{}}
		\toprule
    Measure 						& $\underset{max}{{\mathsf{TPTN}}}$ & $\underset{min}{\overline{\mathsf{FN}}}$ & $\underset{min}{\overline{\mathsf{FP}}}$ & $TP_{\nearrow}$ & $TN_{\nearrow}$ & $\underset{\neq max}{\overline{\mathsf{TN}}}$ & $\underset{\neq max}{\overline{\mathsf{TP}}}$ & $\mathit{ACE}$ & $\mathit{ACH}$ & $\mathit{UnDefs}$ \\
		\midrule
    \textit{Accuracy} 						&$\checkmark$   & $\times$    & $\times$    & $\checkmark^{s}$ & $\checkmark^{s}$ & $\checkmark$   & $\checkmark$   & $\times$    & $\checkmark$   							& none \\
    \textit{Area Under Lift} 		& $\times$    & $\times$    & $\times$    & $\checkmark^{s}$ & $\checkmark^{s}$ & $\checkmark$   & $\checkmark$   & $\checkmark$   & $\times$    							& $\mathsf{TN}$--$\mathsf{FP}$; $\mathsf{TP}$--$\mathsf{FN}$ \\
    \textit{Balanced accuracy} 	& $\checkmark$   & $\times$    & $\times$    & $\checkmark^{s}$ & $\checkmark^{s}$ & $\checkmark$   & $\checkmark$   & $\checkmark$   & $\checkmark$   						& $\mathsf{TN}$--$\mathsf{FP}$; $\mathsf{TP}$--$\mathsf{FN}$ \\
   \textit{F$_1$-score} 						& $\checkmark$   & $\times^{\dagger}$   & $\times$    & $\checkmark^{s*}$ & $\checkmark^{s*}$ & $\checkmark^{*}$  & $\checkmark$   & $\times$    & $\times$    & $\triangle\mathsf{FP}$--$\mathsf{FN}$--$\mathsf{TN}$\\
    \textit{False negative rate} & $\times$    & $\times$    & $\times$    & $\times$ & $\checkmark$  & $\times$    & $\checkmark$   & $\times$    & $\times$    					& $\mathsf{TN}$--$\mathsf{FP}$  \\
    \textit{False positive rate} & $\times$    & $\times$    & $\times$    & $\checkmark$  & $\times$  & $\checkmark$   & $\times$    & $\checkmark$   & $\times$    				& $\mathsf{TP}$--$\mathsf{FN}$ \\
\textit{F$_\beta$}, {\scriptsize $\beta \in [0,\infty)$}& $\checkmark$   & $\times^{\dagger}$   & $\times$    & $\checkmark^{s*}$ & $\checkmark^{s*}$ & $\checkmark^{*}$  & $\checkmark$   & $\times$    & $\times$    & $\triangle\mathsf{FP}$--$\mathsf{FN}$--$\mathsf{TN}$ \\
    \textit{G-mean} 							& $\checkmark$   & $\checkmark$   & $\checkmark$   & $\checkmark$ & $\checkmark$ & $\checkmark$   & $\checkmark$   & $\checkmark$   & $\checkmark$   						& $\mathsf{TN}$--$\mathsf{FP}$; $\mathsf{TP}$--$\mathsf{FN}$ \\
    \textit{IBA$_\alpha$}(\textit{Accuracy}), {\scriptsize $\alpha \in (0,\infty)$} 			& $\times$   & $\times$    & $\times$    & $\checkmark^{s}$ & $\times$    & $\checkmark$   & $\times$    & $\times$    & $\times$    									& $\mathsf{TN}$--$\mathsf{FP}$; $\mathsf{TP}$--$\mathsf{FN}$\\
    \textit{IBA$_\alpha$}(\textit{F$_1$-score}), {\scriptsize $\alpha \in (0,\infty)$}  			& $\times$    & $\times^{\dagger}$   & $\times$    & $\times$ & $\times$    & $\checkmark^{*}$  & $\times$    & $\times$    & $\times$    			& $\triangle\mathsf{FP}$--$\mathsf{FN}$--$\mathsf{TN}$; $\mathsf{TP}$--$\mathsf{FN}$ \\
    \textit{IBA$_\alpha$}(\textit{G-mean}), {\scriptsize $\alpha \in (0,\infty)$}  					& $\times$   & $\checkmark$   & $\checkmark$   & $\checkmark$ & $\times$ & $\checkmark$   & $\times$   & $\checkmark$   & $\times$    							& $\mathsf{TN}$--$\mathsf{FP}$; $\mathsf{TP}$--$\mathsf{FN}$ \\
		\textit{IBA$_\alpha$}(\textit{F$_\beta$}), {\scriptsize $\alpha,\beta \in (0,\infty)$}  			& $\times$    & $\times^{\dagger}$   & $\times$    & $\times$ & $\times$    & $\checkmark^{*}$  & $\times$    & $\times$    & $\times$    			& $\triangle\mathsf{FP}$--$\mathsf{FN}$--$\mathsf{TN}$; $\mathsf{TP}$--$\mathsf{FN}$ \\
    \textit{Jaccard coefficient} & $\checkmark$   & $\checkmark$   & $\times$    & $\checkmark^{s}$ & $\checkmark$ & $\checkmark$   & $\checkmark$   & $\times$    & $\times$    							& $\mathsf{TN}$ \\
    \textit{Kappa} 							& $\checkmark$   & $\times$    & $\times$    & $\checkmark^{s}$ & $\checkmark^{s}$ & $\checkmark$   & $\checkmark$   & $\times$    & $\times$    							& $\mathsf{TN}$; $\mathsf{TP}$ \\
    \textit{Log odds-ratio} 			& $\checkmark$   & $\checkmark^{*}$  & $\checkmark^{*}$  & $\checkmark^{*}$  & $\checkmark^{*}$ & $\times$    & $\times$    & $\times^{\dagger}$ & $\checkmark$		& $\mathsf{TN}$--$\mathsf{FN}$; $\mathsf{TN}$--$\mathsf{FP}$; $\mathsf{TP}$--$\mathsf{FN}$; $\mathsf{TP}$--$\mathsf{FP}$ \\
    \textit{MCC}   							& $\checkmark$   & $\times$    & $\times$    & $\checkmark^{s*}$  & $\checkmark^{s*}$ & $\checkmark^{*}$  & $\checkmark^{*}$  & $\times$    & $\checkmark$   	& $\mathsf{TN}$--$\mathsf{FN}$; $\mathsf{TN}$--$\mathsf{FP}$; $\mathsf{TP}$--$\mathsf{FN}$; $\mathsf{TP}$--$\mathsf{FP}$ \\
  \textit{Neg. predictive value} & $\checkmark$   & $\times$    & $\checkmark^{*}$  & $\checkmark^{*}$  & $\checkmark^{*}$ & $\checkmark$   & $\times$    & $\checkmark$   & $\times$    				& $\mathsf{TP}$--$\mathsf{FP}$ \\
    \textit{OP} & $\checkmark$   & $\times$    & $\times$    & $\times$    & $\times$    & $\checkmark$   & $\checkmark$   & $\times$    & $\checkmark$   									& $\mathsf{TN}$--$\mathsf{FP}$; $\mathsf{FP}$--$\mathsf{FN}$; $\mathsf{TP}$--$\mathsf{FN}$ \\
    \textit{Pointwise AUC-ROC} 	& $\checkmark$   & $\checkmark$   & $\checkmark$   & $\checkmark$ & $\checkmark$ & $\checkmark$   & $\checkmark$   & $\checkmark$   & $\checkmark$   						& $\mathsf{TN}$--$\mathsf{FP}$; $\mathsf{TP}$--$\mathsf{FN}$ \\
    \textit{Precision}  					& $\checkmark$   & $\checkmark^{*}$  & $\times$    & $\checkmark^{*}$ & $\checkmark^{*}$ & $\times$    & $\checkmark$   & $\times$    & $\checkmark$   				& $\mathsf{TN}$--$\mathsf{FN}$\\
    \textit{Recall} 							& $\checkmark$   & $\checkmark$   & $\times$    & $\checkmark^{s}$ & $\checkmark$ & $\checkmark$   & $\times$    & $\checkmark$   & $\times$    					& $\mathsf{TN}$--$\mathsf{FP}$ \\
    \textit{Specificity} 				& $\checkmark$   & $\times$    & $\checkmark$   & $\checkmark$ & $\checkmark^{s}$ & $\times$    & $\checkmark$   & $\times$    & $\times$    					& $\mathsf{TP}$--$\mathsf{FN}$ \\
		\bottomrule
  \end{tabularx}%
\end{table}%

Looking at the entries of Table~\ref{tab:measure-properties}, one can notice that the proposed properties clearly differentiate the analysed measures. Having realized the differences in the measures' behaviour, one can more accurately choose the measures for the application at hand. 

Let us start with having a closer look at one exemplary property listed in Table~\ref{tab:measure-properties}, i.e. the existence and location of undefined values ($\mathit{UnDefs}$). The undefined measure values, usually resulting from division by zero, and commonly neglected, may well occur with imbalanced data, e.g. during unstratified cross-validation procedures when one of the two classes happens to be unrepresented in the learning or the testing set.
The problem becomes aggravated for multi-class problems when the measure is macro-averaged for all classes, since the resulting average becomes undefined if at least one of the averaged values is undefined.

An interesting observation is that, except for \textit{accuracy}, all of the considered measures contain undefined values. 
In particular, the \textit{Kappa} statistic is undefined when there exist only positive or only negative examples in the dataset and none of them is misclassified, which translates to two different locations in the tetrahedron, namely vertex $\mathsf{TP}$ and vertex $\mathsf{TN}$. Even worse, \textit{balanced accuracy} is undefined when there are only positive or only negative examples in the dataset, which translates directly to whole edges $\mathsf{TP}$--$\mathsf{FN}$ and $\mathsf{TN}$--$\mathsf{FP}$ in the tetrahedron. Worst of all, \textit{F$_1$-score} (as well as its generalizations) exhibits undefined values in the whole face $\triangle\mathsf{FP}$--$\mathsf{FN}$--$\mathsf{TN}$, which occurs when all positive examples are misclassified (even when both classes are represented). 

\setlength{\unitlength}{2.15mm}
\newcommand{\labelsImbalanced}{
				\put(-4,1){\tiny $\mathsf{TPFP}$}
				\put(17,1){\tiny $\mathsf{FNFP}$}
				\put(-5,95){\tiny $\mathsf{TPTN}$}
				\put(17,95){\tiny $\mathsf{FNTN}$}
				\put(7,98){\scriptsize $\overline{\mathsf{TN}}$}
				\put(19,50){\scriptsize $\overline{\mathsf{FN}}$}
				\put(7.5,-3){\scriptsize $\overline{\mathsf{FP}}$}
				\put(-4,50){\scriptsize $\overline{\mathsf{TP}}$}
}

\begin{figure}[!htb]
  \centering
	\subfloat[\textit{F$_1$-score}]{
		\centering
		\begin{tabular}{c@{\qquad\quad}c}
			\begin{overpic}[width=0.08\textwidth]{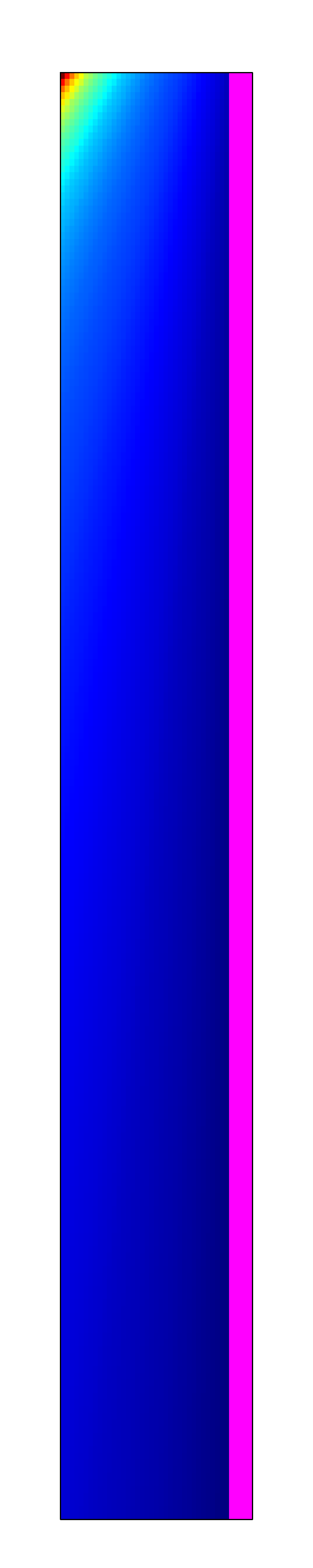}
			\labelsImbalanced
		\end{overpic}
			&
			\includegraphics[width=0.25\textwidth]{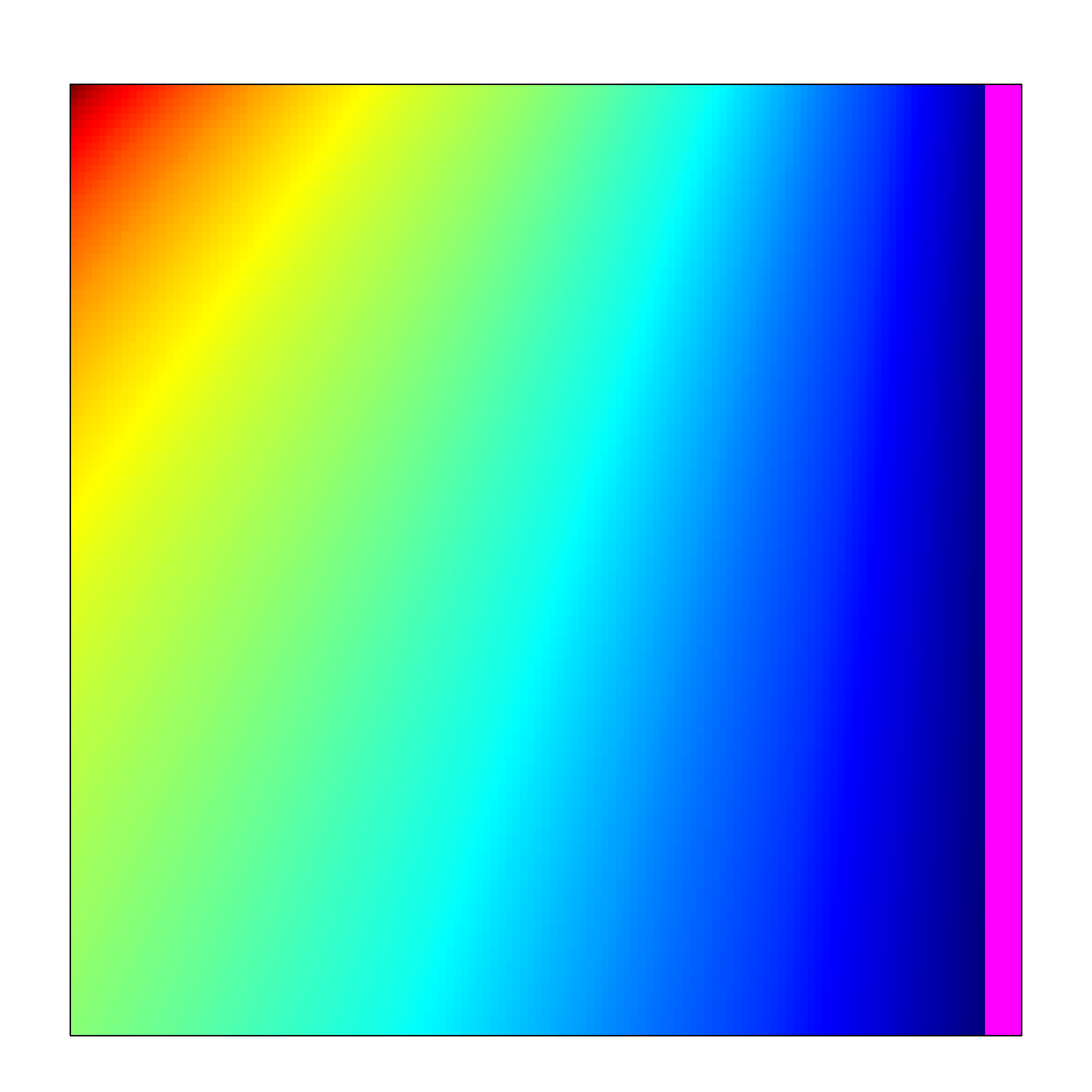}
			\\\\
			$\frac{P}{n} = \frac{1}{6}$ & $\frac{P}{n} = \frac{1}{2}$
		\end{tabular}
		\label{fig:F-cross-sections}}\quad\quad
	\subfloat[\textit{G-mean}]{
		\centering
		\begin{tabular}{c@{\qquad\quad}c}
			\begin{overpic}[width=0.08\textwidth]{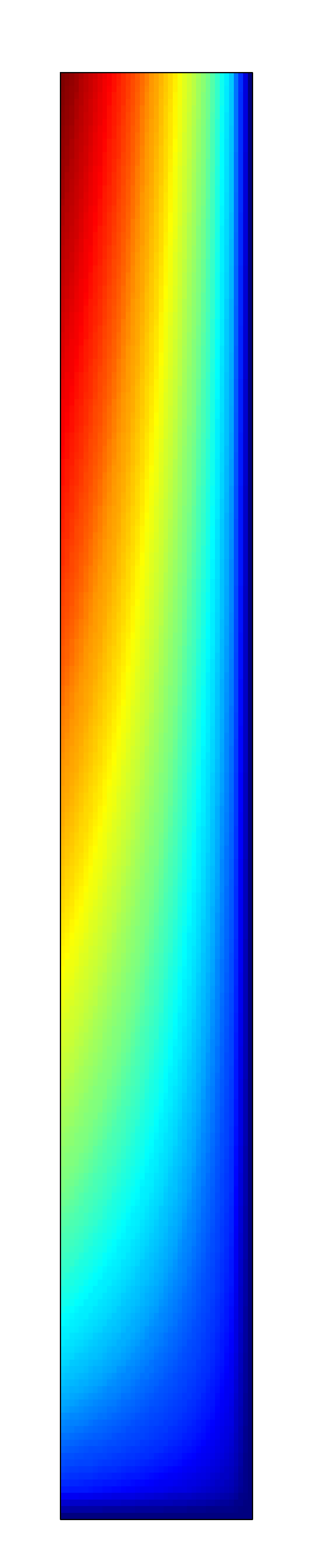}
						\labelsImbalanced
		  \end{overpic}
			&
			\includegraphics[width=0.25\textwidth]{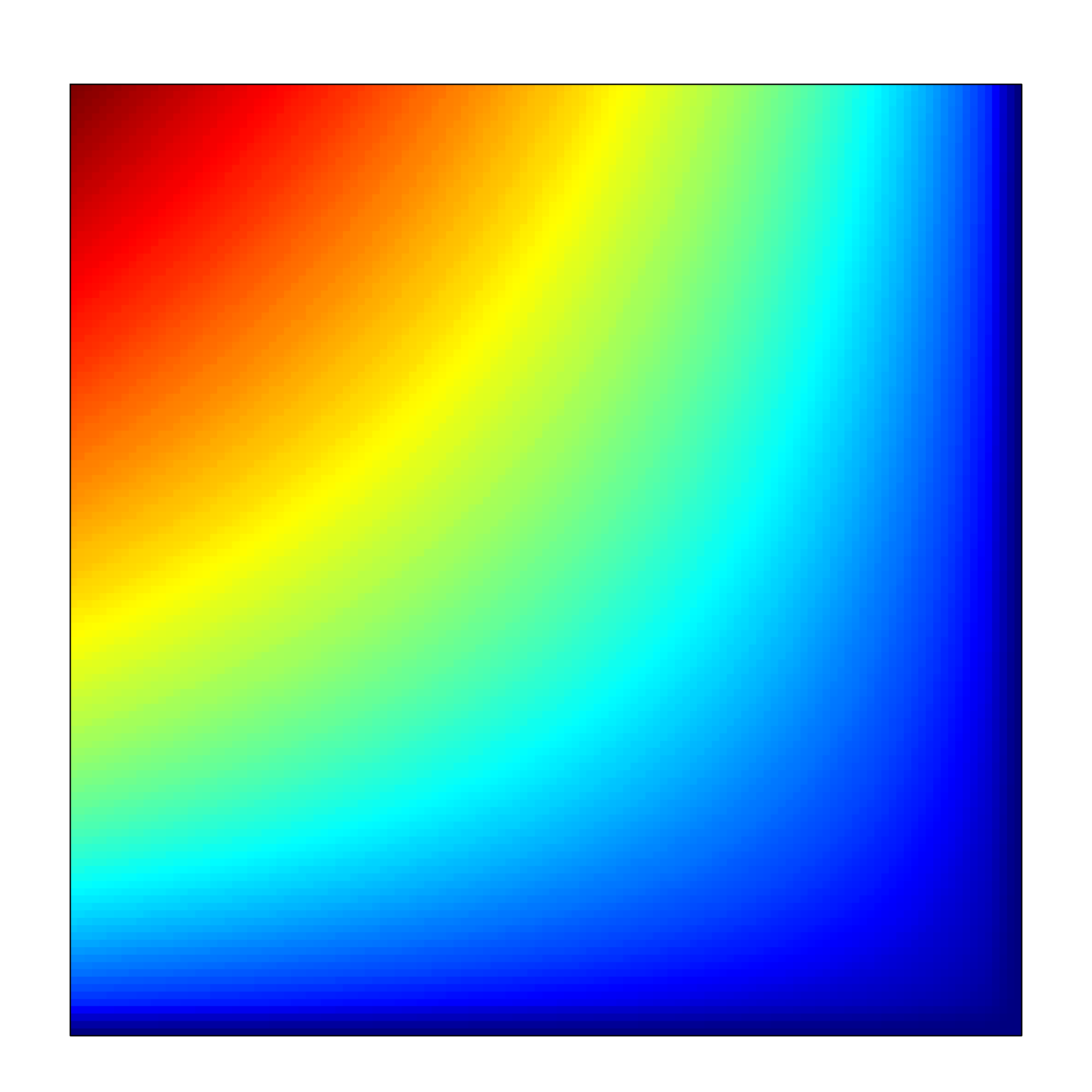}
			\\\\
			$\frac{P}{n} = \frac{1}{6}$ & $\frac{P}{n} = \frac{1}{2}$
		\end{tabular}
		\label{fig:G-mean-cross-sections}}
		
	\subfloat[$MCC$]{
		\centering
		\begin{tabular}{c@{\qquad\quad}c}
			\begin{overpic}[width=0.08\textwidth]{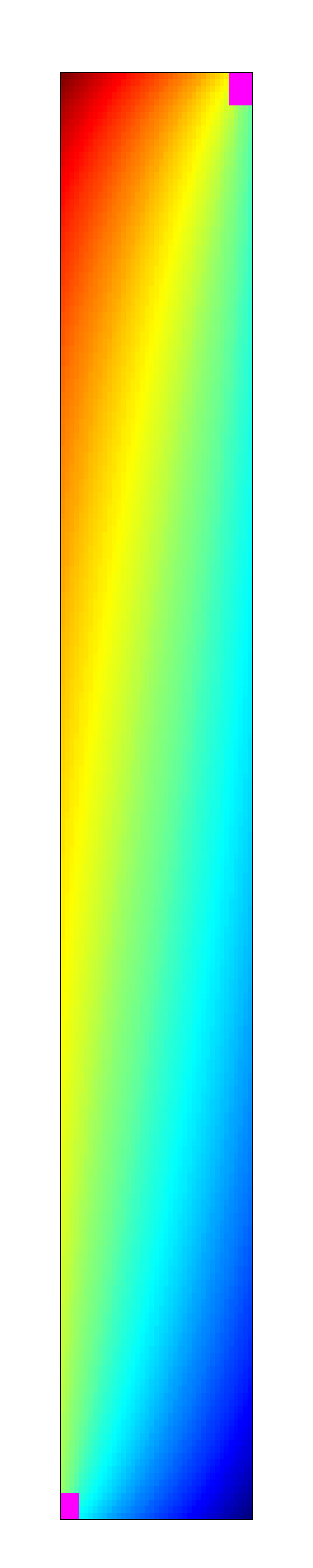}
						\labelsImbalanced
		  \end{overpic}
			&
			\includegraphics[width=0.25\textwidth]{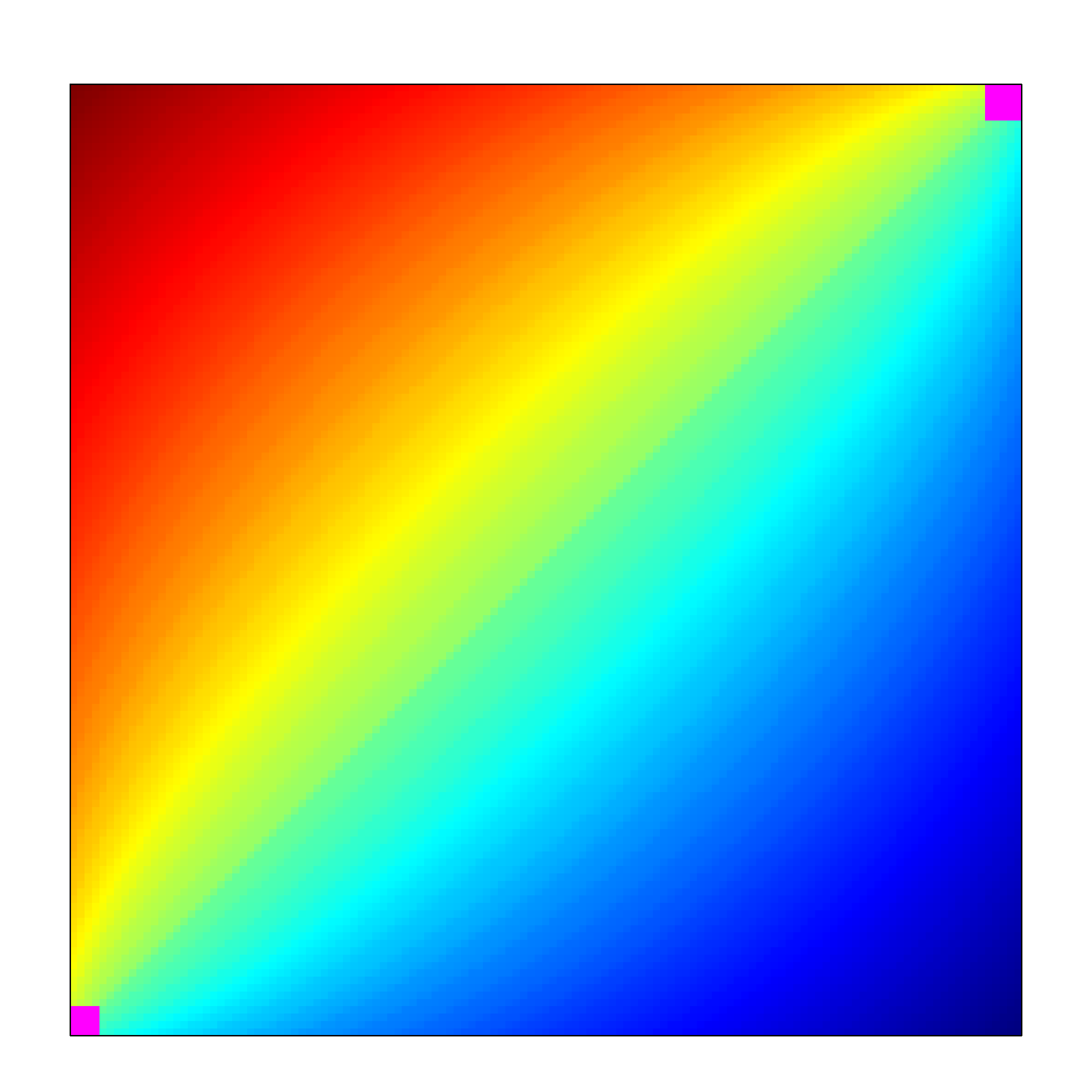}
			\\\\
			$\frac{P}{n} = \frac{1}{6}$ & $\frac{P}{n} = \frac{1}{2}$
		\end{tabular}
		\label{fig:MCC-cross-sections}}\quad\quad
	\subfloat[$OP$]{
		\centering
		\begin{tabular}{c@{\qquad\quad}c}
		\centering
			\begin{overpic}[width=0.08\textwidth]{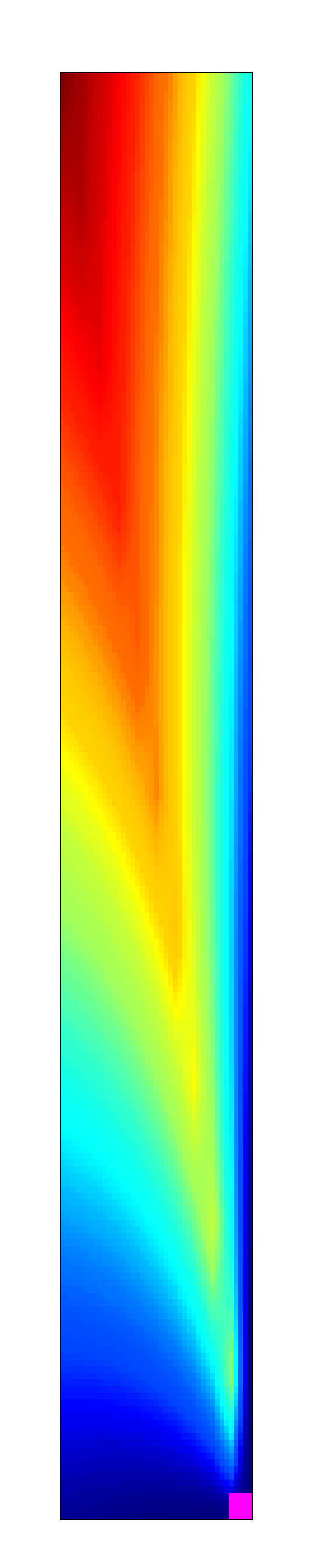}
						\labelsImbalanced
		  \end{overpic}
			&
			\includegraphics[width=0.25\textwidth]{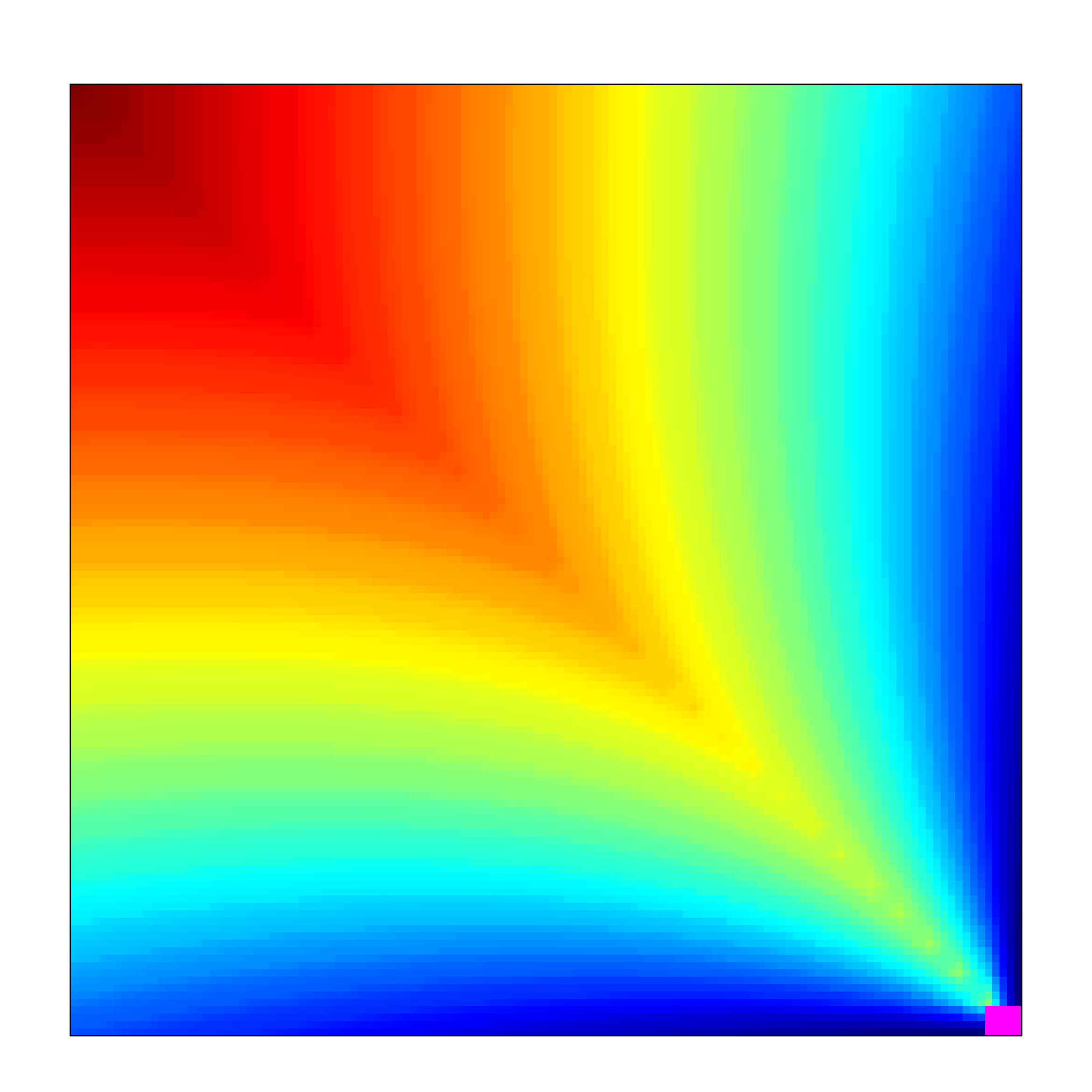}
			\\\\
			$\frac{P}{n} = \frac{1}{6}$ & $\frac{P}{n} = \frac{1}{2}$
		\end{tabular}
		\label{fig:OP-cross-sections}}
	\caption{Cross-sections of selected measures for $P/n=1/6$ and $P/n=1/2$}%
	\label{fig:MeasureComparison}
\end{figure}

\subsection{Non-parametric Measures}
\label{sec:non-parametric}

Let us now conduct a more detailed visual analysis of the four highlighted measures from Section~\ref{sec:preliminaries-related-works}: \textit{F$_1$-score}, \textit{G-mean}, \textit{Mathews Correlation Coefficient} (\textit{MCC}), and \textit{Optimized precision} (\textit{OP}), putting particular emphasis on their behaviour with respect to imbalanced data. Due to the page limit, in this paper we present only cross-sections produced for $P/n=1/6$ and $P/n=1/2$. However, other cross-sections of the tetrahedron, including cross-sections produced for higher levels of class imbalance, can be viewed in the online visualization tool.

The analysis of \textit{F$_1$-score}, visualized in Fig.~\ref{fig:F-cross-sections}, show that the growth (although monotonic) of the measure along side $\overline{\mathsf{TP}}$ is very slow and does not fulfil the $\mathit{ACE}$ property when the data are imbalanced. To illustrate this, consider Fig.~\ref{fig:F-cross-sections}~(left), which corresponds to class imbalance, and a point located in the middle of $\overline{\mathsf{TP}}$. The value there is much lower than the corresponding point on side $\overline{\mathsf{TN}}$. Taking into account the fact that the middle point of $\overline{\mathsf{TP}}$ corresponds to full recognition of the positive class and 50\% recognition of the negative class, this shows that with class imbalance high values corresponding to full recognition of the positive class are harder to obtain. Expressed in terms of values in the confusion matrix: 
$f(\left[ \begin{smallmatrix} \mathit{P} & \enskip \mathit{0} \\ \mathit{\gamma N} & \enskip \mathit{(1-\gamma)N} \end{smallmatrix} \right]) <  f(\left[ \begin{smallmatrix} \mathit{(1-\gamma) P} & \enskip \mathit{\gamma P} \\ \mathit{0} & \enskip \mathit{N} \end{smallmatrix} \right])$, where $\gamma = 1/2$.
Notice that while \textit{F$_1$-score} fulfils $ACE$ for $P/n=1/2$ (\ref{fig:F-cross-sections}~(right)), it does not for the above mentioned $P/n=1/6$ (\ref{fig:F-cross-sections}~(left)), which means that the property is not satisfied in general (i.e. throughout the tetrahedron). Evidently, the property cannot be verified using only one selected cross-section.
As may be observed using the online tool (in particular, by animating $P/n$ from $1/2$ down to $0$), this flawed feature of the measure aggravates for increasing class imbalance (i.e. when $P/n$ drops).
This may be quite surprising as the \textit{F$_1$-score} is often brought out in the literature as especially suited for the positive class. Generalizations of \textit{F$_1$-score} will be discussed in subsection~\ref{sec:parametric} devoted to parametric measures.

The visual-based analysis of \textit{G-mean} (Fig.~\ref{fig:G-mean-cross-sections}) reveals that the measure satisfies the devised properties. In particular, it satisfies some important properties not fulfilled by \textit{F$_1$-score}, \textit{MCC} and \textit{OP}. First, as opposed to the other three measures, \textit{G-mean} features minimal values on whole sides $\overline{\mathsf{FN}}$ and $\overline{\mathsf{FP}}$. Additionally, it enjoys the $ACE$ property, which makes the measure especially useful in the contexts of imbalanced data: for any two corresponding points on sides $\overline{\mathsf{TP}}$ and $\overline{\mathsf{TN}}$, the value on side $\overline{\mathsf{TP}}$ happens to be equal (and thus not smaller) to that on $\overline{\mathsf{TN}}$. 
This means that for any $\gamma$: $f(\left[ \begin{smallmatrix} \mathit{P} & \enskip \mathit{0} \\ \mathit{\gamma N} & \enskip \mathit{(1-\gamma)N} \end{smallmatrix} \right]) =  f(\left[ \begin{smallmatrix} \mathit{(1-\gamma) P} & \enskip \mathit{\gamma P} \\ \mathit{0} & \enskip \mathit{TN} \end{smallmatrix} \right])$.

As to the behaviour of \textit{MCC} (Fig.~\ref{fig:MCC-cross-sections}), one can observe that its values on sides $\overline{\mathsf{FP}}$ and $\overline{\mathsf{FN}}$ are not minimal, which violates properties $\overline{\mathsf{FN}}_{min}$ and $\overline{\mathsf{FP}}_{min}$. Even worse, comparing cross-sections for $P/n = 1/2$ and $P/n = 1/6$, one can observe that small values are harder to obtain with the increase of class imbalance. 
Furthermore, similarly to the \textit{F$_1$-score}, \textit{MCC} does not satisfy property $ACE$. Even though for balanced classes (Fig.~\ref{fig:MCC-cross-sections}(right)) the corresponding points in $\overline{\mathsf{TP}}$ and $\overline{\mathsf{TN}}$ feature equal values, this deteriorates with growing disproportion between classes (Fig.~\ref{fig:MCC-cross-sections}(left)). In other words, for imbalanced data it is easier to obtain undue high values by recognizing the negative class.

Finally, let us consider measure \textit{OP} (Fig.~\ref{fig:OP-cross-sections}). 
The visual-based analysis reveals that \textit{OP} is the only of the selected measures that does not satisfy properties $TP_{\nearrow}$ and $TN_{\nearrow}$. Observe that traversing the cross-sections horizontally right-to-left or vertically bottom-up (thus increasing the recognition of one of the classes while keeping the recognition of the second one constant) the values of the measure first increase and then decrease. 
In fact, the visual analysis discloses that the measure is designed to increase its values monotonically only when the recognition of both classes increases. Undeniably, the increase of the recognition of both classes at the same time is highly desirable and should imply increasing measure values, however, the observed behaviour of \textit{OP} in (acceptable) cases when the classifier increases the recognition of one class, while keeping the recognition of the other constant is rather surprising and counter-intuitive. 

\subsection{Parametric Measures}
\label{sec:parametric}

Recalling that the classifier performance measures are functions of the four entries of the confusion matrix, it may be observed that as far as their analytical forms are concerned, the various measures may be divided into unparametrized (e.g. \textit{G-mean} measure) and parametrized (e.g. \textit{F$_\beta$} measure). This parametrization process has been designed to lend the measures some amount of universality, as is the case with \textit{F$_\beta$}, where the $\beta$ parameter is supposed to control the class bias. As such control is much desired, external parametrization procedures have also been developed to modify the measures' behaviour, e.g. by adapting them to problems with imbalanced data. One such procedure, called \textit{Index of Balanced Accuracy} (\textit{IBA$_\alpha$})~\cite{Garcia09,Garcia10,Garcia-et-al--2014}, 
produces a parametrized measure, in which the $\alpha$ controls the amount by which the original measure is actually modified.

The above approaches allow us to focus on two following parametrization types:
\begin{itemize}
	\item	internal parametrization, (e.g. \textit{F$_\beta$}),
	\item	external parametrization, (e.g. \textit{IBA$_\alpha$}(\textit{G-mean})),
\end{itemize}
though also a kind of a simultaneous parametrization, e.g. \textit{IBA$_\alpha$}(\textit{F$_\beta$}), is feasible.

Observe that measure parametrization actually increases the number of available degrees of freedom, making the inherently complex analyses of such measures even more challenging. The principal question is: how are the particular parameter values to be established? And further, what are their applicability ranges?

Procedures adapted to answer these questions vary from simple trial-and-error approaches to more intricate ones, in which parameter values are possibly gleaned from accessible data.
In all cases visualization seems indispensable, providing valuable insights as to the measures' behaviour throughout their multidimensional, parametrized domains. 

Let us now conduct a more detailed visual analysis of measures, representing both types of parametrization: \textit{F$_\beta$} (internal) and \textit{IBA$_\alpha$}(\textit{G-mean}) (external), which illustrates the impact of the parametrization upon the measures' behaviour with respect to imbalanced data. 
Consistently, we present only cross-sections produced for $P/n=1/6$ and $P/n=1/2$, while other cross-sections as well as the entire tetrahedrons can be viewed in our online visualization tool.

\subsubsection{Internal parametrization: \textit{F$_\beta$}}

While \textit{F$_1$-score} is a regular harmonic mean of \textit{precision} and \textit{recall}, \textit{F$_\beta$} originated as a weighed version of this mean. In \textit{F$_\beta$} $\lambda$ and $1 - \lambda$ act as non-negative ($0 \leq \lambda \leq 1$) weights of \textit{precision} and \textit{recall}, respectively. This means that $\lambda$ may be chosen to produce any convex combination of $\frac{1}{\mathit{precision}}$ and $\frac{1}{\mathit{recall}}$ to be actually used in the mean. Let $p$ denote \textit{precision} and $r$ denote \textit{recall}, the weighted harmonic mean of $p$ and $r$ is:
$
 (\frac{\lambda \frac{1}{p}+(1-\lambda)\frac{1}{r}}{\lambda + (1-\lambda)})^{-1} = 
 \frac{\lambda + (1-\lambda)}{\lambda \frac{1}{p}+(1-\lambda)\frac{1}{r}} = 
 \frac{1}{\lambda \frac{1}{p}+(1-\lambda)\frac{1}{r}} = 
 \frac{1}{\lambda \frac{r}{p r}+(1-\lambda)\frac{p}{p r}} =
 \frac{p r}{\lambda r+(1-\lambda) p} =
 \frac{\frac{1}{\lambda} p r}{r+\frac{1-\lambda}{\lambda} p}
$ (from now on: $\lambda > 0$).

After setting\footnote{Some authors set $\beta = \sqrt{\frac{1-\lambda}{\lambda}}$ instead, resulting in $\frac{1-\lambda}{\lambda} = \beta^2$, which allows for some further interpretation of such $\beta$ \cite{Sasaki07}; not to be pursued in this paper.} $\beta = \frac{1-\lambda}{\lambda}$, one gets $\frac{1}{\lambda} = \beta + 1$, which finally produces: \textit{F$_\beta$} = $\frac{(\beta+1) p r}{\beta p + r}$. Notice that in this scheme:
\begin{itemize}
	\item $\lambda \to 0$ corresponds to $\beta \to \infty $ (emphasis on \textit{precision}),
	\item $\lambda = 0.5$ corresponds to $\beta = 1.0$ (equal emphasis),
	\item $\lambda \to 1$ corresponds to $\beta \to 0$ (emphasis on \textit{recall}).
\end{itemize}
Of course, for $\beta = 1.0$, measure F$_\beta$ becomes $\frac{(1+1) p r}{1 \cdot p + r} = 2 \frac{p r}{p + r}$ = \textit{F$_1$-score}, which is thus the regular (unweighed) harmonic mean of \textit{precision} and \textit{recall}. 

\begin{figure}[!htb]
  \centering
	\subfloat[\textit{precision}]{
		\centering
		\begin{tabular}{c@{\qquad\quad}c}
			\begin{overpic}[width=0.08\textwidth]{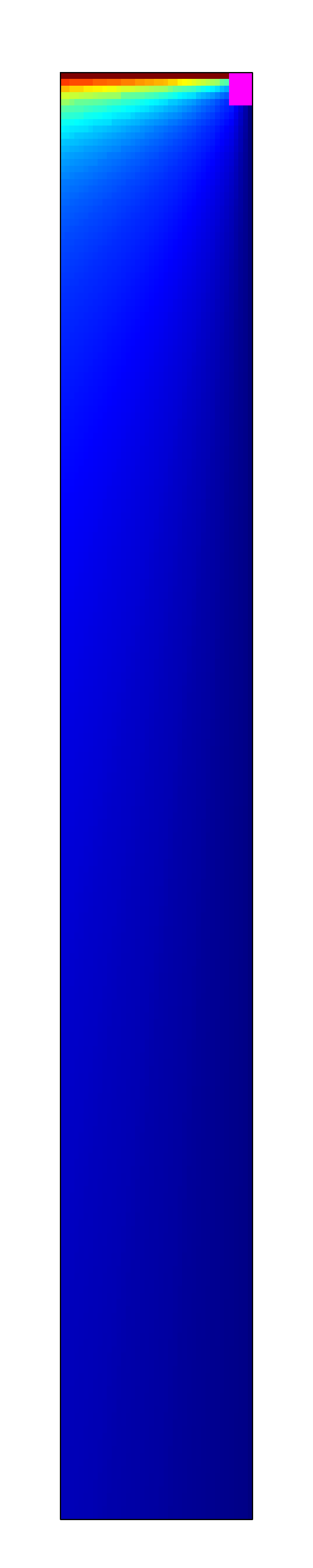}
						\labelsImbalanced
		  \end{overpic}
			&
			\includegraphics[width=0.25\textwidth]{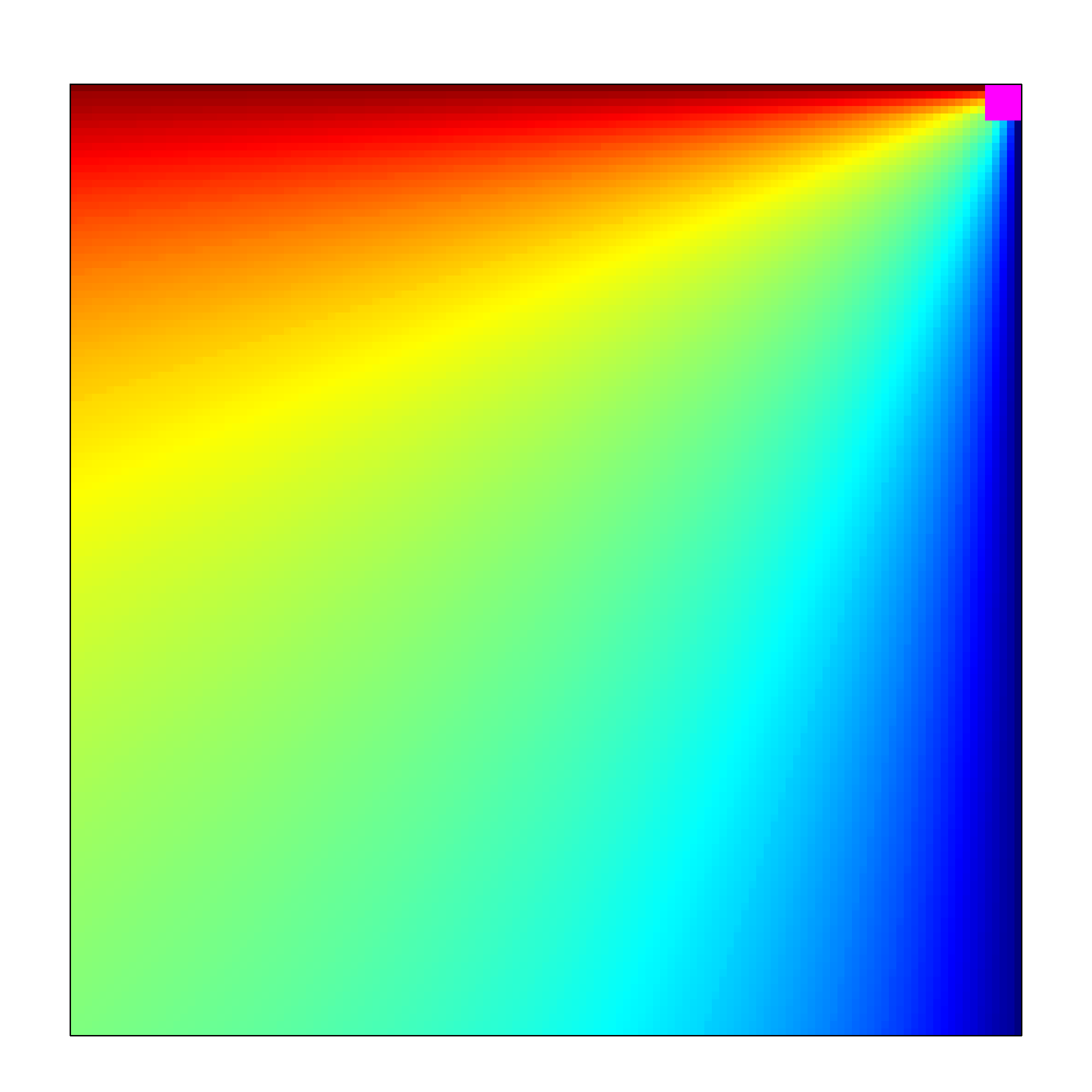}
			\\\\
			$\frac{P}{n} = \frac{1}{6}$ & $\frac{P}{n} = \frac{1}{2}$
		\end{tabular}}\quad\quad
	\subfloat[\textit{recall}]{
		\centering
		\begin{tabular}{c@{\qquad\quad}c}
			\begin{overpic}[width=0.08\textwidth]{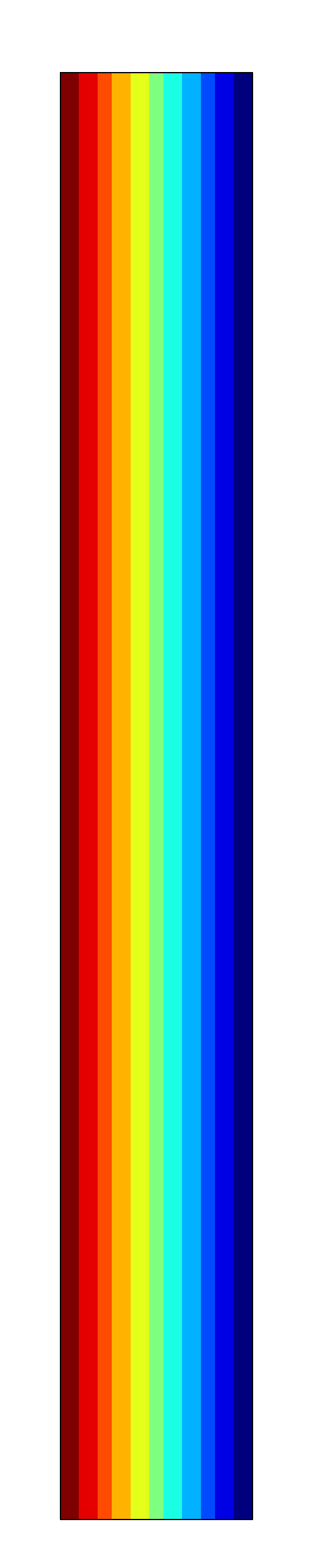}
						\labelsImbalanced
		  \end{overpic}
			&
			\includegraphics[width=0.25\textwidth]{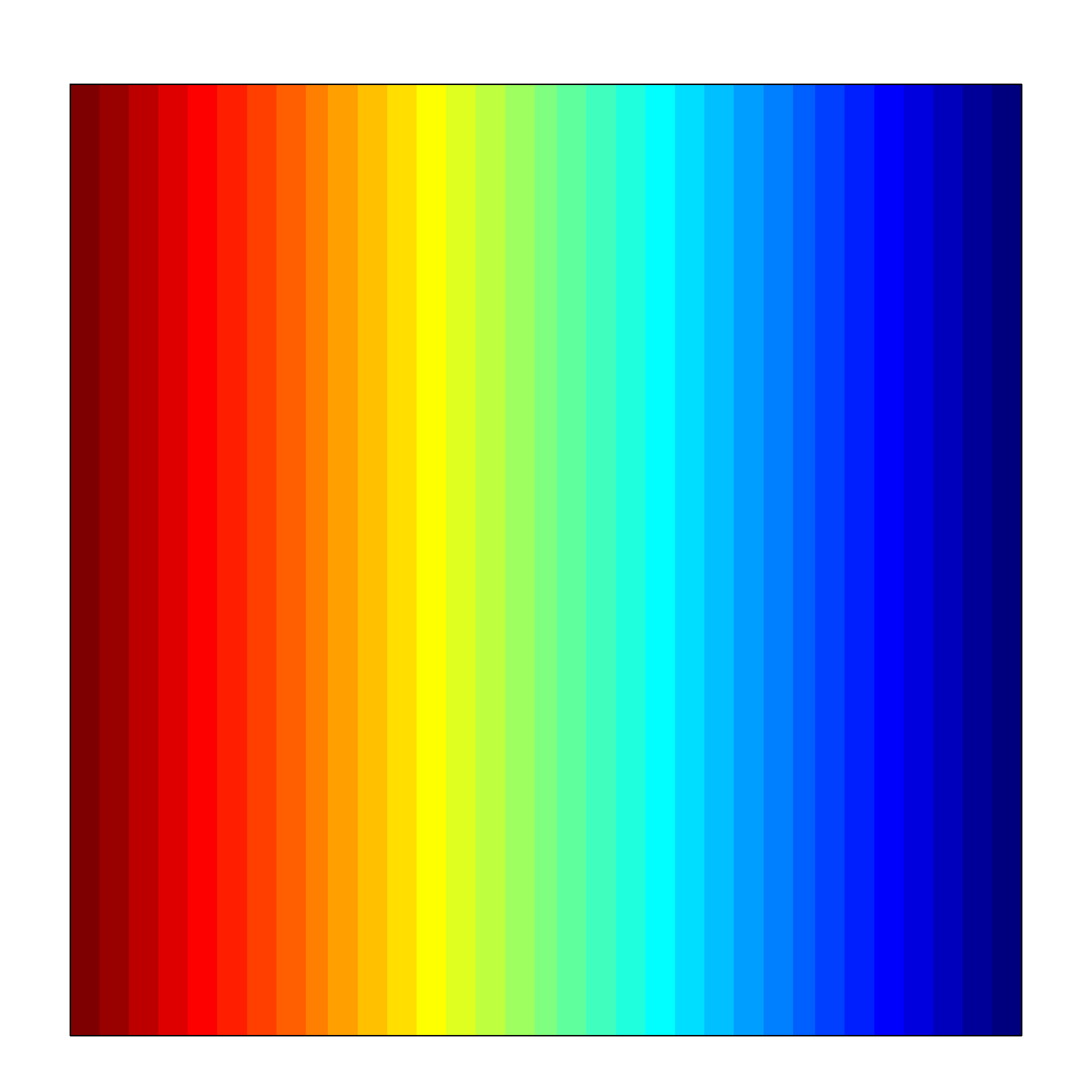}
			\\\\
			$\frac{P}{n} = \frac{1}{6}$ & $\frac{P}{n} = \frac{1}{2}$
		\end{tabular}}
	\caption{Cross-sections of \textit{precision} and \textit{recall}}%
	\label{fig:MeasureComparison--prec-rec}
\end{figure}

\begin{figure}[!htb]
  \centering
		\begin{tabular}{c@{\quad}c@{}c@{}c@{}c@{}c@{}}
			\begin{overpic}[width=0.08\textwidth]{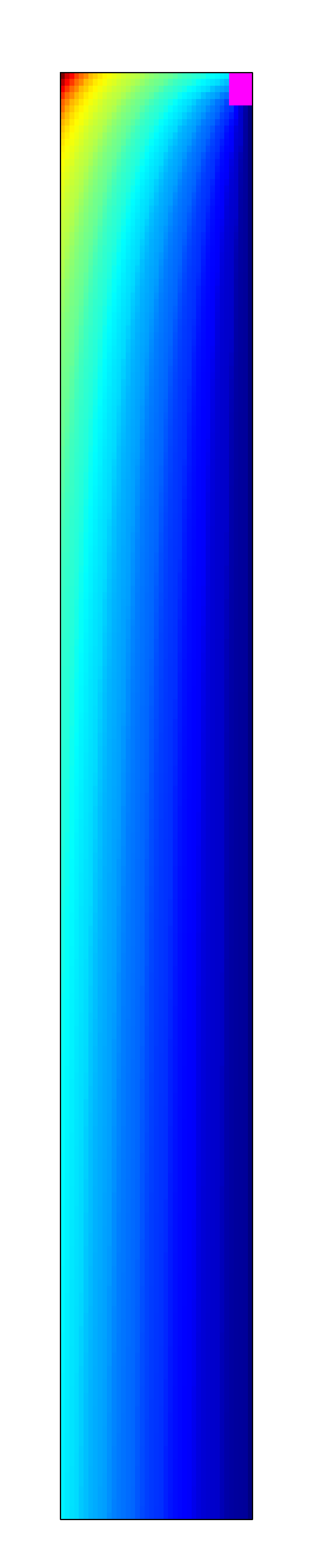}
				\labelsImbalanced
		  \end{overpic}
			&
			\includegraphics[width=0.25\textwidth]{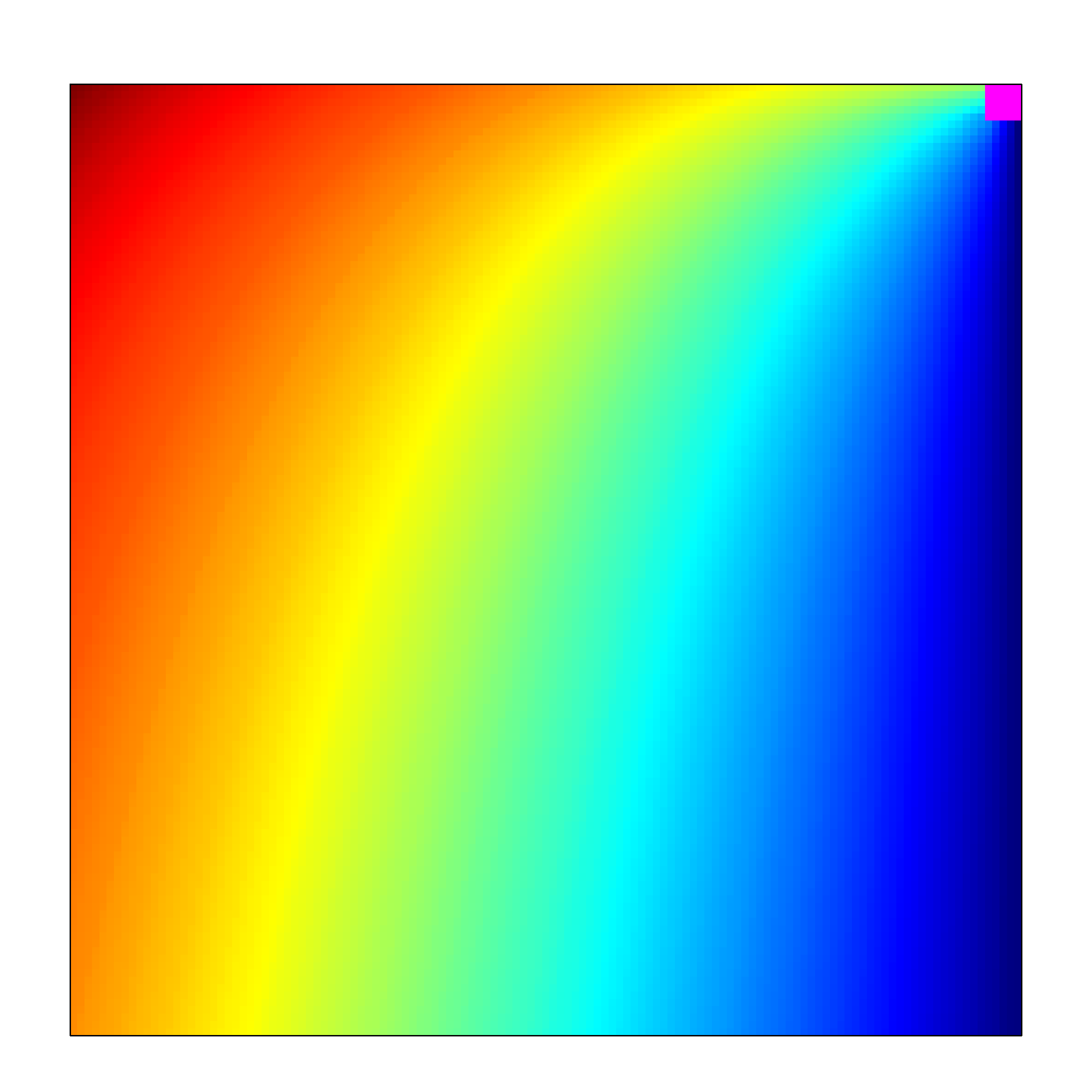}
			&
			\includegraphics[width=0.08\textwidth]{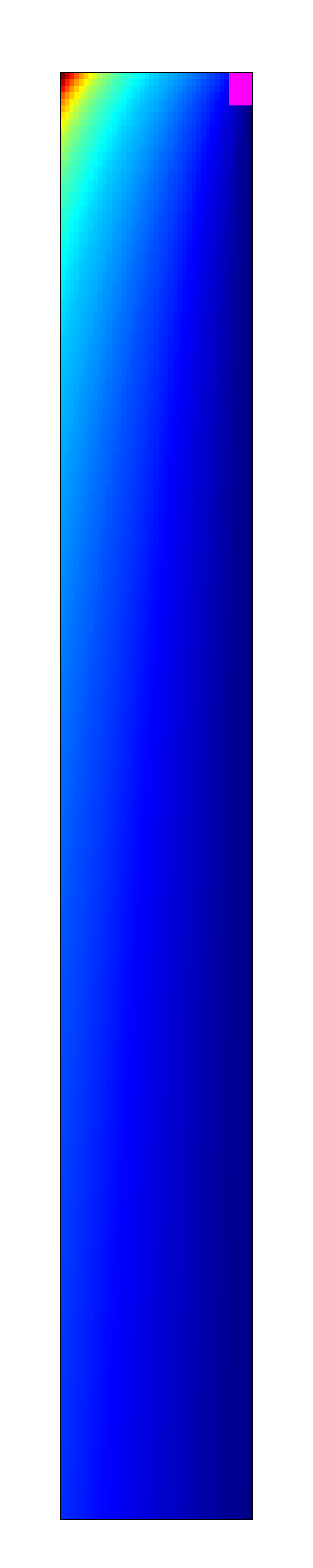}
			&
			\includegraphics[width=0.25\textwidth]{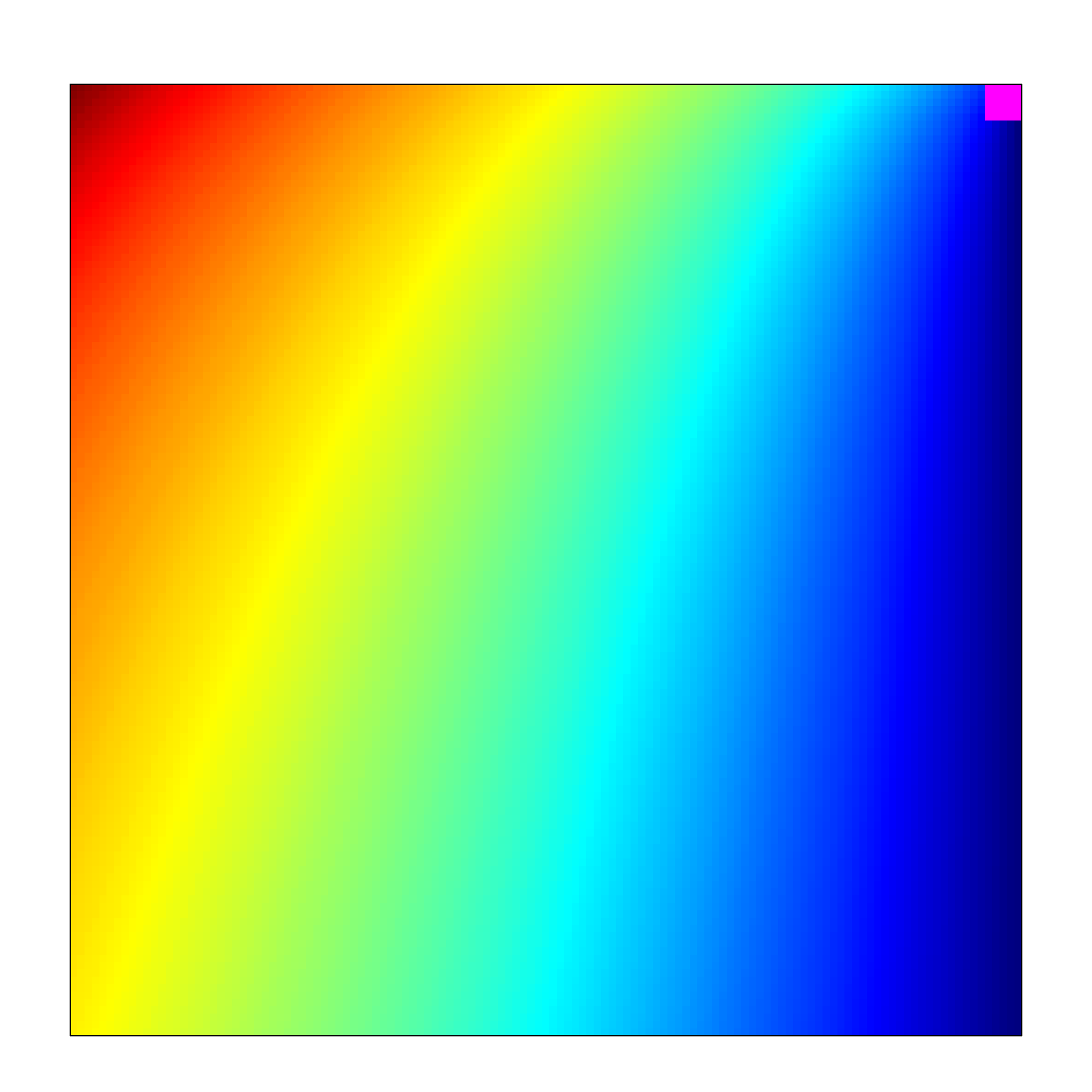}
			&
			\includegraphics[width=0.08\textwidth]{images/F1-score_16.png}
			&
			\includegraphics[width=0.25\textwidth]{images/F1-score_12.png}
			\\
			$A(p,r)$ & $A(p,r)$ & $G(p,r)$ & $G(p,r)$  & $H(p,r)$  & $H(p,r)$
			\\
					 & 			& 			& 			& \textit{F$_1$-score}  & \textit{F$_1$-score}
			\\
			\\
			$\frac{P}{n} = \frac{1}{6}$ & $\frac{P}{n} = \frac{1}{2}$ & $\frac{P}{n} = \frac{1}{6}$ & $\frac{P}{n} = \frac{1}{2}$  & $\frac{P}{n} = \frac{1}{6}$ & $\frac{P}{n} = \frac{1}{2}$
		\end{tabular}
	\caption{Cross-sections of different means of \textit{precision} and \textit{recall}}%
	\label{fig:A-G-H-means}
\end{figure}

The harmonic mean, used in this context
happens to be the most conservative of the three popular Pythagorean means: arithmetic ($A$), geometric ($G$) and harmonic ($H$), as they satisfy $A \geq G \geq H$, but it is also easy to visualize the two others in this role. 
To what extent and in which regions of the domain these three different means of \textit{precision} and \textit{recall} actually diverge from one another may be observed e.g. in Figs~\ref{fig:MeasureComparison--prec-rec} and \ref{fig:A-G-H-means}, where both \textit{precision} and \textit{recall} as well as their three means (arithmetic: $A(p,r)$, geometric: $G(p,r)$ and harmonic: $H(p,r)$) are shown. This visualization illustrates well the concave isolines of $A(p,r)$ and $G(p,r)$, which means that they obtain excessively high values for increasingly divergent recognition of classes, making $H(p,r)$ the best choice out of three in this respect.


Deciding on the mean, however, is not enough, as the remaining problem regards changes in the measure's behaviour across the differing $P/n$. Unfortunately, for the harmonic mean as well as for the other two means, the measure's values gradually shift away from the positive class as $P/n$ decreases, making all the three (regular) means of \textit{precision} and \textit{recall} (and thus the \textit{F$_1$-score} in particular) less and less suited for imbalanced data. This is where the weighed means, in particular \textit{F$_\beta$} (the weighed harmonic mean of \textit{precision} and \textit{recall}) may actually turn out to be more useful.

\begin{figure}[!htb]
  \centering
		\begin{tabular}{c@{\quad}c@{}c@{}c@{}c@{}c@{}}
			\begin{overpic}[width=0.08\textwidth]{images/F1-score_16.png}
				\labelsImbalanced
		  \end{overpic}
			&
			\includegraphics[width=0.25\textwidth]{images/F1-score_12.png}
			&
			\includegraphics[width=0.08\textwidth]{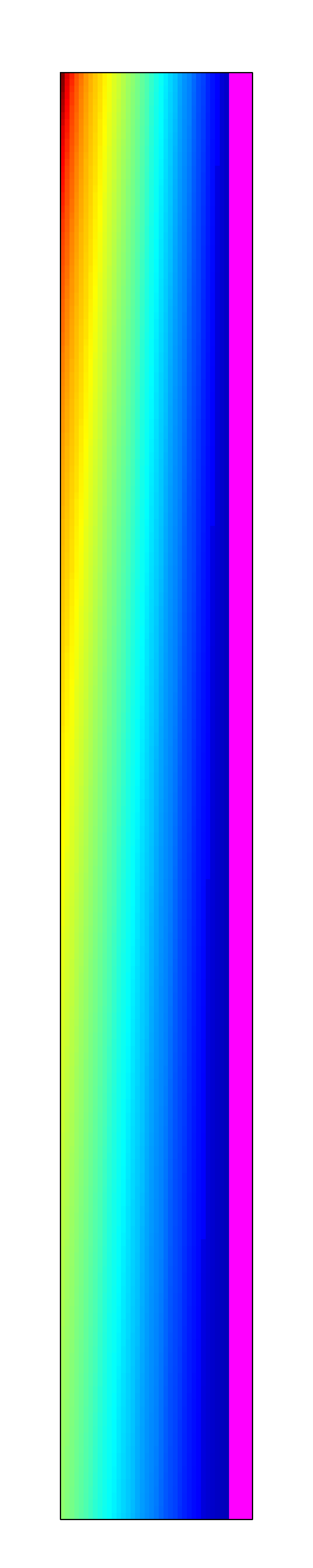}
			&
			\includegraphics[width=0.25\textwidth]{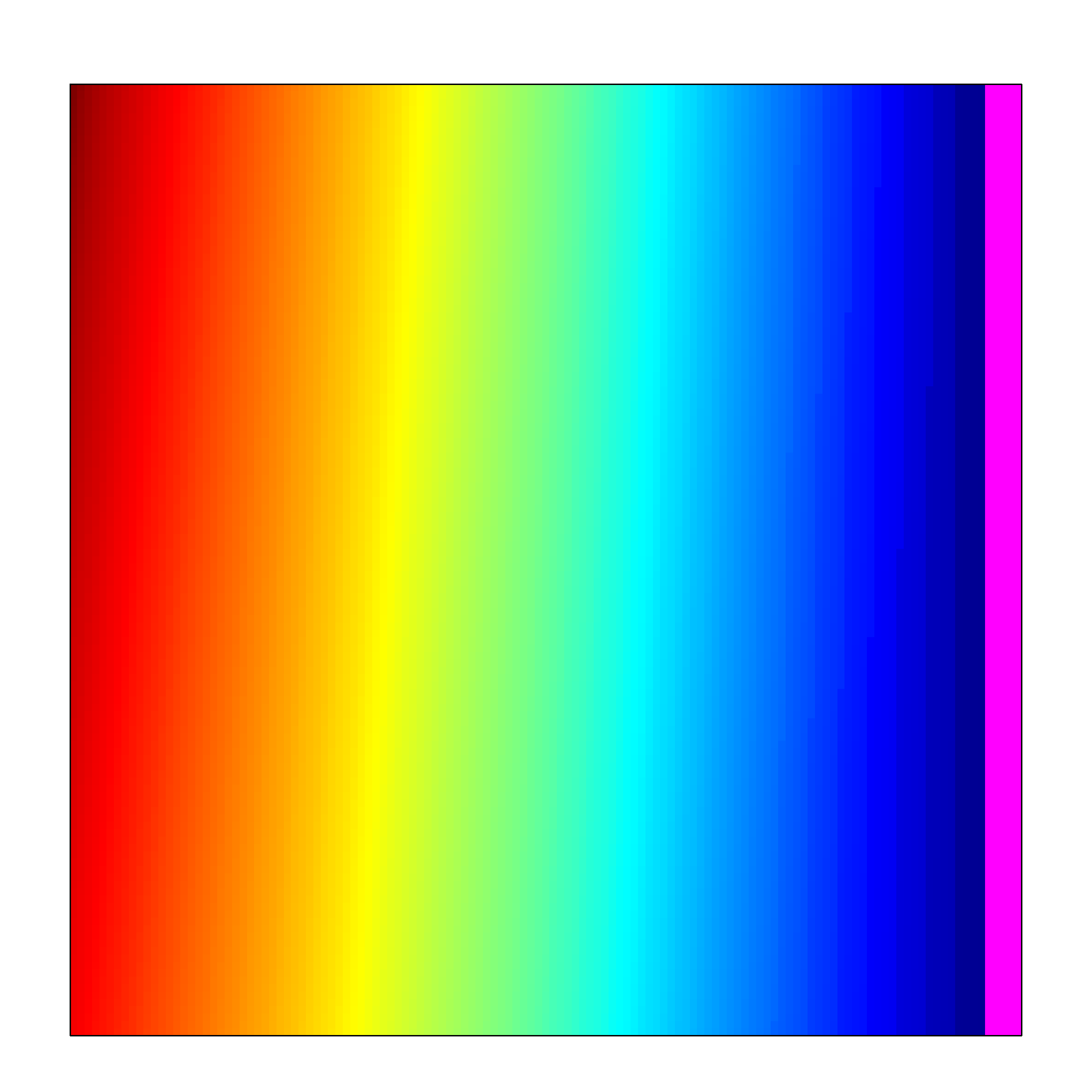}
			&
\includegraphics[width=0.08\textwidth]{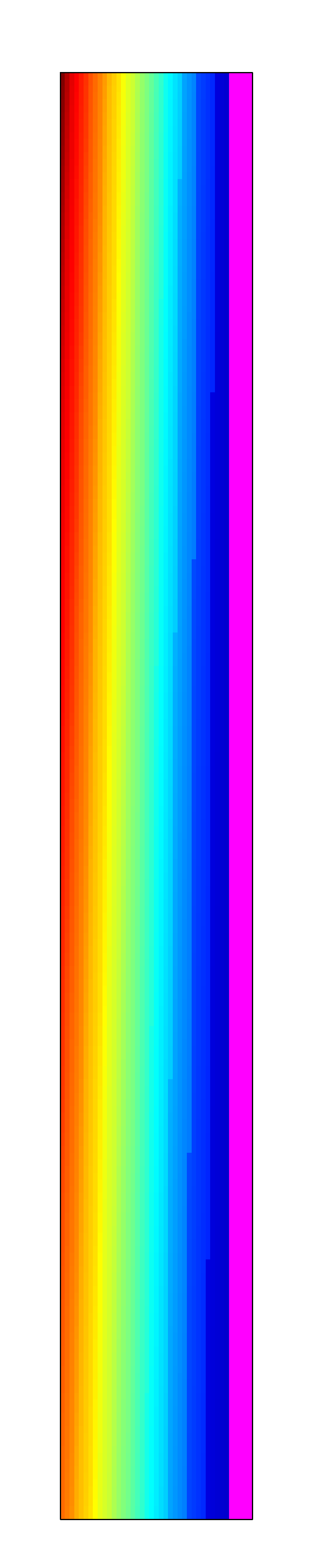}
&
\includegraphics[width=0.25\textwidth]{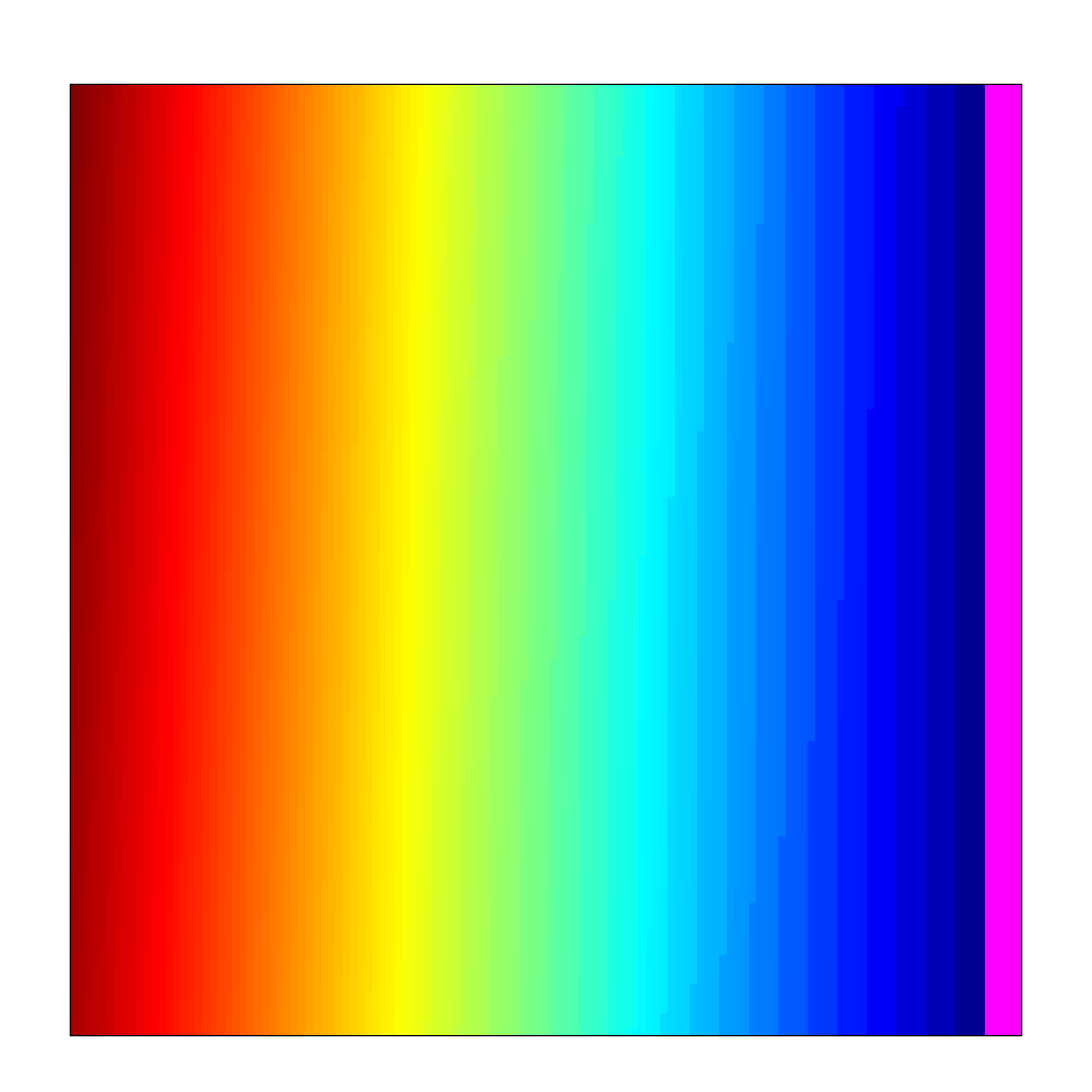}
			\\ 
			$\beta = 1.0$ & $\beta = 1.0$ & $\beta = 3.0$ & $\beta = 3.0$  & $\beta = 5.0$  & $\beta = 5.0$ 
			\\
			\textit{F$_1$-score}		 &  \textit{F$_1$-score}		& 			& 			&   & 
			\\
			\\
			$\frac{P}{n} = \frac{1}{6}$ & $\frac{P}{n} = \frac{1}{2}$ & $\frac{P}{n} = \frac{1}{6}$ & $\frac{P}{n} = \frac{1}{2}$  & $\frac{P}{n} = \frac{1}{6}$ & $\frac{P}{n} = \frac{1}{2}$
		\end{tabular}
	\caption{Cross-sections of \textit{F$_\beta$}}%
	\label{fig:MeasureComparison--Fbeta}
\end{figure}

The arising question regards the appropriate value of $\beta$. Clearly, the desired bias towards the positive class requires $\beta > 1$, which corresponds to applying more weight to \textit{recall}. 
The visual solution to this problem is provided in Fig.~\ref{fig:MeasureComparison--Fbeta}, which shows cross-section visualizations of \textit{F$_\beta$} for three values of $\beta \in \{1, 3, 5\}$. The range of these values has been inspired by the accessible data, in this case the class ratios considered in previous sections: $P/n = 1/6$ and $P/n = 1/2$. These values may be assumed to directly express the $[0,1]$-based weights of \textit{precision} and \textit{recall}, i.e. $\lambda = 1/6$ and $\lambda = 1/2$, which translate to $\beta = 5$ and $\beta = 1$, respectively.

Despite the fact that all ten of the earlier discussed properties of \textit{F$_1$-score} and \textit{F$_\beta$} (for $\beta \in [0, \infty)$) are identical, as exemplified in Table~\ref{tab:measure-properties}, the increasing difference between \textit{F$_1$-score} and \textit{F$_\beta$} resulting from the changing values of $\beta$ is clearly visible in Fig.~\ref{fig:MeasureComparison--Fbeta}, revealing the truly multidimensional complexity of the measures' domains.

A slightly closer explanation may only be due for $ACE$ property, as \textit{F$_\beta$}'s particular visualization for $\beta=5$ in Fig.~\ref{fig:MeasureComparison--Fbeta} suggests that the measure satisfies the requirements of $ACE$ (its values on the $\overline{\mathsf{TP}}$ side are not lower than their counterparts on the $\overline{\mathsf{TN}}$ side for both $P/n=1/6$ and $P/n=1/2$), whereas Table~\ref{tab:measure-properties} states that $ACE$ is not met by \textit{F$_\beta$}.
This is because the ten proposed properties are of general character, i.e. they concern the whole tetrahedron, which means that they must be satisfied in cross-sections corresponding to all feasible class proportions. In case of \textit{F$_{\beta=5}$}, for some class ratios that are lower than those considered in the presented visualizations, e.g. for $P/n = 1/10$ (easily reproducible in the online visualization tool), the $ACE$ conditions are actually not satisfied, thus justifying the contents of Table~\ref{tab:measure-properties}. 

Nevertheless, for cases when the class ratio is known or predictable, the visualizations are of utmost practical value. In the discussed situation, the visual-based analysis may suggest non-trivial values of $\beta$ for which \textit{F$_\beta$} certainly satisfies selected properties, in this case the conditions of $ACE$ for a particular $P/n$.
A thorough analysis of cross-sections clearly suggested the existence of a particular dependency between $\beta$ and the class proportion, which influences the $ACE$ property. This observation inspired us to derive analytically the borderline value of $\beta$ that ensures that $ACE$ is met by \textit{F$_\beta$}. 
\begin{prop}
\textit{F$_\beta$} satisfies $ACE$ property for $\beta \geq N/P$ (for proof see the Appendix).
\end{prop}
Practically this means that the user must bear in mind the class proportions and may use it to make \textit{F$_\beta$} satisfy $ACE$ property, if needed.

\subsubsection{External parametrization: \textit{IBA$_\alpha$}(\textit{G-mean})}
Applying any external parametrization, e.g. the \textit{IBA$_\alpha$} scheme~\cite{Garcia09,Garcia10,Garcia-et-al--2014}, to different measures evokes the usual problems, first of all related to establishing the values of required parameters. Visualization provides a very practical solution to these issues, as shall be demonstrated in this section.

Given a classifier performance measure $M$, a parameter $\alpha \ge 0$ and a tentative measure $Dom = sensitivity - specificity$, the formula:
$$\textit{IBA$_\alpha$}(M) = (1+\alpha Dom)M$$ 
defines the parametrization of $M$, in which this measure is multiplicatively combined with $(1+\alpha Dom)$. Of course, $\textit{IBA$_\alpha$}(M) = M$ for $\alpha = 0$. Simultaneously, when $Dom \in [-1,+1]$ and $\alpha \leq 1$ then $1+\alpha Dom \geq 0$, which, together with $M \geq 0$, implies $\textit{IBA$_\alpha$}(M) \geq 0$.  

The scheme has been conceived to increase the measure orientation towards the positive class, which makes it a good choice in the imbalanced contexts. Notice, however, that neither $Dom$ is a classic classifier performance measure (as its domain includes negative values), nor is $\textit{IBA$_\alpha$}(M)$ a simple convex combination of $Dom$ and $M$. This renders strictly analytical (without any visualization tool) analysis of $\textit{IBA$_\alpha$}(M)$ very hard, especially for larger values of $\alpha$.
In result, while the general goal of reorienting the measure towards the positive class is certainly achieved by  \textit{IBA$_\alpha$}, it is not instantly clear how this reorientation is practically manifested. In particular, one might be interested in identifying whether measure $M$ subjected to $\textit{IBA$_\alpha$}(M)$ satisfies any of the postulated properties, or not (and, if it does, which ones and for what ranges of $\alpha$).

\begin{figure}[!htbp]
  \centering
	\subfloat[\textit{G-mean}]{
		\centering
		\begin{tabular}{c@{\qquad\quad}c}
			\begin{overpic}[width=0.08\textwidth]{images/G-mean_16.png}
						\labelsImbalanced
		  \end{overpic}
			&
			\includegraphics[width=0.25\textwidth]{images/G-mean_12.png}
			\\\\
			$\frac{P}{n} = \frac{1}{6}$ & $\frac{P}{n} = \frac{1}{2}$
		\end{tabular}
		\label{fig:G-mean-rec-spec-cross-sections}}\quad\quad
	\subfloat[\textit{Dom}]{
		\centering
		\begin{tabular}{c@{\qquad\quad}c}
			\begin{overpic}[width=0.08\textwidth]{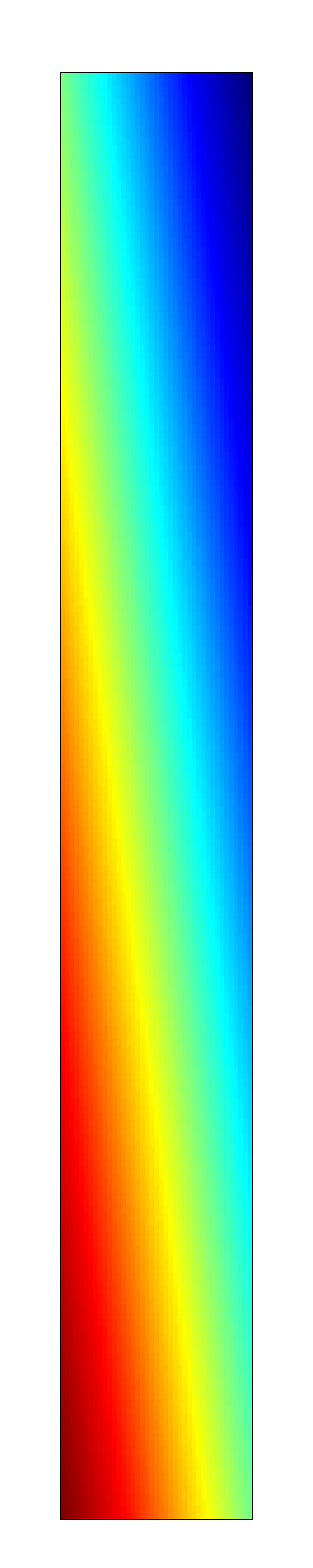}
						\labelsImbalanced
		  \end{overpic}
			&
			\includegraphics[width=0.25\textwidth]{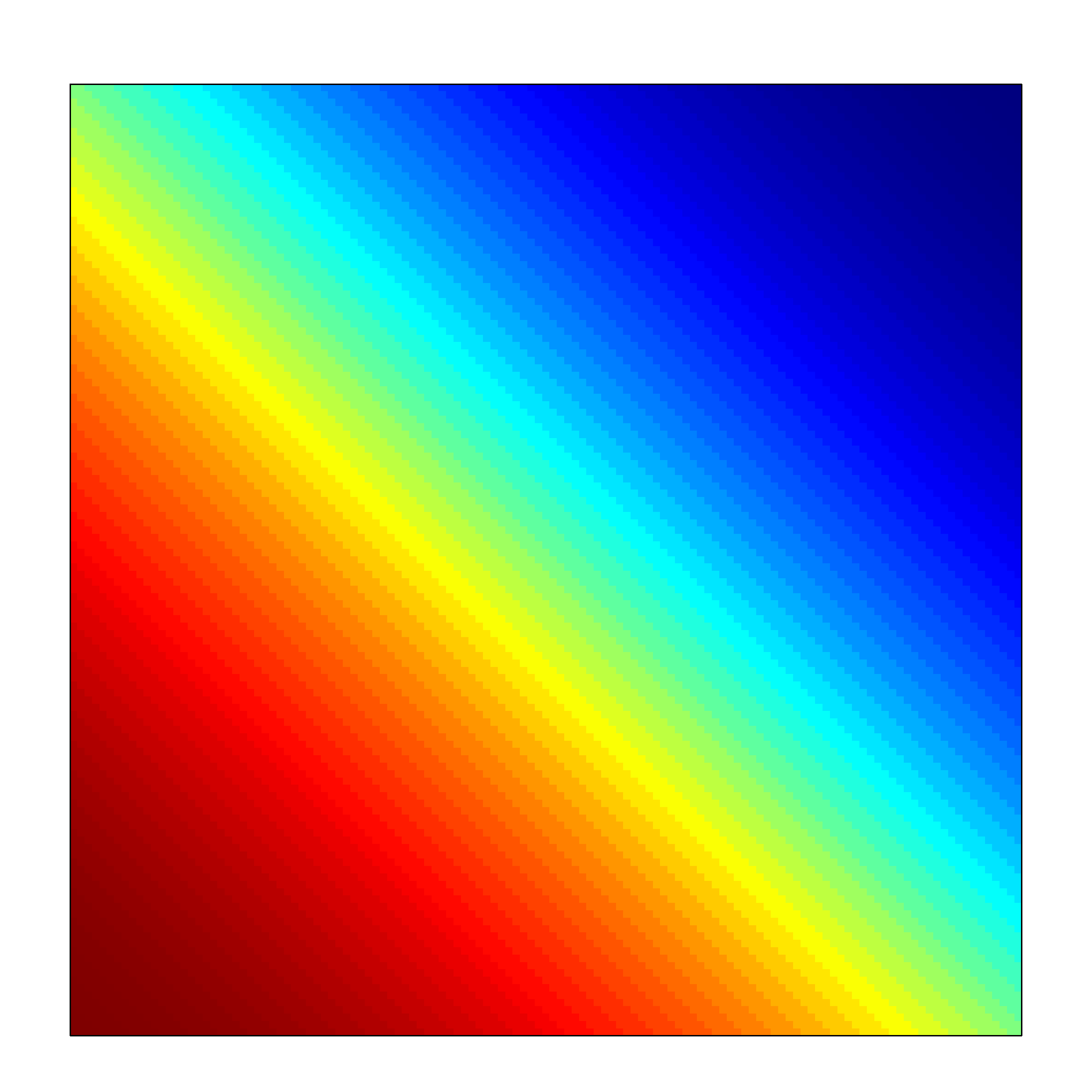}
			\\\\
			$\frac{P}{n} = \frac{1}{6}$ & $\frac{P}{n} = \frac{1}{2}$
		\end{tabular}
		\label{fig:Dom-rec-spec-cross-sections}}
	\caption{Cross-sections of \textit{G-mean} and \textit{Dom}}%
	\label{fig:MeasureComparison--G-mean--Dom}
\end{figure}

\begin{figure}[!htbp]
  \centering
		\begin{tabular}{c@{\quad}c@{}c@{}c@{}c@{}c@{}}
			\begin{overpic}[width=0.08\textwidth]{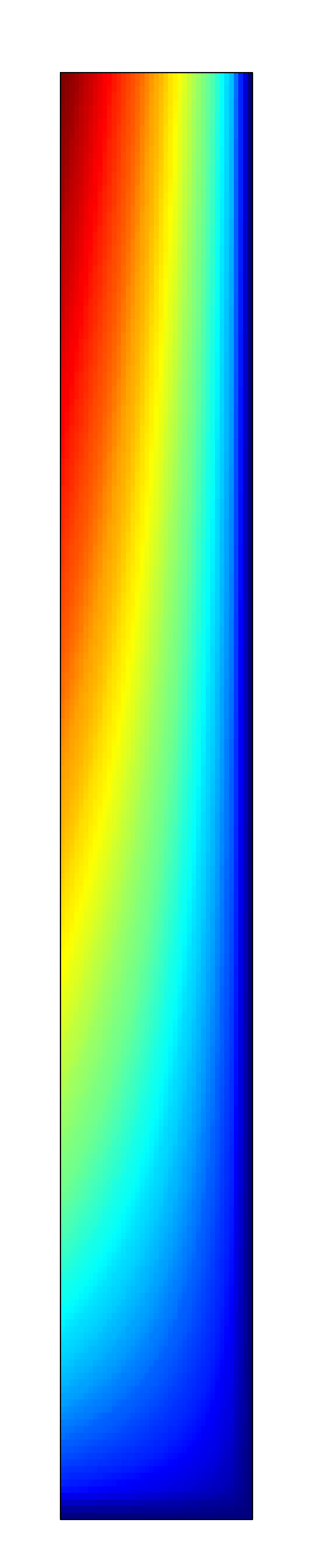}
				\labelsImbalanced
		  \end{overpic}
			&
			\includegraphics[width=0.25\textwidth]{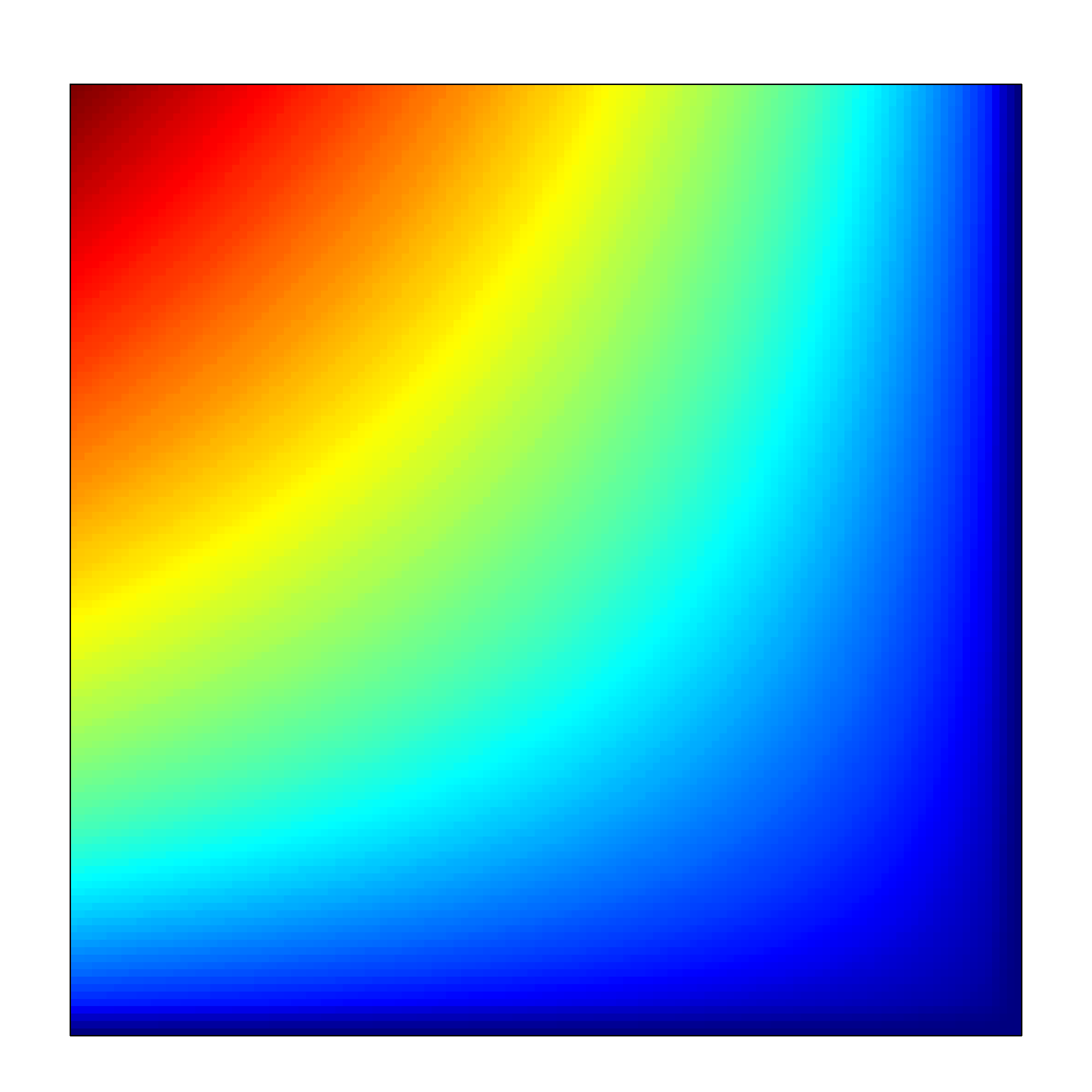}
			&
			\includegraphics[width=0.08\textwidth]{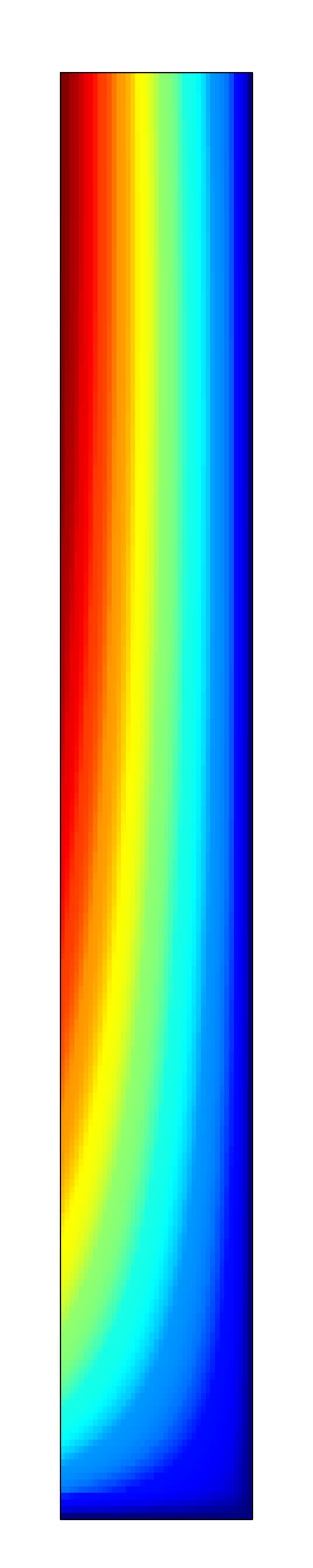}
			&
			\includegraphics[width=0.25\textwidth]{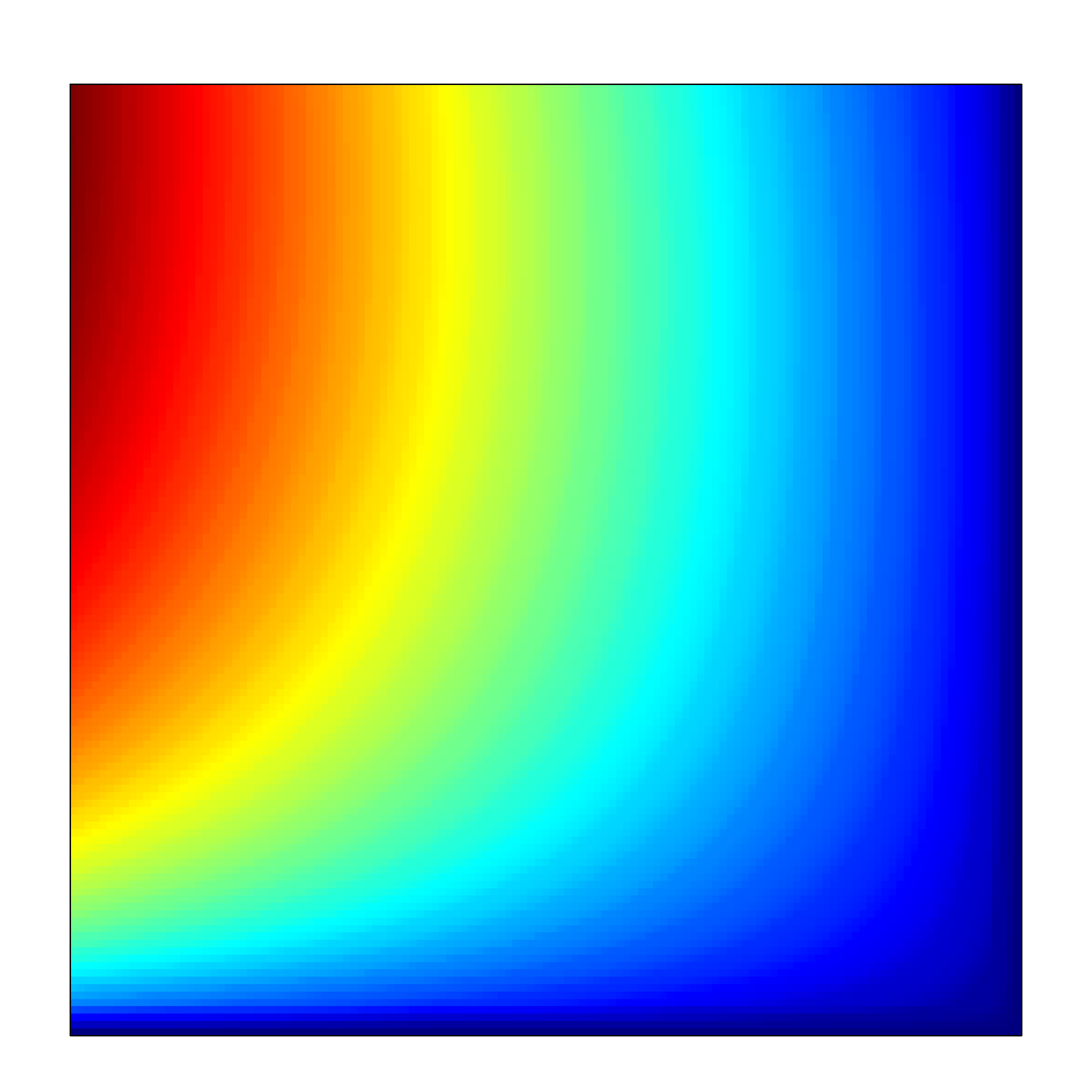}
			&
			\includegraphics[width=0.08\textwidth]{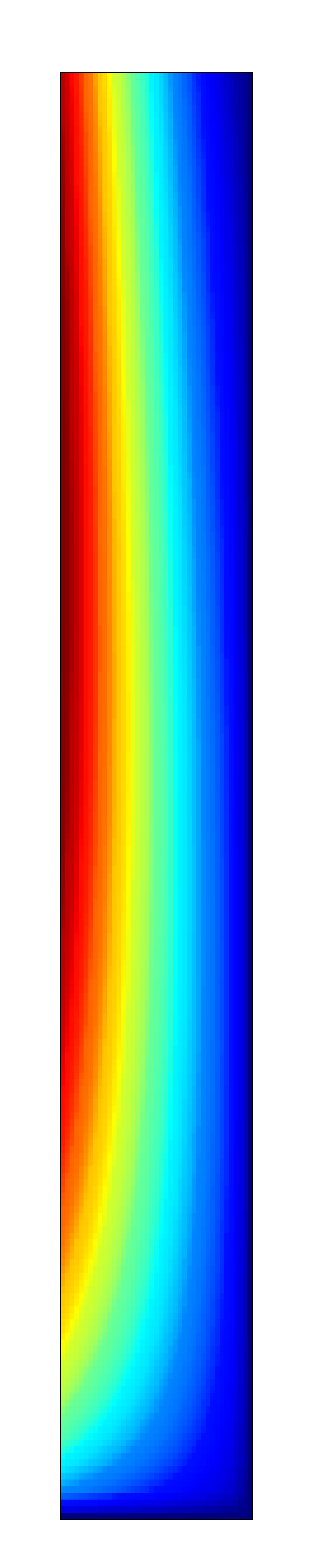}
			&
			\includegraphics[width=0.25\textwidth]{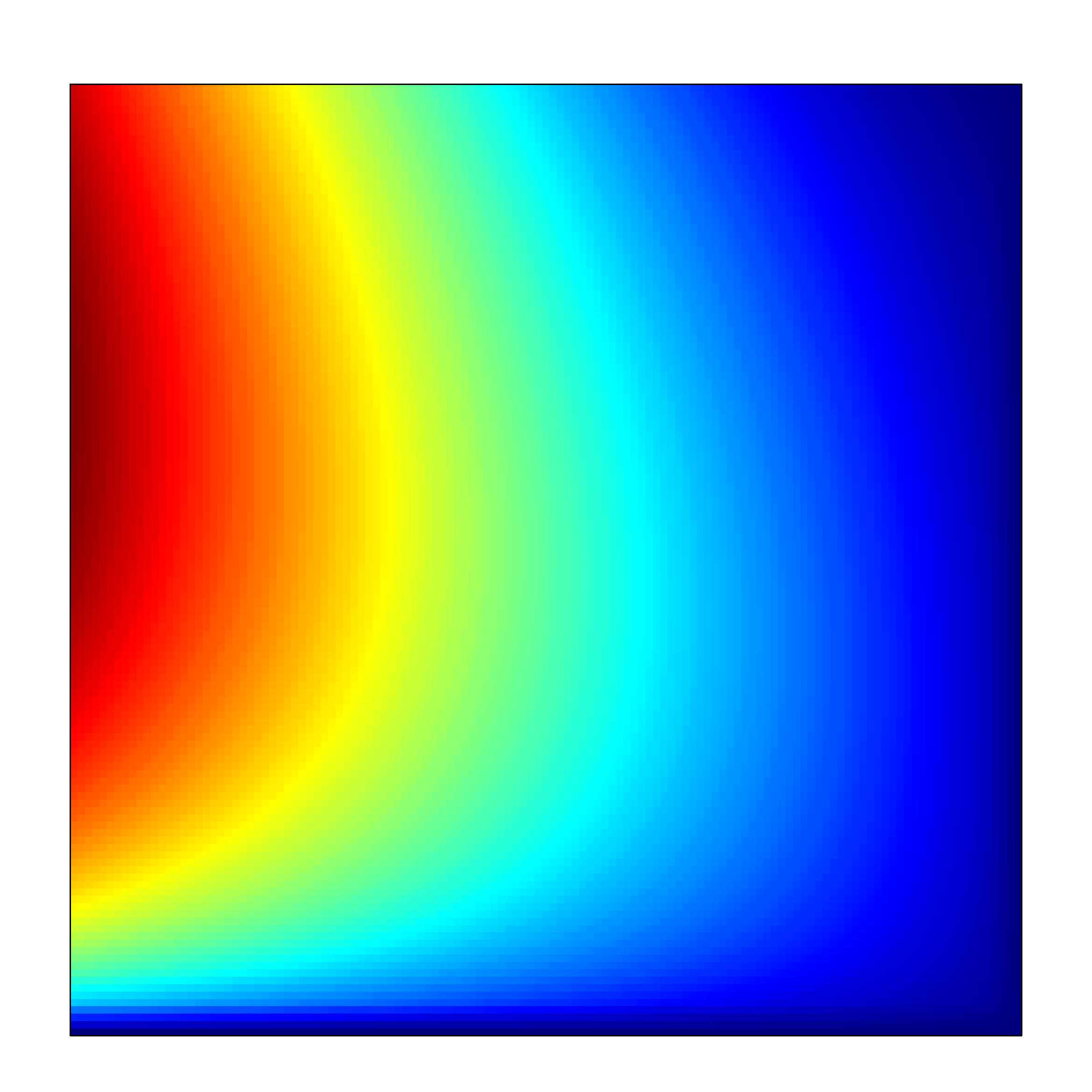}
			\\ 
			$\alpha = 0.0$ & $\alpha = 0.0$ & $\alpha = 0.5$ & $\alpha = 0.5$  & $\alpha = 1.0$  & $\alpha = 1.0$ 
			\\
			\textit{G-mean} & \textit{G-mean} & & & &
			\\\\
			$\frac{P}{n} = \frac{1}{6}$ & $\frac{P}{n} = \frac{1}{2}$ & $\frac{P}{n} = \frac{1}{6}$ & $\frac{P}{n} = \frac{1}{2}$  & $\frac{P}{n} = \frac{1}{6}$ & $\frac{P}{n} = \frac{1}{2}$ 
		\end{tabular}
	\caption{Cross-sections of \textit{IBA$_\alpha$}(\textit{G-mean})}%
	\label{fig:MeasureComparison--IBA-G-mean}
\end{figure}

Below, we visualize and analyse \textit{G-mean} externally parametrized according to \textit{IBA$_\alpha$} for $\alpha \in \{0, 0.5, 1\}$. The combination \textit{IBA$_\alpha$}(\textit{G-mean}) was particularly recommended and analytically studied for the aforementioned $\alpha$ values by Garc{\'{\i}}a et al.~\cite{Garcia09}.  
Tracing the influence of $Dom$ on \textit{G-mean} within the \textit{IBA$_\alpha$} approach may well be started with the visualization of the components of the parametrization procedure, see Fig.~\ref{fig:MeasureComparison--G-mean--Dom}, as only having realized the behaviour of \textit{G-mean} and $Dom$, can one infer how the changing $\alpha$ impacts the parametrized measure.  
Clearly, for $\alpha \to 0$, \textit{IBA$_\alpha$}(\textit{G-mean})$ \to $ \textit{G-mean}, so only $\alpha > 0$ exerts any influence on the result.
Notice that $Dom$ features a rather unexpected growth towards vertex $\mathsf{TPFP}$, implying the specific behaviour of \textit{IBA$_\alpha$}(\textit{G-mean}), see Fig.~\ref{fig:MeasureComparison--IBA-G-mean}. Because the combination is multiplicative, the values of \textit{G-mean} are being `amplified' by the corresponding values of $(1+\alpha Dom)$, in particular: increased for $(1+\alpha Dom) > 1$, and decreased for $(1+\alpha Dom) < 1$. 

As stated in Table~\ref{tab:measure-properties} \textit{G-mean} satisfies all of the proposed properties. The important question is how the application of external parametrization to the measure influences its properties, e.g. the $ACE$ property (thoroughly discussed for internal parametrization). Unsurprisingly, \textit{IBA$_\alpha$}\text{(\textit{G-mean})} may be proven to satisfy the $ACE$ property for all assumed values of $\alpha$. Notably, this comes with a cost, as this parametrization of \textit{G-mean} is not equally stable with respect to all other properties.
  
\begin{prop}
\textit{IBA$_\alpha$}\text{(\textit{G-mean})} satisfies $ACE$ property for $\alpha \geq 0$ (for proof see the Appendix).
\end{prop}

Practically, this means that the \textit{IBA$_\alpha$}\text{(\textit{G-mean})} does not depart from the original \textit{G-mean} in terms of $ACE$. 
However, 
a thorough visual-based analysis of the impact of the $\alpha$ parameter on satisfying $TN_{\nearrow}$ by \textit{IBA$_\alpha$}(\textit{G-mean}) suggested a border-line value of $\alpha$. Inspired thereby, we derived the exact value analytically. 

\begin{prop}
\textit{IBA$_\alpha$}(\textit{G-mean}) satisfies $TN_{\nearrow}$ property for $\alpha \leq 1/3$ (for proof see the Appendix).
\end{prop}

Table~\ref{tab:param-G-mean-properties} gathers the results concerning the ten devised properties for particular intervals of the $\alpha$ parameter implied by its border-line value. 

\begin{table}[htb]
  \centering
  \caption{Properties of \textit{G-mean} and its parametrizations; $^{*}$: contains NaN (undefined value); $^{\dagger}$: NaN side, $^{s}$: strong monotonicity}
	\label{tab:param-G-mean-properties}%
  \begin{tabularx}{\textwidth}{@{}p{4.3cm}@{}C@{}C@{}C@{}C@{}C@{}C@{}C@{}C@{}C@{}p{2.5cm}@{}}
		\toprule
    Measure 						& $\underset{max}{{\mathsf{TPTN}}}$ & $\underset{min}{\overline{\mathsf{FN}}}$ & $\underset{min}{\overline{\mathsf{FP}}}$ & $TP_{\nearrow}$ & $TN_{\nearrow}$ & $\underset{\neq max}{\overline{\mathsf{TN}}}$ & $\underset{\neq max}{\overline{\mathsf{TP}}}$ & $\mathit{ACE}$ & $\mathit{ACH}$ & $\mathit{UnDefs}$ \\
		\midrule
     \textit{IBA$_0$}(\textit{G-mean}) = \textit{G-mean} & $\checkmark$   & $\checkmark$   & $\checkmark$   & $\checkmark$ & $\checkmark$ & $\checkmark$   & $\checkmark$   & $\checkmark$   & $\checkmark$   						& $\mathsf{TN}$--$\mathsf{FP}$; $\mathsf{TP}$--$\mathsf{FN}$ \\
    \textit{IBA$_\alpha$}(\textit{G-mean}), {\scriptsize $\alpha \in (0,1/3]$} 			& $\checkmark$   & $\checkmark$   & $\checkmark$   & $\checkmark$ & $\checkmark$ & $\checkmark$   & $\checkmark$   & $\checkmark$   & $\times$    							& $\mathsf{TN}$--$\mathsf{FP}$; $\mathsf{TP}$--$\mathsf{FN}$ \\
		\textit{IBA$_\alpha$}(\textit{G-mean}), {\scriptsize $\alpha \in (1/3,\infty)$} 					& $\times$   & $\checkmark$   & $\checkmark$   & $\checkmark$ & $\times$ & $\checkmark$   & $\times$   & $\checkmark$   & $\times$    							& $\mathsf{TN}$--$\mathsf{FP}$; $\mathsf{TP}$--$\mathsf{FN}$ \\
		\bottomrule
  \end{tabularx}%
\end{table}%

In particular, the entries for \textit{G-mean} and for \textit{IBA$_{\alpha \in (0, 1/3]}$}(\textit{G-mean}) state that such external parametrization eliminates the symmetry of handling both classes. It is the result of incorporating the class-asymmetric $Dom$ component in the \textit{IBA$_\alpha$} parametrization procedure. The parametrized \textit{G-mean} becomes slightly (as $\alpha$ does not exceed $1/3$) more oriented towards the positive class (see also Fig.~\ref{fig:MeasureComparison--IBA-G-mean}), and thus does not satisfy the $ACH$ property any more. Nevertheless, other properties remain satisfied as long as $\alpha \leq 1/3$.  
The behaviour of \textit{IBA$_\alpha$}(\textit{G-mean}) changes drastically, however, when $\alpha$ exceeds $1/3$ (see Table~\ref{tab:param-G-mean-properties} and  Fig.~\ref{fig:MeasureComparison--IBA-G-mean}).
On one hand, for $\alpha > 0$ one gets the much desired focus on the positive class, reflected by the $ACE$ property (for any two corresponding points on sides $\overline{\mathsf{TP}}$ and $\overline{\mathsf{TN}}$, the value on side $\overline{\mathsf{TP}}$ is strictly greater than that on $\overline{\mathsf{TN}}$), however, for $\alpha > 1/3$ this comes with the inevitable cost of losing not only the above-mentioned $ACH$ property, but also the $TN_{\nearrow}$ property (manifested by non-monotonic growth of the measure from $\overline{\mathsf{FP}}$ to $\overline{\mathsf{TN}}$). 
Additionally, for $\alpha > 1/3$ the maximal value of \textit{IBA$_\alpha$}(\textit{G-mean}) drifts away from vertex $\mathsf{TPTN}$ (i.e. the full recognition of both classes is no longer rewarded with the maximal measure value) 
violating the $\mathsf{TPTN}_{max}$ and $\overline{\mathsf{TP}}_{\neq max}$ properties. In this context, the usability of \textit{IBA$_\alpha$}(\textit{G-mean}) for $\alpha > 1/3$ becomes questionable.

\section{Conclusions}
\label{sec:conclusions}
In this paper, we proposed a new visualization technique for analysing classification performance measures
and contributed an interactive tool implementing it in the form of a web application. 
The technique uses a barycentric coordinate system by projecting values from the confusion matrix into a three-dimensional figure --- a tetrahedron.
Unlike simpler visualizations this technique:
\begin{itemize}
	\item	provides general interpretations in terms of the four values of the two-class confusion matrix,
	\item	involves exclusively linear, and thus easily interpretable, 4D $\to$ 3D transformations,
	\item	allows for analysing full ranges of measure values with respect to all possible combinations of confusion matrix entries,
	\item	naturally illustrates the $TP+FN+FP+TN = n$ constraint, manifested in the shape of the space (i.e. tetrahedron), 
	\item	remains defined for all possible combinations of the matrix entries,
	\item	admits multiple cross-sections with natural interpretations in terms of simple measures, e.g. horizontal cross-sections, which correspond to the proportion of actual classes (i.e. the positive ($\frac{P}{n}$) and the negative ($\frac{N}{n}$) class) and are thus especially well suited for analysis of imbalanced data.
\end{itemize}

Using this visualization technique, we analysed 22 classifier performance measures in terms of ten purposefully defined properties, which can help assess the measures in the context of class imbalanced data. The analysis included non-parametric as well as parametric measures, which led to discovering property changes upon certain parametrizations for the latter. In particular, we have derived threshold values for selected properties of \textit{F$_\beta$} and \textit{IBA$_\alpha$}\text{(\textit{G-mean})}. The detection of these non-trivial thresholds would be difficult without the proposed visualization technique. 


The analysis of the selected measures illustrates how the proposed visualization can depict individual characteristics and potential caveats of each measure. 
It is worth stressing that it was not our intention to promote any single measure as the best, since the measure choice always finally depends on the user and the application at hand. Nevertheless, our visualization tool and the results gathered in Table~\ref{tab:measure-properties} should support making this choice.  

As future work, we plan to consider also other properties, such as gradients of measure as functions of the four arguments. Moreover, it would be interesting to analyse the effects of applying cost matrices to the visualized measures. Similarly, the effects of micro- and macro-averaging of binary measures in multi-class scenarios are worth studying. Finally, we hope that the visualization technique may be helpful in defining new classifier performance measures.

\section*{Acknowledgement}
{This research was partly supported by the FNP START scholarship (first author)
and Institute of Computing Science Statutory Funds.}

\newpage

\section*{Appendix: Proofs of Propositions}
\label{sec:appendix}

\subsection*{Proof of Proposition 1}


\noindent
For $P > 0$ (the positive class) and $N \geq 0$ (the negative class), the (positive) class ratio is expressed as $N/P$.
Given that, $F_\beta$ satisfies the $ACE$ property if 
$F_\beta(\left[ \begin{smallmatrix} \mathit{P} & \enskip \mathit{0} \\ \mathit{\gamma N} & \enskip \mathit{(1-\gamma)N} \end{smallmatrix} \right]) \geq  F_\beta(\left[ \begin{smallmatrix} \mathit{(1-\gamma) P} & \enskip \mathit{\gamma P} \\ \mathit{0} & \enskip \mathit{N} \end{smallmatrix} \right])$ for every $\gamma \in [0, 1]$, provided both sides of the inequality are defined.

Because \textit{F}$_\beta$, a function of \textit{precision} (everywhere below in this subsection: $p$) and \textit{recall} (everywhere below in this subsection: $r)$, is defined as  \textit{F}$_\beta = \frac{(1+\beta) p  r}{\beta p + r}$ with $\beta \geq 0$, and
\begin{itemize}
	\item on the left-hand side: 
	\begin{itemize}
		\item $p = \frac{P}{P+\gamma N}$,
		\item $r = \frac{P}{P+0}$, so under the assumed $P > 0$, $r = 1$,
	\end{itemize}
	\item on the right-hand side: 
	\begin{itemize}
		\item $p = \frac{(1-\gamma)P}{(1-\gamma)P+0}$, so under the assumed $P > 0$, $= 1$ for $\gamma \neq 1$, 
		\item $r = \frac{(1-\gamma)P}{(1-\gamma)P+\gamma P} = \frac{(1-\gamma)P}{(1-\gamma+\gamma)P} = \frac{(1-\gamma)P}{P}$, so under the assumed $P > 0$, $r = 1-\gamma$,
	\end{itemize}
\end{itemize}
the inequality $F_\beta(\left[ \begin{smallmatrix} \mathit{P} & \enskip \mathit{0} \\ \mathit{\gamma N} & \enskip \mathit{(1-\gamma)N} \end{smallmatrix} \right]) \geq  F_\beta(\left[ \begin{smallmatrix} \mathit{(1-\gamma) P} & \enskip \mathit{\gamma P} \\ \mathit{0} & \enskip \mathit{N} \end{smallmatrix} \right])$ is expressed as: \newline
\noindent
$\frac{(1+\beta)\frac{P}{P+\gamma N}\cdot 1}{\beta\frac{P}{P+\gamma N} + 1} \geq \frac{(1+\beta)(1-\gamma)}{\beta\cdot 1+(1-\gamma)}$ \newline
\noindent
$\frac{(1+\beta)\frac{P}{P+\gamma N}}{\beta\frac{P}{P+\gamma N} + 1} \geq \frac{(1+\beta)(1-\gamma)}{\beta+(1-\gamma)}$ \newline
\noindent
The assumed $P > 0$, $N \geq 0$, $\gamma \geq 0$, $\beta \geq 0$ ensure $\beta\frac{P}{P+\gamma N} + 1 > 0$, so: \newline
\noindent
$(1+\beta)\frac{P}{P+\gamma N} \geq \frac{(1+\beta)(1-\gamma)}{\beta+(1-\gamma)}(\beta\frac{P}{P+\gamma N} + 1)$ \newline
The assumed $\gamma \in [0,1]$ and $\beta \geq 0$ ensure $\beta + (1-\gamma) \geq 0$, so assuming additionally $\gamma < 1$ ensures $\beta + (1-\gamma) > 0$, so: \newline
\noindent
$(\beta+(1-\gamma))(1+\beta)\frac{P}{P+\gamma N} \geq (1+\beta)(1-\gamma)(\beta\frac{P}{P+\gamma N} + 1)$ \newline
\noindent
The assumed $\beta \geq 0$ ensures $1 + \beta > 0$, so: \newline
\noindent
$(\beta+(1-\gamma))\frac{P}{P+\gamma N} \geq (1-\gamma)(\beta\frac{P}{P+\gamma N} + 1)$ \newline
\noindent
$\beta\frac{P}{P+\gamma N} + (1-\gamma)\frac{P}{P+\gamma N} \geq (1-\gamma)(\beta\frac{P}{P+\gamma N} + 1)$ \newline
\noindent
$\beta\frac{P}{P+\gamma N} + (1-\gamma)\frac{P}{P+\gamma N} \geq (1-\gamma)\beta\frac{P}{P+\gamma N} + (1-\gamma)$ \newline
\noindent
$\beta\frac{P}{P+\gamma N} - (1-\gamma)\beta\frac{P}{P+\gamma N} \geq (1-\gamma)(1 - \frac{P}{P+\gamma N})$ \newline
\noindent
$\beta\frac{P}{P+\gamma N}(1 - (1-\gamma)) \geq (1-\gamma)(1 - \frac{P}{P+\gamma N})$ \newline
\noindent
$\beta\frac{P}{P+\gamma N}\gamma \geq (1-\gamma)(1 - \frac{P}{P+\gamma N})$ \newline
\noindent
$\beta\frac{P}{P+\gamma N}\gamma \geq (1-\gamma)(\frac{P+\gamma N}{P+\gamma N} - \frac{P}{P+\gamma N})$ \newline
\noindent
$\beta\frac{P}{P+\gamma N}\gamma \geq (1-\gamma)\frac{P+\gamma N - P}{P+\gamma N}$ \newline
\noindent
$\beta\frac{P}{P+\gamma N}\gamma \geq (1-\gamma)\frac{\gamma N}{P+\gamma N}$ \newline
\noindent
The assumed $P > 0$, $N \geq 0$, $\gamma \geq 0$ ensure $P + \gamma N > 0$, so: \newline
\noindent
$\beta P \gamma \geq (1-\gamma)\gamma N$ \newline
\noindent
The assumed $P > 0$ allows for: \newline
\noindent
$\beta \gamma \geq (1-\gamma)\frac{\gamma N}{P}$ \newline
\noindent
Assuming additionally $\gamma > 0$ allows for: \newline
\noindent
$\beta \geq (1-\gamma)\frac{N}{P}$ \newline

Intermediate conclusion: given $P > 0$ (the positive class) and $N \geq 0$ (the negative class) the inequality: 
$F_\beta(\left[ \begin{smallmatrix} \mathit{P} & \enskip \mathit{0} \\ \mathit{\gamma N} & \enskip \mathit{(1-\gamma)N} \end{smallmatrix} \right]) \geq  F_\beta(\left[ \begin{smallmatrix} \mathit{(1-\gamma) P} & \enskip \mathit{\gamma P} \\ \mathit{0} & \enskip \mathit{N} \end{smallmatrix} \right])$ is fully defined and holds for $\gamma \in (0,1)$ if $\beta$ is taken to satisfy $\beta \geq (1-\gamma)\frac{N}{P}$.

The two remaining border cases (resulting from additionally assuming $\gamma < 1$ and $\gamma > 0$) are:
\begin{itemize}
	\item $\gamma = 1$: \newline
	\text{} \newline
	\noindent
$F_\beta(\left[ \begin{smallmatrix} \mathit{P} & \enskip \mathit{0} \\ \mathit{\gamma N} & \enskip \mathit{(1-\gamma)N} \end{smallmatrix} \right]) \geq  F_\beta(\left[ \begin{smallmatrix} \mathit{(1-\gamma) P} & \enskip \mathit{\gamma P} \\ \mathit{0} & \enskip \mathit{N} \end{smallmatrix} \right])$ \newline
	\noindent
$F_\beta(\left[ \begin{smallmatrix} \mathit{P} & \enskip \mathit{0} \\ \mathit{N} & \enskip \mathit{0} \end{smallmatrix} \right]) \geq  F_\beta(\left[ \begin{smallmatrix} \mathit{0} & \enskip \mathit{P} \\ \mathit{0} & \enskip \mathit{N} \end{smallmatrix} \right])$ \newline
	\noindent
	which cannot be established, as $p = \frac{0}{0+0}$, and thus \textit{F}$_\beta$, is undefined on the right-hand side. 
	\item $\gamma = 0$: \newline
	\text{} \newline
	\noindent
$F_\beta(\left[ \begin{smallmatrix} \mathit{P} & \enskip \mathit{0} \\ \mathit{0} & \enskip \mathit{N} \end{smallmatrix} \right]) \geq  F_\beta(\left[ \begin{smallmatrix} \mathit{P} & \enskip \mathit{0} \\ \mathit{0} & \enskip \mathit{N} \end{smallmatrix} \right])$ \newline
	\noindent
	which holds trivially, as the argument on both sides is the same \newline 
	\noindent	
	(so, on both sides, either \textit{F}$_\beta$ is undefined or it is defined and equal).
\end{itemize}
Final conclusion: given $P > 0$ (the positive class) and $N \geq 0$ (the negative class) the inequality: 
$$F_\beta(\left[ \begin{smallmatrix} \mathit{P} & \enskip \mathit{0} \\ \mathit{\gamma N} & \enskip \mathit{(1-\gamma)N} \end{smallmatrix} \right]) \geq  F_\beta(\left[ \begin{smallmatrix} \mathit{(1-\gamma) P} & \enskip \mathit{\gamma P} \\ \mathit{0} & \enskip \mathit{N} \end{smallmatrix} \right])$$ is fully defined and holds for every $\gamma \in [0,1]$ if $\beta$ is taken to satisfy $\beta \geq (1-\gamma)\frac{N}{P}$.

This result may be further simplified, because $(1-\gamma)\frac{N}{P}$ changes linearly with $\gamma$ and for $\gamma = 1$: $(1-\gamma)\frac{N}{P} = 0$ (which means that $\beta$ is required to satisfy condition $\beta \geq 0$), while for $\gamma = 0$: $(1-\gamma)\frac{N}{P} = \frac{N}{P}$ (which means that $\beta$ is required to satisfy condition $\beta \geq \frac{N}{P}$). 
Notice that the assumed $P > 0$ and $N \geq 0$ ensure $(1-\gamma)\frac{N}{P} \geq 0$, which subsumes the assumed $\beta \geq 0$.
Setting $\beta$ to satisfy $\beta \geq \frac{N}{P}$ ensures satisfying both conditions.

Summarizing all the considered cases, $\beta \geq \frac{N}{P}$ ensures
$F_\beta(\left[ \begin{smallmatrix} \mathit{P} & \enskip \mathit{0} \\ \mathit{\gamma N} & \enskip \mathit{(1-\gamma)N} \end{smallmatrix} \right]) \geq  F_\beta(\left[ \begin{smallmatrix} \mathit{(1-\gamma) P} & \enskip \mathit{\gamma P} \\ \mathit{0} & \enskip \mathit{N} \end{smallmatrix} \right])$ for every $\gamma \in [0, 1]$ for which both sides of the inequality are defined, which proves that \textit{F}$_{\beta \geq \frac{N}{P}}$ satisfies $ACE$.


\subsection*{Proof of Proposition 2}

Let $P > 0$ (the positive class) and $N > 0$ (the negative class).
\textit{IBA}$_\alpha$(\textit{G-mean}) satisfies the $ACE$ property if 
\textit{IBA}$_\alpha$(\textit{G-mean})$(\left[ \begin{smallmatrix} \mathit{P} & \enskip \mathit{0} \\ \mathit{\gamma N} & \enskip \mathit{(1-\gamma)N} \end{smallmatrix} \right]) \geq$ \textit{IBA}$_\alpha$(\textit{G-mean})$(\left[ \begin{smallmatrix} \mathit{(1-\gamma) P} & \enskip \mathit{\gamma P} \\ \mathit{0} & \enskip \mathit{N} \end{smallmatrix} \right])$ for every $\gamma \in [0, 1]$, provided both sides of the inequality are defined.

\textit{IBA}$_\alpha$(\textit{G-mean}) is a function of \textit{recall} (everywhere below in this subsection: $r$) and \textit{specificity} (everywhere below in this subsection: $s$), and
\begin{itemize}
	\item on the left-hand side: 
	\begin{itemize}
		\item $r = \frac{P}{P+0}$, so under the assumed $P > 0$, $r = 1$,
		\item $s = \frac{(1-\gamma)N}{\gamma N + (1-\gamma)N} = \frac{(1-\gamma)N}{(\gamma + 1 - \gamma)N} = \frac{(1-\gamma)N}{(\gamma + 1 - \gamma)N} = \frac{(1-\gamma)N}{N}$, so under the assumed $N > 0$, $s = 1 - \gamma$,
	\end{itemize}
	\item on the right-hand side: 
	\begin{itemize}
		\item $r = \frac{(1-\gamma)P}{(1-\gamma)P+\gamma P} = \frac{(1-\gamma)P}{(1-\gamma+\gamma)P} = \frac{(1-\gamma)P}{P}$, so under the assumed $P > 0$, $r = 1-\gamma$,
		\item $s = \frac{N}{N+0}$, so under the assumed $N > 0$, $s = 1$.
	\end{itemize}
\end{itemize}
Given $r \in [0,1]$, $s \in [0,1]$ and $\alpha \geq 0$, \textit{IBA}$_\alpha$(\textit{G-mean}) is defined in terms of $r$ and $s$ as:
\textit{IBA}$_\alpha$(\textit{G-mean}) = $(1+\alpha(r - s))r^{\frac{1}{2}}s^{\frac{1}{2}}$.


In result, \textit{IBA}$_\alpha$(\textit{G-mean}) satisfies the $ACE$ property if 
\textit{IBA}$_\alpha$(\textit{G-mean})$(\left[ \begin{smallmatrix} \mathit{P} & \enskip \mathit{0} \\ \mathit{\gamma N} & \enskip \mathit{(1-\gamma)N} \end{smallmatrix} \right]) \geq$ \textit{IBA}$_\alpha$(\textit{G-mean})$(\left[ \begin{smallmatrix} \mathit{(1-\gamma) P} & \enskip \mathit{\gamma P} \\ \mathit{0} & \enskip \mathit{N} \end{smallmatrix} \right])$, which is also expressed as: \newline
\noindent
$(1+\alpha(1-(1-\gamma)))1^{\frac{1}{2}}(1-\gamma)^{\frac{1}{2}} \geq (1+\alpha((1-\gamma)-1)))(1-\gamma)^{\frac{1}{2}}1^{\frac{1}{2}}$ \newline
\noindent
$(1+\alpha\gamma)(1-\gamma)^{\frac{1}{2}} \geq (1-\alpha\gamma)(1-\gamma)^{\frac{1}{2}}$ \newline
\noindent
Assuming additionally $\gamma < 1$, which implies $(1-\gamma)^{\frac{1}{2}} > 0$, and dividing by $(1-\gamma)^{\frac{1}{2}}$ \newline
\noindent
$1+\alpha\gamma \geq 1-\alpha\gamma$ \newline
\noindent
$\alpha\gamma \geq -\alpha\gamma$ \newline
\noindent
$2\alpha\gamma \geq 0$ \newline
Assuming additionally $\gamma > 0$, which implies $2\gamma > 0$, and dividing by $2\gamma$ \newline
\noindent
$\alpha \geq 0$ \newline

Intermediate conclusion: given $P > 0$ (the positive class) and $N \geq 0$ (the negative class) the inequality: 
\textit{IBA}$_\alpha$(\textit{G-mean})$(\left[ \begin{smallmatrix} \mathit{P} & \enskip \mathit{0} \\ \mathit{\gamma N} & \enskip \mathit{(1-\gamma)N} \end{smallmatrix} \right]) \geq$ \textit{IBA}$_\alpha$(\textit{G-mean})$(\left[ \begin{smallmatrix} \mathit{(1-\gamma) P} & \enskip \mathit{\gamma P} \\ \mathit{0} & \enskip \mathit{N} \end{smallmatrix} \right])$ is fully defined and holds for every $\gamma \in (0,1)$ if $\alpha$ is taken to satisfy $\alpha \geq 0$.

The two remaining border cases (resulting from additionally assuming $\gamma < 1$ and $\gamma > 0$) are:
\begin{itemize}
	\item $\gamma = 1$: \newline
	\text{} \newline
	\noindent
	\textit{IBA}$_\alpha$(\textit{G-mean})$(\left[ \begin{smallmatrix} \mathit{P} & \enskip \mathit{0} \\ \mathit{\gamma N} & \enskip \mathit{(1-\gamma)N} \end{smallmatrix} \right]) \geq$ \textit{IBA}$_\alpha$(\textit{G-mean})$(\left[ \begin{smallmatrix} \mathit{(1-\gamma) P} & \enskip \mathit{\gamma P} \\ \mathit{0} & \enskip \mathit{N} \end{smallmatrix} \right])$ \newline
	\noindent
\textit{IBA}$_\alpha$(\textit{G-mean})$(\left[ \begin{smallmatrix} \mathit{P} & \enskip \mathit{0} \\ \mathit{N} & \enskip \mathit{0} \end{smallmatrix} \right]) \geq$ \textit{IBA}$_\alpha$(\textit{G-mean})$(\left[ \begin{smallmatrix} \mathit{0} & \enskip \mathit{P} \\ \mathit{0} & \enskip \mathit{N} \end{smallmatrix} \right])$ \newline
	\noindent
$(1+\alpha(1-(1-1)))1^{\frac{1}{2}}(1-1)^{\frac{1}{2}} \geq (1+\alpha((1-1)-1)))(1-1)^{\frac{1}{2}}1^{\frac{1}{2}}$ \newline
	\noindent
	$0 \geq 0$
	which holds trivially for every $\alpha$,
	\item $\gamma = 0$: \newline
	\text{} \newline
	\noindent
\textit{IBA}$_\alpha$(\textit{G-mean})$(\left[ \begin{smallmatrix} \mathit{P} & \enskip \mathit{0} \\ \mathit{0} & \enskip \mathit{N} \end{smallmatrix} \right]) \geq$ \textit{IBA}$_\alpha$(\textit{G-mean})$(\left[ \begin{smallmatrix} \mathit{P} & \enskip \mathit{0} \\ \mathit{0} & \enskip \mathit{N} \end{smallmatrix} \right])$ \newline
	\noindent
	which holds trivially, as the argument on both sides is the same \newline 
	\noindent	
	(so, on both sides, either \textit{IBA}$_\alpha$(\textit{G-mean}) is undefined or it is defined and equal).
\end{itemize}
Final conclusion: given $P > 0$ (the positive class) and $N > 0$ (the negative class) the inequality: 
\textit{IBA}$_\alpha$(\textit{G-mean})$(\left[ \begin{smallmatrix} \mathit{P} & \enskip \mathit{0} \\ \mathit{\gamma N} & \enskip \mathit{(1-\gamma)N} \end{smallmatrix} \right]) \geq$ \textit{IBA}$_\alpha$(\textit{G-mean})$(\left[ \begin{smallmatrix} \mathit{(1-\gamma) P} & \enskip \mathit{\gamma P} \\ \mathit{0} & \enskip \mathit{N} \end{smallmatrix} \right])$ is fully defined and holds for every $\gamma \in [0,1]$ if $\alpha$ is taken to satisfy $\alpha \geq 0$.

Summarizing all the considered cases, $\alpha \geq 0$ ensures
$$\textit{IBA}_\alpha(\textit{G-mean})(\left[ \begin{smallmatrix} \mathit{P} & \enskip \mathit{0} \\ \mathit{\gamma N} & \enskip \mathit{(1-\gamma)N} \end{smallmatrix} \right]) \geq \textit{IBA}_\alpha(\textit{G-mean})(\left[ \begin{smallmatrix} \mathit{(1-\gamma) P} & \enskip \mathit{\gamma P} \\ \mathit{0} & \enskip \mathit{N} \end{smallmatrix} \right])$$ for every $\gamma \in [0, 1]$ for which both sides of the inequality are defined, which proves that \textit{IBA}$_{\alpha \geq 0}$(\textit{G-mean}) satisfies $ACE$.

\subsection*{Proof of Proposition 3}


Let $M(\alpha,r,s)$, where $M$ is a function of \textit{recall} (everywhere below in this subsection: $r$), \textit{specificity} (everywhere below in this subsection: $s$) and $\alpha$, denote \textit{IBA}$_\alpha$(\textit{G-mean}).

Given $r \in [0,1]$, $s \in [0,1]$ and $\alpha \geq 0$: 

$$M(\alpha,r,s) = 
(1+\alpha(r - s))(r s)^{\frac{1}{2}} = 
(1+\alpha r)r^{\frac{1}{2}}s^{\frac{1}{2}} - \alpha r^{\frac{1}{2}}s^{\frac{3}{2}} =
r^{\frac{1}{2}}((1+\alpha r)s^{\frac{1}{2}} - \alpha s^{\frac{3}{2}})$$.

$M(\alpha,r,s)$ satisfies the $TP_{\nearrow}$ property if it features a weakly monotonic value growth along vertical lines in its cross-sections for $P/n \in (0,1)$, which is equivalent to it being a weakly increasing function of $s \in [0,1]$.

Calculating 
$\frac{\partial M}{\partial s} = 
r^{\frac{1}{2}}(\frac{1}{2}(1+\alpha r)s^{-\frac{1}{2}} - \frac{3}{2}\alpha s^{\frac{1}{2}})$ 
allows for: \newline
\noindent
$\frac{\partial M}{\partial s} \geq 0$ \newline
\noindent
$r^{\frac{1}{2}}(\frac{1}{2}(1+\alpha r)s^{-\frac{1}{2}} - \frac{3}{2}\alpha s^{\frac{1}{2}}) \geq 0$ \newline
\noindent
Assuming additionally $r > 0$, which implies $r^{\frac{1}{2}} > 0$, and dividing by $r^{\frac{1}{2}}$ \newline
\noindent
$\frac{1}{2}(1+\alpha r)s^{-\frac{1}{2}} - \frac{3}{2}\alpha s^{\frac{1}{2}} \geq 0$ \newline
\noindent
Assuming additionally $s > 0$, which implies $2s^{\frac{1}{2}} > 0$, and multiplying by $2s^{\frac{1}{2}}$ \newline
\noindent
$2\frac{1}{2}(1+\alpha r)s^{-\frac{1}{2}}s^{\frac{1}{2}} - 2\frac{3}{2}\alpha s^{\frac{1}{2}}s^{\frac{1}{2}} \geq 0$ \newline
\noindent
$1 + \alpha r - 3\alpha s \geq 0$ \newline
\noindent
$1 + \alpha(r - 3s) \geq 0$

Let $F(\alpha,r,s) = 1 + \alpha r - 3\alpha s$. $F(\alpha,r,s)$ is defined and continuous for $r \in [0,1]$, $s \in [0,1]$ and $\alpha \geq 0$, and treats $r$ and $s$ independently (as indicated by $\frac{\partial F}{\partial s} = \alpha$ and $\frac{\partial F}{\partial s} = -3\alpha$, which are independent of $r$ and $s$). Thus,

\noindent
$F(\alpha,r,s) \geq 0$ \newline
\noindent
$1 + \alpha(r - 3s) \geq 0$ \newline
\noindent
$\alpha(r - 3s) \geq -1$ \newline

Consider $r - 3s$:
\begin{itemize}
\item case $r - 3s = 0$ produces $0 \geq -1$ (holds trivially),
\item case $r - 3s > 0$ produces $\alpha \geq \frac{-1}{r - 3s}$, with $\frac{-1}{r - 3s} < 0$, 
	\noindent
\item case $r - 3s < 0$ produces $\alpha \leq \frac{-1}{r - 3s}$, with $\frac{-1}{r - 3s} > 0$, further resolved into:
	\begin{itemize}
		\item for $r \to 0$ and $s \to 0$: $\frac{-1}{r - 3s} \to \infty$, in which sub-case $\alpha \leq \infty$
		\item for $r \to 0$ and $s = 1$: $\frac{-1}{r - 3s} \to 1/3$, in which sub-case $\alpha \leq 1/3$
		\item for $r = 1$ and $s = 1$: $\frac{-1}{r - 3s} = 1/2$, in which sub-case $\alpha \leq 1/2$
	\end{itemize}
\end{itemize}
The resulting conditions on $\alpha$ are: 
$\alpha \geq \frac{-1}{r - 3s}$ with $\frac{-1}{r - 3s} < 0$,
$\alpha \leq \infty$,
$\alpha \leq 1/3$ and 
$\alpha \leq 1/2$, 
while the assumed condition is $\alpha \geq 0$ (with some of them subsuming some others). 
Setting $\alpha$ to satisfy $\alpha \in [0,1/3]$ ensures satisfying all those conditions.

Intermediate conclusion: given $r \in (0,1]$: $F(\alpha,r,s)$ is non-negative function of $s \in (0,1]$ and $\frac{\partial M}{\partial s}$ is non-negative function of $s \in (0,1]$ and $M(\alpha,r,s)$ is a weakly increasing function of $s \in (0,1]$ if $\alpha$ is taken to satisfy $\alpha \in [0,1/3]$.

The two remaining border cases (resulting from additionally assuming $r > 0$ and $s > 0$) are:
\begin{itemize}
	\item $r = 0$: $M(\alpha,0,s) = 0$ for $\alpha \in [0,1/3]$ and $s \in [0,1]$, so $M(\alpha,r,s)$ is a weakly increasing function of $s \in (0,1]$ (thus also for $r = 0$),
	\item $s = 0$: 
	$M(\alpha,r,0) = 0$ for $\alpha \in [0,1/3]$ and $r \in [0,1]$ (including the above considered $r = 0$), 
	while simultaneously  
	$M(\alpha,r,s) \geq 0$ for $\alpha \in [0,1/3]$ and $r \in [0,1]$ (including the above considered $r = 0$) and $s \in [0,1]$\footnote{Proving $M \geq 0$ is analogous to proving $\frac{\partial M}{\partial s} \geq 0$}, so $M(\alpha,r,s)$ is a weakly increasing function of $s \in [0,1]$ (thus also for $s = 0$).
\end{itemize}
Final conclusion: given $r \in [0,1]$, $M(\alpha,r,s)$ is a weakly increasing function of $s \in [0,1]$ if $\alpha$ is taken to satisfy $\alpha \in [0,1/3]$.

Summarizing all the considered cases, $\alpha \in [0,1/3]$ ensures the weakly increasing character of $M(\alpha,r,s)$ = \textit{IBA}$_\alpha$(\textit{G-mean}) as a function of $s \in [0,1]$ for any $r \in [0,1]$, being equivalent to featuring a weakly monotonic value growth along vertical lines in its cross-sections for $P/n \in (0,1)$, which proves that \textit{IBA}$_\alpha$(\textit{G-mean}) satisfies $TP_{\nearrow}$.


\end{document}